# OSIRIS-REx: Sample Return from Asteroid (101955) Bennu

D.S. Lauretta[1], S.S. Balram-Knutson[1], E. Beshore[1], W.V. Boynton[1], C. Drouet d'Aubigny[1], D.N. DellaGiustina[1], H.L. Enos[1], D.R. Gholish[1], C.W. Hergenrother[1], E.S. Howell[1], C.A. Johnson[1], E.T. Morton[1], M.C. Nolan[1], B. Rizk[1], H.L. Roper[1], A.E. Bartels[2], B.J. Bos[2], J.P. Dworkin[2], D.E. Highsmith[2], D.A. Lorenz[2], L.F. Lim[2], R. Mink[2], M.C. Moreau[2], J.A. Nuth[2], D.C. Reuter[2], A.A. Simon[2], E.B. Bierhaus[3], B.H. Bryan[3], R. Ballouz[4], O.S. Barnouin[5], R.P. Binzel[6], W.F. Bottke[7], V.E. Hamilton[7], K.J. Walsh[7], S.R. Chesley[8], P.R. Christensen[9], B.E. Clark[10], H.C. Connolly[11], M.K. Crombie[12], M.G. Daly[13], J.P. Emery[14], T.J. McCoy[15], J.W. McMahon[16], D.J. Scheeres[16], S. Messenger[17], K. Nakamura-Messenger[17], K. Righter[17], S.A. Sandford[18]

[1]*Lunar and Planetary Laboratory, University of Arizona, Tucson, AZ, USA*

(lauretta@lpl.arizona.edu)

[2]*NASA Goddard Space Flight Center, Greenbelt, MD, USA*

[3]*Lockheed Martin Space Systems, Littleton, CO, USA*

[4]*Department of Astronomy, University of Maryland, College Park, MD, USA*

[5]*The Johns Hopkins University, Applied Physics Laboratory, Laurel, MD, USA*

[6]*Massachusetts Institute of Technology, Cambridge, MA, USA*

[7]*Southwest Research Institute, Boulder, CO, USA*

[8]*Jet Propulsion Laboratory, Pasadena, CA, USA*

[9]*Arizona State University, Tempe, AZ, USA*

[10]*Ithaca College, Ithaca, NY, USA*

[11]*Rowan University, Glassboro, New Jersey, USA*

[12]*Indigo Information Services, Tucson, AZ, USA*

[13]*Centre for Research in Earth and Space Science, York University, Toronto, Ontario, M3J1P3, Canada*

[14]*University of Tennessee, Knoxville, TN, USA*

[15]*Smithsonian National Museum of Natural History, Washington, DC, USA*

[16]*University of Colorado, Boulder, CO, USA*

[17]*NASA Johnson Space Center, Houston, TX, USA*

[18]*NASA Ames Research Center, Moffett Field, CA, USA*

**Abstract** In May of 2011, NASA selected the **O**rigins, **S**pectral **I**nterpretation, **R**esource **I**dentification, and **S**ecurity–**R**egolith **Ex**plorer (OSIRIS-REx) asteroid sample return mission as the third mission in the New Frontiers program. The other two New Frontiers missions are *New Horizons*, which explored Pluto during a flyby in July 2015 and is on its way for a flyby of Kuiper Belt object 2014 MU69 on Jan. 1, 2019, and *Juno*, an orbiting mission that is studying the origin, evolution, and internal structure of Jupiter. The spacecraft departed for near-Earth asteroid (101955) Bennu aboard an United Launch Alliance Atlas V 411 evolved expendable launch vehicle at 7:05 p.m. EDT on September 8, 2016, on a seven-year journey to return samples from Bennu. The spacecraft is on an outbound-cruise trajectory that will result in a rendezvous with Bennu in August 2018. The science instruments on the spacecraft will survey Bennu to measure its physical, geological, and chemical properties, and the team will use these data to select a site on the surface to collect at least 60 g of asteroid regolith. The team will also analyze the remote-sensing data to perform a detailed study of the sample site for context, assess Bennu's resource potential, refine estimates of its impact probability with Earth, and provide ground-truth data for the extensive astronomical data set collected on this asteroid. The spacecraft will leave Bennu in 2021 and return the sample to the Utah Test and Training Range (UTTR) on September 24, 2023.

# Acronyms

| | |
|---|---|
| AAM | Asteroid Approach Maneuver |
| APID | Application Process Identifier Definition |
| ASIST | Advanced System for Integration and Spacecraft Test |
| ATLO | Assembly, Test, and Launch Operations |
| AU | Astronomical Unit: the average distance between the Earth and the Sun |
| C3 | Characteristic energy of launch |
| Delta-DOR | Delta-Differential One-way Ranging |
| DLA | Declination of the Launch Asymptote |
| DRA | Design Reference Asteroid |
| DRM | Design Reference Mission |
| DSM | Deep Space Maneuver |
| DSN | Deep Space Network |
| ECAS | Eight-Color Asteroid Survey |
| EGA | Earth gravity assist |
| FDS | Flight Dynamics System |
| FEDS | Front End Data System |
| FOB | Flight Operations Bucket |
| GM | universal coefficient of gravity (G) multiplied by the mass of a planetary object (M) |
| GN&C | Guidance, Navigation, and Control |
| GRAIL | Gravity Recovery and Interior Laboratory mission |
| I/F | Ratio of reflected energy to incoming energy (i.e. irradiance/solar flux) |
| IEST | Institute of Environmental Sciences and Technology |
| ISO | International Organization for Standardization |
| ISVM | Integrated Global Science Value Map |
| JPL | Jet Propulsion Laboratory |
| JSC | NASA Johnson Space Center |
| LIDAR | Light Detection and Ranging |
| LINEAR | Lincoln Near-Earth Asteroid Research survey |
| MAVEN | Mars Atmosphere and Volatile EvolutioN mission |
| MIC | Microparticle Impact Collection |
| MRO | Mars Reconnaissance Orbiter |
| MSA | Mission Support Area |
| NASA | National Aeronautics and Space Administration |
| NEA | Near-Earth asteroid |
| NEAR | Near-Earth Asteroid Rendezvous mission |
| NFT | Natural Feature Tracking |
| OCAMS | OSIRIS-REx Camera Suite |
| OLA | OSIRIS-REx Laser Altimeter |
| OSIRIS-REx | Origins, Spectral Interpretation, Resource Identification, and Security–Regolith Explorer |
| OTES | OSIRIS-REx Thermal Emission Spectrometer |
| OVIRS | OSIRIS-REx Visible and Infrared Spectrometer |
| PDS | Planetary Data System |
| PET | Preliminary-examination Team |
| PSFD | Particle Size Frequency Distribution |
| REXIS | Regolith X-ray Imaging Spectrometer |
| RGB | Red, Green, Blue |
| RLA | Right Ascension of the Launch Asymptote |
| ROI | Region of Interest |
| SAP | Sample Analysis Plan |
| SARA | Sample Acquisition and Return Assembly |
| SPC | stereophotoclinometry |
| SPICE | S- Spacecraft ephemeris; P- Planet, satellite, comet, or asteroid ephemerides; I- Instrument description kernel; C- Pointing kernel; E- Events kernel |
| SPK | Spacecraft and Planet Kernel |
| SPOC | Science Processing and Operations Center |
| SRC | Sample Return Capsule |
| TAG | Touch and Go |
| TAGCAMS | Touch and Go Camera System |
| TAGSAM | Touch and Go Sample Acquisition Mechanism |
| UTTR | Utah Test and Training Range |
| Vis-NIR | Visible and Near-Infrared |
| YORP | Yarkovsky–O'Keefe–Radzievskii–Paddack effect |

# 1 Introduction

OSIRIS-REx is led by Principal Investigator Dante Lauretta at the University of Arizona in partnership with Lockheed Martin Space Systems, the NASA Goddard Spaceflight Center, the Canadian Space Agency, Arizona State University, KinetX Aerospace, the Massachusetts Institute of Technology, and the NASA Johnson Space Center. Science team participation involves researchers from the United States, Canada, France, Japan, the United Kingdom, Spain, Italy, and the Czech Republic.

[Insert Figure 1 here]

The spacecraft launched (Figure 1) on an outbound-cruise trajectory that will result in a rendezvous with Bennu in August 2018 (Figure 2). The science instruments on the spacecraft will survey Bennu to measure its physical, geological, and chemical properties, and the team will use these data to select a site on the surface to collect at least 60 g of asteroid regolith. The team will also analyze the remote-sensing data to perform a detailed study of the sample site for context, assess Bennu's resource potential, refine estimates of its impact probability with Earth, and provide ground-truth data for the extensive astronomical data set collected on this asteroid. The spacecraft will leave Bennu in 2021 and return the sample to the Utah Test and Training Range (UTTR) on September 24, 2023 (Figure 3).
[Insert Figure 2 here]
[Insert Figure 3 here]

# 2 Mission Objectives

Asteroids are geologic remnants from the early Solar System. The prime objective of the OSIRIS-REx mission is to return pristine carbonaceous regolith from Bennu to understand both the role that primitive asteroids may have played in the origin of life on Earth and how they served as one of the fundamental "building blocks" of planet formation.

The mission has several secondary science objectives. One of the great values of sample return lies in the knowledge of sample context. The mission will provide an extensive global data set of Bennu along with thorough documentation of the sample site. Surface processes that affect the origin of the sample, and which act across the asteroid, will be investigated at unprecedented resolutions. The mission team will also study the Yarkovsky effect, improving our ability to predict the long-term ephemerides of objects like Bennu, whose orbits cross the path of the Earth. This work can help better predict its future trajectory and determine whether an impact is a likely outcome within the next several hundreds of years. In addition, near-Earth asteroids (NEAs) harbor resources such as water and organic molecules that can support future missions of Solar System exploration. OSIRIS-REx will assess the resource potential of Bennu and extrapolate this knowledge to other accessible NEAs. Finally, the OSIRIS-REx mission provides ground-truth data for the extensive telescopic observations of Bennu, permitting an assessment of the fidelity of these astronomical techniques for asteroid characterization. These science investigations are codified in five mission science objectives.

## 2.1 Objective 1—Return and Analyze a Sample

*Return and analyze a sample of pristine carbonaceous asteroid regolith in an amount sufficient to study the nature, history, and distribution of its constituent minerals and organic material*

One of the most important questions in planetary science today relates to the sources of Earth's organic materials and water, which may have influenced the origin and early evolution of life. Were these materials present in the terrestrial planet region during the earliest planet formation era? Were they delivered to the Earth after the Moon-forming impact event by a late phase of bombardment, or can we rule out their contribution to Earth's earliest history after a thorough study of their properties? The carbonaceous asteroids, suspected to be chemically unaltered since the formation of the Solar System, may hold the answers to these questions (Lauretta and McSween 2006). Definitive answers will result from direct examination of samples using a variety of laboratory analytical methods. Sample return also provides an archive of material for the next generation of scientists to study using laboratory techniques not available today.

## 2.2 Objective 2—Map Bennu's Global Properties

*Map the global properties, chemistry, and mineralogy of a primitive carbonaceous asteroid to characterize its geologic and dynamic history and provide context for the returned samples*

The first goal of any explorer upon arriving at a new location is a global survey and mapping campaign to put subsequent observations in proper context. During the early phases of asteroid proximity operations, the science team will develop a series of global maps to characterize the geology, mineralogy, surface processes, and dynamic state of Bennu. Such a survey will also help the team constrain how Bennu found its way onto an Earth-like orbit from the main asteroid belt, and provide important clues to what happened to it during the dynamical evolution into its current orbit. Most importantly, these global maps guide sample-site selection and allow the team to place the returned samples into geological context.

## 2.3 Objective 3—Document the Sample Site

*Provide sample context by documenting the regolith at the sampling site in situ at scales down to the sub-centimeter*

To select a sample site, the global mapping campaign will be followed by detailed observations of candidate sites to determine if a sampling attempt will be safe, that material for sampling is present, and that the site represents an area of high scientific interest. Careful observations of prospective sampling sites at high resolution will allow the team to extensively document the nature of the region where the returned sample was collected. In addition, the nature of asteroid regolith and how it has evolved via collisions, downslope movement, and nongravitational forces is an area of active research (e.g., Walsh et al. 2012; Mazrouei et al. 2014; Rozitis et al. 2014; Sánchez and Scheeres 2014). By characterizing the surface of Bennu at sub-cm scales, the team will gain a deeper understanding of how material behaves in the unique microgravity environment of a small asteroid.

## 2.4 Objective 4—Study the Yarkovsky Effect

*Understand the interaction between asteroid thermal properties and orbital dynamics by measuring the Yarkovsky effect on a potentially hazardous asteroid and constraining the asteroid properties that contribute to this effect*

Rotating asteroids absorb sunlight and reradiate it in the thermal infrared. For sub-km-sized asteroids like Bennu, this radiation absorption and reemission results in a small but continuous force applied in a preferential direction. This force is the result of the Yarkovsky effect, and its magnitude and direction depend on several factors that OSIRIS-REx will directly measure: asteroid size, shape, mass, obliquity, spin state, albedo, and surface thermal conductivity. The Yarkovsky effect is important because, over tens to hundreds of Myr or more, it can drive Bennu-sized bodies within the main asteroid belt into orbital resonances with the planets. From there, planetary gravitational perturbations can increase the eccentricity of the objects enough that they reach planet-crossing orbits (Figure 4). It is now recognized that this thermal drift force is the dominant factor in moving small asteroids from the main asteroid belt to the Earth-crossing object population (e.g., Vokrouhlický et al. 2015). This force also limits our ability to predict the future position of these objects, which in turn increases the uncertainties we have to cope with in making impact predictions. For this reason, understanding how the Yarkovsky effect is currently modifying Bennu's orbit will improve our ability to precisely determine how this force affects the orbits of other potentially hazardous asteroids.
[Insert Figure 4 here]

### 2.5 Objective 5—Improve Asteroid Astronomy

*Improve asteroid astronomy by characterizing the astronomical properties of a primitive carbonaceous asteroid to allow for direct comparison with ground-based telescopic data of the entire asteroid population.*
The extensive telescopic observations of Bennu between 1999 and 2012 have made it one of the best-characterized NEAs not yet visited by spacecraft (Clark at al. 2011; Müller et al. 2012; Hergenrother et al. 2013, 2014; Nolan et al. 2013; Chesley et al. 2014; Emery et al. 2014; Binzel et al. 2015; Lauretta et al. 2015; Takir et al. 2015; Yu and Ji 2015; Scheeres et al. 2016). The OSIRIS-REx mission thus affords an exceptional opportunity to compare our Earth-based observations with spacecraft-based measurements. During mission development, the team compiled all measured properties in a document called the Design Reference Asteroid (DRA) (Hergenrother et al. 2014). The DRA contains our best estimates for many physical parameters that span a number of categories including: orbital parameters, bulk properties, rotation state, radar scattering properties, photometric properties, spectroscopic signatures, thermal state, surface analogs, and environmental characteristics. In total, the team constrained 111 parameters that describe the properties of Bennu. The parameters were directly measured from telescopic observations of Bennu, based on models, or inferred from the properties of analogous small Solar System bodies and meteorites. The remote-sensing data that this mission will obtain allows for a comparison of the knowledge of the properties of Bennu with post-encounter knowledge. This comparison will help the team to improve the techniques used for astronomical characterization of asteroids and better interpret the observations of other asteroids made over the last century.

## 3 Target: Asteroid Bennu

Selecting a suitable target for the OSIRIS-REx mission began with the 500,000 or so asteroids known at the time New Frontiers proposals were submitted in 2010 (Figure 5). Consideration of available launch-vehicle capabilities and mission timelines led the team to conclude that NEAs were the only feasible options for a sample-return mission target. At the time of target selection, over 7,000 NEAs were known. Further analysis revealed that of these, only 192 had orbits that were optimal for a sample-return mission within the constraints of the New Frontiers program, spacecraft design, and return capsule reentry velocity. Once the orbital constraints had winnowed the list, the team looked to asteroid size to further refine the options. Objects smaller than 200 m often have fast rotation rates (<2 hours) or exhibit non-principal-axis rotation (tumbling) that could complicate proximity operations. The team also thought that smaller, rapidly rotating objects might be deficient in the kind of surface regolith material needed for sampling. Of the remaining objects, only 26 had diameters greater than 200 m. Once all engineering constraints were met, the science objective of returning pristine carbon-rich material was the deciding factor. Of the final 26 candidate objects, only five had spectral features consistent with a carbon-rich (presumably carbonaceous chondrite-like) composition. These were likely primitive asteroids that had seen minimal alteration within their parent bodies since Solar System formation 4.5 billion years ago. It was these objects that were likely to yield the answers sought to the fundamental question this mission seeks to address about the earliest stages of planetary formation. Bennu rose to the top of the list based on the extensive data set obtained from ground- and space-based astronomical characterization (Hergenrother et al. 2014; Lauretta et al. 2015). The detailed knowledge on the size, shape, and rotation state obtained from ground-based radar characterization was a critical deciding factor in the selection of Bennu as the OSIRIS-REx mission target (Nolan et al. 2013).

[Insert Figure 5 here]

Asteroid (101955) Bennu was discovered on September 11, 1999 (Williams 1999) by the Lincoln Near-Earth Asteroid Research (LINEAR) survey using a 1.0-m telescope located in Socorro, New Mexico (Stokes et al. 2000). Bennu is a member of the rare B-type spectral subgroup of carbonaceous asteroids (Clark et al. 2011). Bennu is also one of the most potentially hazardous of all the currently known NEAs, based on its size and calculable nonzero probability of future impacts with Earth (Chesley et al. 2014). The name "Bennu" was contributed in 2013 by then 9-year-old Michael Puzio, who won a contest led by the OSIRIS-REx mission, the LINEAR survey, and the Planetary Society.

Bennu is one of the best characterized NEAs due to the significant number of optical and radar observations that have been collected since its discovery in 1999. Bennu has been extensively studied in the visible and infrared spectra (Clark et al. 2011; Müller et al. 2012; Emery et al. 2014; Binzel et al. 2015). Ground-based radar has been used to characterize its shape, spin state, and surface roughness (Figure 6) (Nolan et al. 2013). These observations have led to precise measurement of its rotational period (4.2978 hours) and confirmed that its rotation is retrograde with an obliquity of 176±4°. No non-principal-axis rotation has been detected. The radar-derived shape model and measurements of the Yarkovsky force on Bennu have led to the current estimate for Bennu's bulk density of 1260±70 $kg/m^3$, which yields a range of possible GM values of 5.2±0.6 $km^3/s^2$ (Chesley et al. 2014).

[Insert Figure 6 here]

Bennu's relatively low-inclination, Apollo-type Earth-crossing orbit makes it accessible with relatively low energy (Figure 7) (Delta-V = 5.087 km/s, according to the Jet Propulsion Laboratory (JPL) Echo website). Bennu's orbit is slightly inclined to the ecliptic plane by 6 degrees, and its orbit semimajor axis is 1.13 AU, yielding an orbit period of 1.20 years and a synodic period with respect to Earth of 6 years. Bennu's orbit is somewhat elongated, with an eccentricity of 0.204 and perihelion and aphelion distances of 0.897 AU and 1.356 AU, respectively.

[Insert Figure 7 here]

Bennu is an exciting target for an asteroid sample return mission. It is likely the fragment of an ancient object larger than 100 km in diameter that formed over 4.5 Ga (Lauretta et al. 2015). Its chemistry and mineralogy were established within the first 10 Ma of the Solar System. Its spectrum, low albedo, and low density (Clark et al. 2011; Hergenrother et al. 2013; Chesley et al. 2014) suggest a primitive nature that is relatively unprocessed since formation. It is likely a direct remnant of the original building blocks of the terrestrial planets and may represent the type of object that served as an important source of volatiles and organic matter for Earth during its formative era.

Bennu is different from all other NEAs previously visited by spacecraft. Asteroid (433) Eros, target of the NEAR-Shoemaker mission, and (25143) Itokawa, target of the Hayabusa mission, are higher-albedo, S-type asteroids with irregular shapes. In contrast, Bennu has a very low albedo, is spectral B-type, and has a distinct spheroidal shape. While Eros and Itokawa are similar to ordinary chondrite meteorites (McCoy et al. 2002; Foley et al. 2006; Nakamura et al. 2011), Bennu is likely related to carbonaceous chondrites, meteorites that preserve records of volatiles and organic compounds in the early Solar System (Clark et al. 2011). The closest match to Bennu visited so far by spacecraft is (253) Mathilde, a ~60 km diameter C-type main belt asteroid possessing a Bennu-like bulk density near ~1300 kg/m$^3$,

Bennu was probably part of a larger (>100-km), carbonaceous parent asteroid in the inner main belt that was catastrophically disrupted 0.7–2 Ga ago (Walsh et al. 2013; Bottke et al. 2015). The fragment that became Bennu was eventually delivered to near-Earth space via a combination of Yarkovsky-induced drift and interaction with giant-planet gravitational resonances (Campins et al. 2010; Bottke et al. 2015). During its journey, solar thermal torques and possibly planetary close encounters may have modified Bennu's spin state, potentially reshaping and resurfacing the asteroid (Scheeres et al. 2016). This inferred history and dynamical evolution provides the driving hypotheses for scientific analysis of the remote-sensing data and laboratory investigation of the returned samples. Both of these complementary science investigations are enabled by the highly capable OSIRIS-REx spacecraft.

# 4 Spacecraft Overview

The spacecraft is composed of the following principal components: the spacecraft bus (containing the spacecraft structure and all supporting subsystems for the operation and control of the vehicle), the Touch and Go Sample Acquisition Mechanism (TAGSAM), the Sample Return Capsule (SRC), and the five science instruments responsible for the remote-sensing campaign at Bennu (Figure 8).

[Insert Figure 8 here]

The spacecraft (Bierhaus et al. 2017) builds on Lockheed Martin Space Systems Company hardware, software, technology, and processes established from the Stardust, Odyssey, MRO (Mars Reconnaissance Orbiter), Juno, GRAIL (Gravity Recovery and Interior Laboratory), and MAVEN (Mars Atmosphere and Volatile EvolutioN) spacecraft (Figure 9). The flight system architecture was developed in concert with science, instruments, subsystems, mission operations, navigation, and management. The resulting design was populated with heritage subsystems and components (Figure 10). The spacecraft is single-fault tolerant with block, functional, and subsystem internal redundancies with appropriate cross-strapping, autonomous fault detection, isolation, and recovery. Instrument accommodations meet all pointing, power, thermal, and data handling requirements with significant margins. The flight system design was evaluated for its ability to execute the Design Reference Mission (DRM) within the constraints of the launch vehicle capability, trajectory, communications, and ground systems. The spacecraft design pulls from previous spacecraft designs to create a flight system that is fully capable of achieving the mission at a low level of risk. This ability to select the "best of the best" enabled the team to select heritage components and designs without overconstraining the mission.

[Insert Figure 9 here]

[Insert Figure 10 here]

The spacecraft structure consists of aluminum honeycomb sandwiched between graphite composite face sheets. The spacecraft body frame coordinate system is shown in Figure 9. The core of the structure is a 1.3-meter diameter cylinder that encloses the propellant tank of the spacecraft. Two decks, the forward and aft deck, are installed on the top and bottom of the central cylinder with the upper panel supporting the Sample Acquisition and Return Assembly (SARA), the science instruments, navigation equipment, and antennas. The aft deck houses the batteries, medium gain antenna, reaction wheels, and solar array gimbals. The high-gain antenna is centered on the +X axis of the spacecraft.

The spacecraft hosts a total of 28 engines divided into four groups: a bank of four high-thrust main engines, six medium-thrust engines, 16 attitude control thrusters, and a pair of specialized low-thrust engines. All thrusters installed on the spacecraft are fed from a central propellant tank holding the hydrazine supply needed for the mission.

The spacecraft has solar panels affixed to the aft deck of the spacecraft. Covered with gallium-arsenide solar cells, the arrays deliver between 1,226 and 2,500 watts of electrical power depending on the spacecraft's distance from the sun, which varies over the course of the seven-year mission. The solar arrays are attached to the spacecraft structure with two-axis gimbals, allowing the arrays to be moved into a range of configurations depending on the mission phase. During sample acquisition, the arrays are translated into a Y-wing configuration to avoid dust accumulating on the solar cells and increase the ground clearance in case the spacecraft rotates during contact with the asteroid surface.

The spacecraft is designed to ensure that it delivers the TAGSAM sampling head to within 25 m of a selected point on the surface with greater than 98% probability, assuming the asteroid environmental conditions described in the Design Reference Asteroid document (Hergenrother et al. 2014). To meet this requirement, the spacecraft is equipped with two independent autonomous guidance systems for the final closure with the asteroid surface during sample acquisition: LIDAR-guided TAG and Natural Feature Tracking (NFT). The LIDAR-guided TAG technique relies on a LIDAR (Light Detection and Ranging) system, which is part of the Guidance, Navigation, and Control (GN&C) system, to detect the time of a range-threshold crossing, followed by a range measurement at a specific time to make corrections to the propulsive maneuvers required to safely reach the asteroid surface (Berry et al. 2013).

NFT uses a catalog of known features built from an asteroid shape model that is produced during flight as more information about Bennu is collected (Olds et al. 2015; Mario et al. 2016). During the sample collection phase of the mission, these features are rendered using a predicted camera pointing and Sun position, and correlated against real-time images of the asteroid surface. The results of this correlation are then used to provide a state update of the spacecraft's position and velocity relative to the asteroid surface, which are then used to calculate the required corrections to the sampling propulsive maneuvers. Since LIDAR-guided TAG and NFT have complementary capabilities, the team will evaluate which combination of these techniques provides for the lowest risk approach to sample acquisition as part of the sample-acquisition sequence design process.

# 5 Science Payload

Key to achieving the science requirements is a set of scientific data products derived from observations made by a specially designed suite of instruments. These include an imaging camera suite, a visible and near-infrared spectrometer, a thermal-emission spectrometer, an imaging LIDAR system, and an X-ray-emission spectrometer that was designed and built by students (Figure 11).

[Insert Figure 11 here]

## 5.1 Sample Acquisition and Return Assembly (SARA)

The sample is acquired using the Touch-and-Go Sample Acquisition Mechanism, or TAGSAM (Figure 12) (Bierhaus et al. 2017). TAGSAM is a mechanical sampling device that consists of two major components: a sampler head and an articulated positioning arm. The arm extends 2.8 meters from the spacecraft. TAGSAM acquires the bulk sample by releasing a jet of high-purity nitrogen gas that excites and “fluidizes” at least 60 g of regolith into the collection chamber. A number of ground tests and “reduced-gravity” flight tests (where a reduced-gravity flight involves the research airplane flying a parabolic trajectory, providing periods of reduced acceleration), using asteroid surface simulants, have routinely demonstrated collection of over 600 g. The baseplate of the TAGSAM head contains 24 contact-pad samplers made of stainless steel Velcro®. These pads collect small grains up to 1-mm diameter upon contact with the asteroid surface. The TAGSAM subsystem has three separate bottles of nitrogen gas, providing the capacity to make three separate sampling attempts. However, the baseline plan is to execute one successful sampling event. If sufficient sample is collected on the first attempt, the team will not make an additional attempt to collect more or a different type of sample.

[Insert Figure 12 here]

The SRC employed by OSIRIS-REx is based on that of Stardust. The SRC is a blunt-nosed cone, 81 centimeters in diameter, 50 centimeters tall, and composed of five principal components: the heat shield, backshell, sample canister, parachute system, and avionics. The total mass of the capsule is 46 kg. Relative to the Stardust design, the team made changes in the ballast, added contamination witness plates, and changed the spring seal. The SRC also features changes to the avionics and the sample stowage deck (Figure 13). Instead of the Stardust aerogel grid, the SRC contains a capture ring to lock in the TAGSAM head during stowage.

The heatshield and backshell provide thermal protection of the SRC during atmospheric entry. A seal between the heatshield and backshell protects against a heat leak during entry. Two vents on the side of the backshell allow depressurization during ascent and repressurization during descent to preclude unacceptable pressure loading within the SRC. Mechanisms provide the capability to open the clamshell and deploy the TAGSAM.

The sample canister is designed as an integral part of the SRC structure. The lower part of the sample canister is attached to the heatshield, while the upper part doubles as the avionics deck. When the SRC is opened, the sample canister is opened as well. When the SRC is closed, a seal between the avionics deck and the lower part of the sample canister protects the enclosed TAGSAM from external contamination. A vent with a multilayer gas filter is located in the base of the lower part of the sample canister. An inlet is provided on the avionics deck to allow attachment of a purge line to the closed sample canister upon recovery. After attachment of a suitable gas source, ultrapure gaseous nitrogen can be made to flow at a low rate into the sample canister. It will then flow out of the canister through the filter and exit the SRC through the backshell vents.

The parachute system includes a drogue chute and its deployment mortar, a main chute, and a parachute canister that houses the entire parachute system. The drogue chute is deployed at Mach 1.4 and approximately 30-km altitude to stabilize the SRC through the transonic and subsonic regimes. At approximately 3-km altitude the main chute is deployed, the drogue is released, and the SRC slows its descent to 4.6 m/s. A cutter in the riser is commanded at ground impact by a 10-*g* G-switch, separating the main chute from the SRC to prevent surface winds from dragging the SRC across the ground.
[Insert Figure 13 here]
The TAGSAM and SRC were fabricated with stringent contamination-control protocols to ensure that the sample will be collected and kept pristine (Dworkin et al. 2017). The OSIRIS-REx Contamination Control plan established procedures to limit the total contamination burden on the sample via proxies to species of scientific interest by limiting sensitive surfaces to cleanliness levels established in IEST-STD-CC1246D at the 100A/2 level (Borson 2005). In addition, due to the unique mission science objectives related to trace organic molecular analysis, TAGSAM and SRC cleanliness was monitored to ensure that no surface accumulated more than 180 ng/cm$^2$ total amino acids and hydrazine. Furthermore, throughout development an extensive catalog of witness plates was collected during spacecraft assembly, testing, and launch operations (ATLO). The witness plates, as well as an extensive collection of materials coupons (>300 items), are archived at the Astromaterials Acquisition and Curation Office at NASA Johnson Space Center. Finally, both the TAGSAM and the SRC have witness plates mounted in strategic locations to document any contamination collected in flight during different phases of the mission.

## 5.2 The OSIRIS-REx Camera Suite (OCAMS)

OCAMS consists of three cameras and a shared, redundant camera control module (Figure 14) (Rizk et al. 2017). The three cameras meet all imaging requirements over a distance ranging from 500K km down to 2 meters from the surface. Built by the University of Arizona, all OCAMS cameras employ 1K × 1K detectors and identical, passively cooled focal plane electronics.
[Insert Figure 14 here]
MapCam is a medium-field imager, with a 125-mm-focal-length f/3.3 optical system that provides a ~70 mrad (4 deg) field of view. MapCam will be used to search for natural satellites and dust plumes during approach and will provide images needed for base maps, global shape model development, and spin-state measurements. It is equipped with a filter wheel that contains four filters (b′, v, w, and x, centered at 470, 550, 770, and 860 nm wavelengths, respectively) that map to the Eight-Color Asteroid Survey (ECAS) filter system, the standard for ground-based broad-band spectrophotometry of asteroids. These wavelength regions characterize the broad spectral features observed on a wide variety of carbonaceous asteroids and provide direct comparison with ground-based observations. MapCam filters will also enable the production of color images and, with its ability to produce color indices associated with some key spectral features, can act as a partial backup for the loss of the visible–infrared spectrometer. In addition, one of the filter positions is a refocusing plate that allows MapCam to view the surface in focus from a range of 30 m.
PolyCam is a narrow-field, 630-mm-focal-length (at ∞ range) f/3.15 Ritchey-Chretien telescope that will provide the images used for producing boulder maps, and for close scrutiny of the sample sites during the Orbit-B and Reconnaissance phases of the mission. During orbital and reconnaissance phases, the close proximity to, and the changing distance of the spacecraft from, Bennu's surface during imaging require that PolyCam have a variable refocusing ability. This feature is implemented by translating one of a doublet of field-correcting lenses in front of the detector.

SamCam is a wide-angle camera that provides context imaging during reconnaissance passes and records the sampling event. It has a 24-mm-focal-length f/5.6 optical system and, like MapCam, is equipped with a filter wheel. The filter wheel contains three clear filters that protect the objective from dust and damage during sampling, and can be changed in the event a second or third attempt is required. In addition, a simple optical diopter in the fourth filter-wheel position can change focus for inspection of the TAGSAM head before stowage.

### 5.3 The OSIRIS-REx Laser Altimeter (OLA)

LIDAR instruments provide direct measurement of spacecraft range to a planetary surface. OLA is a scanning LIDAR system provided by the Canadian Space Agency and built by MacDonald, Dettwiler and Associates of Canada (Figure 15) (Daly et al. 2017). OLA will scan the entire surface of the asteroid to create a highly accurate, 3D model of Bennu, which will provide fundamental and unprecedented information on the asteroid's shape, topography, surface processes, and evolution. With two laser transmitters, OLA can range to Bennu from beyond 7 km at a 100-Hz measurement rate and as close as 50 m at a 10-kHz measurement rate. The unique OLA scanning capability provides the flexibility to provide dense, global coverage of the entire surface, including ≤5-cm resolution maps of proposed sample sites.
[Insert Figure 15 here]

### 5.4 The OSIRIS-REx Visible and Infrared Spectrometer (OVIRS)

OVIRS was built by the Goddard Space Flight Center (Figure 16) (Reuter et al. 2017). It is a point spectrometer with a 4-mrad field of view that provides coverage in the range of 0.4 to 4.3 μm. This wavelength range covers spectral bands needed to identify key mineral species, water, and organic materials. When it executes its global survey observations during Detailed Survey, surface resolutions for an OVIRS spot will be 20 m. OVIRS is passively cooled and uses a novel linear-variable filter design to cover its wavelength range. At 1 μm, resolving power is about 200 ($\Delta\lambda/\lambda$). OVIRS spectra will be used to identify volatile- and organic-rich regions of Bennu's surface and guide sample-site selection.
[Insert Figure 16 here]

### 5.5 The OSIRIS-REx Thermal Emission Spectrometer (OTES)

Built by Arizona State University, OTES is a compact, spot Fourier-transform infrared interferometric spectrometer and is based on similar designs used for instruments flown on the Mars Exploration Rovers and Mars Global Surveyor (Figure 17) (Christensen et al. 2017). OTES will measure the absolute flux of thermally emitted radiation from 5.5 to 50 μm with an accuracy of better than 3 percent. From the Detailed Survey distance of 5 km, the OTES 8-mrad field of view will translate to a spot size of 40 m on the surface.
[Insert Figure 17 here]

### 5.6 The Regolith X-Ray Imaging Spectrometer (REXIS)

REXIS is a student experiment whose primary goal is the education of science and engineering students (Figure 18) (Binzel et al. 2017). REXIS was designed and built through a collaborative effort between students and faculty at the Massachusetts Institute of Technology and Harvard University. REXIS will measure solar X-ray-induced fluorescence from Bennu to map the distribution of elements across its surface. REXIS is capable of detecting fluorescence photons of 0.5–7.5 keV energy, enabling the detection of O, Fe, and Mg. REXIS employs coded aperture imaging, which is similar to the operation of a pinhole camera. When light shines on a pinhole, the direction of the incident light can be determined by the position of the pinhole's light. Coded aperture imaging uses many holes arranged in a known pattern on a mask, allowing the direction of the reemitted X-rays to be determined and mapped to a position on the asteroid surface.
[Insert Figure 18 here]

### 5.7 Radio Science Observations

Like many missions, the team will use the radio transmitter onboard the spacecraft to make sensitive Doppler, ranging, and Delta-DOR (Delta-Differential One-way Ranging) measurements through the Deep Space Network (DSN) (McMahon et al. 2017). Radio science observations will provide the information necessary to make estimates of the gravity field, mass, Bennu ephemeris, Yarkovsky and YORP effects, and other important physical characteristics of the asteroid. Particularly interesting to our site-selection team will be maps showing the angular deviation from the local downward normal of the total acceleration vector due to gravity and rotation. These "slope" maps will inform sampleability assessments by showing where regolith is likely to accumulate.

### 5.8 The Touch and Go Camera System (TAGCAMS)

The main purpose of the TAGCAMS suite of imagers is to aid in navigation around Bennu (Figure 19). They will also contribute significantly to the photo-documentation of Bennu (Bos et al. 2017). The purpose of TAGCAMS is to provide imagery during the mission to facilitate navigation to the target asteroid, acquisition of the asteroid sample, and confirmation of sample stowage. The cameras were designed and built by Malin Space Science Systems based on requirements developed by Lockheed Martin and the OSIRIS-REx project. All three of the cameras are mounted to the spacecraft nadir deck and provide images in the visible part of the spectrum, 400–700 nm. Two of the TAGCAMS cameras, NavCam 1 and NavCam 2, serve as fully redundant navigation cameras to support optical navigation and NFT. Their boresights are approximately aligned in the nadir direction with small angular offsets for operational convenience. The third TAGCAMS camera, StowCam, provides imagery to assist with and confirm proper stowage of the sample. Its boresight is pointed at the SRC located on the spacecraft deck. All three cameras have a 2592 × 1944-pixel detector array that can provide up to 12-bit pixel depth. All three cameras also share the same lens design which produces a camera field of view of roughly 44° × 32° with a pixel scale of 0.28 mrad/pixel. The StowCam lens is focused to image features on the spacecraft deck, while both NavCam lens focus positions are optimized for imaging at infinity.
[Insert Figure 19 here]

## 6 Ground System and Concept of Operations

The OSIRIS-REx ground system is designed to meet the unique challenges of an asteroid sample-return mission. These challenges include accurate spacecraft navigation in the microgravity environment, precision delivery of the spacecraft to the asteroid surface, tight coupling and interdependence between the science team and spacecraft operations, and data product production on a tactical timeline to enable sample-site selection. In order to meet these challenges, the project implemented a comprehensive ground-system engineering approach, starting with the mission science objectives and flowing these down to all ground-system elements.

The OSIRIS-REx baseline mission is achieved by satisfying the five primary objectives described in Section 2. The mission science plan maps these five objectives to 15 "Level-1" science requirements, which define the baseline mission (Figure 20). Following a rigorous systems-engineering process, the Level-1 science requirements are broken down into 86 "Level-2" science data-product requirements. When developing the Level-2 requirements, the team's strategy was to identify every individual science data product that is needed to fully satisfy the corresponding Level-1 requirement. In order to determine the operational timeline needed to collect the science data to fulfill those requirements, the team developed a DRM that outlined the sequence of observations, DSN downlink schedule, and delivery of required operational science data products. This information was used to guide the design and construction of the flight and ground systems.

[Insert Figure 20 here]

The DRM is broken up into distinct mission phases, each designed to permit spacecraft operations at progressively lower altitudes to the asteroid surface. Each Level-2 requirement is mapped to a specific observing opportunity within a phase of the DRM, the relevant mission elements, science working groups, and a co-investigator with primary responsibility for producing the data product. This level of detail provides the ability to properly estimate the flight-system capabilities and ground-system level of effort required to meet all the Level-2 requirements. In addition, this list clearly defines the higher-level data products that will be developed by the OSIRIS-REx team, ensuring that resources are available to meet the Planetary Data System (PDS) archiving requirements. The DRM is also a validation mechanism to ensure that our mission addresses the science and engineering objectives of the mission, and forms the foundation for developing detailed case studies of observation planning and commanding efforts by our operations teams. The sophisticated mission observing plan, along with navigational challenges, drives the design of our ground operations—navigation, planning and commanding, downlink, data verification, and generation of science products.

The ground system is composed of three elements: The Flight Dynamics System (FDS), the Mission Support Area (MSA), and the Science Processing and Operations Center (SPOC) (Figure 21). The FDS uses spacecraft and instrument data to produce mission design and navigation products. The MSA commands the spacecraft and processes returned telemetry. Science data is processed from instrument telemetry to high-level derived data sets at the SPOC.

[Insert Figure 21 here]

## 6.1 Flight Dynamics

Accurate navigation of the spacecraft in the microgravity environment around Bennu is one of the key challenges of this mission. The difficulties in navigation arise from the weak and complex asteroid gravity field, imperfect knowledge of the position and trajectory of the spacecraft, and nongravitational forces on the spacecraft. Overcoming this challenge requires the use of multiple navigation techniques. During the course of the mission, the navigation team will be preparing orbit-determination solutions based on radio-tracking data, star-field optical navigation, and asteroid landmark–based optical-navigation techniques. To enable the landmark-tracking navigation, the science team will develop a series of asteroid shape models, at increasing higher spatial resolutions, along with a core set of surface landmarks. The shape models will contain a comprehensive set of information about the asteroid that includes the asteroid shape, pole orientation, rotation state, and the coordinate system. As the mission progresses, the navigation team will enhance the set of landmarks with additional optical navigation observations taken by the NavCam as well as OCAMS. The science and navigation teams will frequently exchange assessments of the gravity field, rotation state, surface landmarks, shape model, and topographic information to ensure both teams are working with consistent, high-fidelity information.

The spacecraft will experience nongravitational forces from solar radiation pressure, reemitted infrared radiation from the spacecraft and the asteroid, spacecraft outgassing, and other sources. These forces will be comparable to the force from the asteroid's gravity, and as a result, they lead to substantial uncertainties in the prediction of the spacecraft trajectory. Frequent orbit determination and ephemeris updates can reduce these uncertainties. This solution requires a sophisticated late-update process that impacts the entire ground system to revise planned science and optical navigation observations to incorporate the latest, high-fidelity trajectory prediction. Another consequence of working in the small gravity field of the asteroid is the small Delta-V needed for propulsive maneuvers. Indeed, during sample acquisition, maneuver Delta-Vs range between 1 and 20 cm/sec. The small sizes of these burns drive the proportional errors to be larger than typical maneuver execution errors.

To deal with these challenges, our operations team has taken a multipronged approach, including careful analysis, calibration, and spacecraft management to understand small-force errors; formulating operations plans and cadence to deliver late-update ephemerides to science planning; revising science planning approaches and tools to optimize observations during the times when ephemerides are accurate; and using the spacecraft reaction wheels to scan a large enough area to accommodate the positional uncertainties and complete the required observations. The paper by Williams et al. (2017) in this volume describes these processes in greater detail.

## 6.2 Spacecraft Operations

Primary spacecraft operations are performed at Lockheed Martin's dedicated MSA in Littleton, CO. The MSA is the system of software, hardware, and people that enables engineers to command the spacecraft and analyze its telemetry. This facility has a long history of deep-space operations including missions such as Stardust, MRO, Mars Odyssey, Juno, and MAVEN, allowing the OSIRIS-REx team to leverage existing infrastructure, facilities, hardware, and software for operations and testing. The MSA radiates commands and command sequences through the DSN to the spacecraft. Primary mission control and telemetry processing is performed at the MSA. This ground data system was employed during the spacecraft development and instrument integration and testing in both the software-only SoftSIM and the software/hardware Spacecraft Test Lab, giving the development and flight operations teams valuable hands-on experience prior to launch. Lockheed Martin has an integrated command and telemetry database delivery process, as well as comprehensive telemetry analysis and trending tools.

Command sequence generation development is performed on MSA development workstations and is reviewed and tested in the Spacecraft Simulation Laboratory using a combination of software simulators, hardware emulators, and engineering development units. The command sequences are placed on the project flight operations file system and passed through the Front End Data System (FEDS) for uplink at the DSN. The command and sequence generation process has over twenty years of operational heritage.

All spacecraft data, engineering telemetry, and science data comply with the Consultative Committee for Space Data Systems format for data packets. Spacecraft data are first received on the ground at the DSN and are then transmitted to Building 230 at JPL. Building 230 performs two functions: the first is to forward the data to the MSA, and the second is to retain a short-term backup copy. Upon receipt at the MSA, data are processed by the FEDS, identified as either engineering or science data based on the Application Process Identifier Definition (APID) number. Each producer of data packets on the spacecraft has one or more APIDs to identify the type of data. All engineering and science data are assigned an appropriate APID relative to their data type. The FEDS reads the data packets, determines the APID number, and distributes the packet to the appropriate end user (i.e., MSA, FDS, SPOC). Additionally, all data are stored onto a project server at the MSA where they are available to the various mission teams for the duration of the project. Data packets containing engineering telemetry are further processed into channelized data that are displayed for real-time spacecraft health and status monitoring. Additionally, the MSA performs telemetry trending and analysis that is used for future spacecraft performance predictions. All science data and instrument engineering data are distributed to the SPOC for further processing.

## 6.3 Science Operations

From approach through TAG, members of the science team will work from the SPOC, housed in the Michael J. Drake building near the University of Arizona campus in Tucson, AZ. At this facility, the science operations team will plan science observations, generate instrument command sequences, review downlinked data, generate science products needed for mission operations, and work together to interpret the exciting new views of Bennu.

The SPOC is responsible for hosting the software used to develop science data products. The SPOC will provide the communications and IT infrastructure, tools, and frameworks to support the science processing, planning, instrument commanding, instrument downlink and calibration pipelines, and a repository for the science data. The SPOC will also ensure that science data is delivered to the PDS, the public archive for all NASA planetary missions.

The SPOC generates instrument command sequences in coordination with science team and the cognizant engineers for each instrument. Instrument commanding is done with a combination of onboard (frequently executable) blocks, as well as command sequences for one-time or unique commanding activities. The SPOC operations engineers check syntax and verify that none of the instrument flight rules or constraints are violated. They pass instrument command sequences to the MSA for incorporation into the spacecraft command and sequence development process. There, the sequences are again checked for syntax and given final scrutiny to ensure that they do not violate flight rules. If necessary, they are run through ground-based simulations before being routed to the DSN for radiation to the spacecraft. Downlinked telemetry and observations are likewise routed to the MSA, and raw data are placed in a repository where they are fetched by SPOC personnel for ingest, calibration, and delivery to the science team. The SPOC continuously monitors instrument health using both a work station located in the Drake Building and also in-house tools. Key instrument housekeeping telemetry is trended. Watch items are generated and tracked for any housekeeping telemetry trends that are different than expected. Watch items may be elevated to an Incident, Surprise, or Anomaly to be addressed at the project level.

As noted, one unique challenge for this mission is maintaining consistency between planned observations driven by science requirements and the spacecraft navigational uncertainties. The use of onboard blocks will support the science late-update process and timeline, whereby changes to instrument pointing, timing, and parameters such as exposure time can be quickly iterated by SPOC operations personnel and updated command products can be delivered to the MSA for validation, uplink, and execution.

The science team will develop data products using instrument data and scientific analysis software. Many of these products are needed to develop the site-selection maps and other operational data products, placing the science team in a time-sensitive operational role. The team has carefully mapped the Level-2 requirements to identify the products that are needed for site selection and navigation. Science data production is tracked using a resource-loaded schedule that traces the observational dependencies of each product, the time to develop them, and the personnel who will prepare them. These management tools allow the team to accurately identify when maps will be available and to estimate when site-selection decisions can be made.

Science team members responsible for the production of map and navigation products will be familiarized and practiced in their role, and will participate in postlaunch science operations tests to verify execution timing and interfaces with other products. In addition, critical phases of the mission will be the subject of comprehensive operational readiness tests, many of which will include members of the science team.

During operations, science team and SPOC members will have access to a number of software products to help them with their work. Approved flight system and science product status are available through an internal science operations website. After all relevant observations have been downlinked and calibrated, the science team and SPOC personnel responsible for products that will be generated from these observations will be notified. If important observations are lost, the planning schedule allows the team to identify the affected products and notifies mission management to implement contingency operations.

### *6.3.1 Sample Site Selection*

OSIRIS-REx is unique among NASA planetary missions in that remote sensing is performed primarily to support the sample-return objective. Prior to the sampling event, the science instruments will be used to survey Bennu for almost two years to select and document the best candidate sample sites. During this period, the team will combine coordinated observations from OLA, OCAMS, OVIRS, OTES, TAGCAMs, and radio science into four thematic maps of decision-making properties: deliverability, safety, sampleability, and science value (Figure 22). Since REXIS is a student experiment, it is not part of the baseline plan for site-selection product development. However, if interesting data are available from this instrument, the science team will include them in their evaluation of potential sample sites.

[Insert Figure 22 here]

The plans to produce data products from observations are a consequence of some basic mission requirements:

- Deliverability: The spacecraft must be delivered to within 25 m of a selected location.
- Safety: When the TAGSAM head contacts the surface, the spacecraft must avoid any damage that would prevent a successful return to Earth.
- Sampleability: A sample site must have characteristics that permit TAGSAM to obtain the required 60 g or more of regolith.
- Science Value: The sample site should contain material that provides the most value toward meeting the mission objective of returning pristine carbonaceous regolith.

Each of these requirements has been assigned to the appropriate mission element (flight dynamics for deliverability, project and flight-system engineering for safety, TAGSAM engineering and regolith science for sampleability, and sample site science for science value). In order to evaluate various locations on the surface of Bennu, each of these parameters will be mapped onto the asteroid shape model.

**Deliverability Map**

The deliverability map indicates the probability that the flight dynamics team can deliver the spacecraft to a desired location. Deliverability maps are scaled maps that indicate areas of high probability of accurate spacecraft delivery (green) to low probability for successful delivery (red). This map product is a numerical measure of navigation-accuracy capability based on site-specific Monte Carlo analysis. Initially, the team will generate a coarse global map by simulating spacecraft delivery to ~100 evenly distributed locations across Bennu. The deliverability of regions between these locations will be estimated using an interpolation technique. The objective of a global map is to provide preliminary information describing which regions may be suitable for consideration as candidate TAG sites. If the team can learn early that certain regions are not accessible, then more of the eventual candidate sites could survive the vetting process.

Once specific candidate sample sites have been identified, more refined site-specific deliverability maps will be generated. The site-specific deliverability maps represent the contact probability density in the vicinity of the TAG site. This probability is derived from a 2×2 covariance matrix describing the distribution of Monte Carlo cases on the surface at the nominal TAG site. Each site-specific map is thus derived from the targeting error covariance matrix obtained from the Monte Carlo runs for that TAG site. These maps are used to quantify the ability to deliver the spacecraft to the surface with the required accuracy, to estimate the probability of being affected by local hazards in the vicinity of the TAG site, and to show where the accuracy may be better than required, allowing a finer choice of TAG location.

**Safety Map**

Safety maps are scaled maps that quantify the probability of safety during TAG operations, including the dispersions in the spacecraft trajectory. The safety maps include an assessment of the probability that physical hazards will be present during approach to the asteroid. In addition, these maps indicate whether the spacecraft operational constraints will be met at the TAG site. These constraints include telecommunications link, surface reflectance (for LIDAR operation), proper imaging conditions (for NFT guidance), gravity-field uncertainties, and thermal conditions. Maps for individual safety parameters will also be generated to understand which parameters are the driving concerns. Global versions of the safety map will be produced to identify and prioritize favorable sampling sites. Site-specific versions of the maps use similar information in conjunction with the deliverability map to provide the overall probability the spacecraft will be safe during sampling.

**Sampleability Map**

The sampleability map quantifies the expected collected mass from the surface of Bennu, based on metrics associated with the size–frequency distribution of the regolith, and the presence of obstacles. The sampleability map is designed to provide relative assessments of different regions of interest (ROI) during the Detailed Survey and Orbit-B phases of the mission. Individual facets on the global shape model will adopt values of tilt and average grain size based on the global shape model and the thermal inertia map, respectively (where average grain size is estimated from thermal inertia using the algorithm of Gundlach and Blum 2013). The global sampleability map also incorporates the dust cover index to adjust the confidence of the correlation between thermal inertia and grain size.

At the site-specific level, sampleability will be quantified for each of the candidate sample sites that are inspected during the Reconnaissance phase of the mission. The sampleability algorithm is empirically derived based on experiments of TAGSAM collection efficiency as a function of regolith properties. This connection to TAGSAM head performance is based on tilt, maximum and minimum grain size, and the slope of the particle size frequency distribution. These parameters are determined for candidate sample sites based on remote-sensing characterization, allowing for a sampleability value to be assigned to each shape model facet. To calculate the probability of successful sampling, these data are convolved with the deliverability map for each candidate sample site.

**Science Value Map**

A primary goal of the OSIRIS-REx mission is to test the many hypotheses on the origin, geological history, and dynamical evolution of Bennu through coordinated analytical studies of the returned samples. Bennu is spectrally a B-type asteroid and is thought to be similar to CI and CM carbonaceous chondrite meteorites (Clark et al. 2011). The returned samples are thus expected to contain primitive Solar System materials that formed both before and after asteroid accretion. The science value of an area should be the omnibus of the diversity of chemistry, mineralogy, and geology. In other words, sample sites with the highest scientific value should contain as many key materials as possible to address the questions and hypotheses posed by the sample analysis team [see Sample Analysis section]. Overall science value assessment will be based on the integrated global science value map (ISVM). This map will rank the science value of candidate sampling locations within the 2-sigma deliverability targeting ellipse on the surface of Bennu according to a semiquantified science value scale (Nakamura-Messenger et al. 2014). The ISVM consists of four sub-science value maps in order of importance of the mission: (1) chemical composition to emphasize the nature and abundances of organic matter, (2) mineralogical diversity, (3) geological features, which serve as proxy for surface age, and (4) temperature of the surface for the best preservation of the volatiles and organic matter. The ISVM will rank the overall science value of candidate sampling locations from 5 to less than 2 cm resolution to help select the sample site. The sample-site selection board will rely on the “Safety”, “Deliverability”, and “Sampleability” maps to identify permissible areas for sampling. After this down-select process, the ISVM will be of critical importance for final sample-site selection.

## 6.4 Mission Plan

The mission has been designed with several distinct operational phases that include hyperbolic passes at 7, 5, and 3.5 km, two orbital phases at 1.5 and 1 km, and reconnaissance overflights at 225 and 525 m. All operations are designed to support a cautious approach to accumulating expertise and confidence for operation of the spacecraft around a body with low gravity, and where nongravitational forces and low Delta-V maneuvers create a significant challenge to navigation and maneuver design. The approach to sample-site selection also drives mission design, in that it identifies the need to obtain incremental understanding of the asteroid on a global level, and then again at the scale of the candidate sample sites, and ultimately at the scale of the TAGSAM sampling head. These assessments require observations made from a number of geometries relative to Bennu.
Science goals for the OSIRIS-REx mission are met through the production of a number of science products built from observations taken over several mission phases (Figure 23). These phases have been carefully documented in the DRM document, which, along with the DRA, forms a comprehensive resource around which the flight system and payload engineers performed their design, verification, and validation activities.
[Insert Figure 23 here]

### *6.4.1 Launch*

The spacecraft launched at 7:05 p.m. EDT on September 8, 2016, on an Atlas V rocket in the 411 configuration: a standard common core booster with a single strap-on solid rocket motor (Figure 24). The upper-stage was a Single-Engine Centaur. The spacecraft was protected in a 4-m fairing. The first stage was powered by the RD-180 main engine, which delivered more than 860,000 pounds of thrust at liftoff. The upper stage relied on the RL10 propulsion system, which harnessed the power of liquid hydrogen rocket fuel and employed a precision control system to accurately target the spacecraft. With a total spacecraft mass of 2105 kg, OSIRIS-REx left the Earth on a direct outbound trajectory with a Delta-V of 1400 m/s. The hyperbolic injection C3 was 29.29678 $km^2/s^2$, RLA was 177.00097°, and DLA was 0.081643°. The FDS team performed a reconstruction of the launch injection state, which verified that the Atlas V launch vehicle delivered the spacecraft almost perfectly to the specified target conditions (0.5426 sigma for C3, 0.4721 sigma for RLA, and 0.1469 sigma for DLA).
[Insert Figure 24 here]

### *6.4.2 Outbound Cruise*

The Outbound Cruise trajectory includes a one-year Earth-to-Earth trajectory and one large Deep Space Maneuver (DSM) that occurs on December 28, 2016. The Earth-to-Earth trajectory culminates in an Earth gravity assist (EGA) for which the perigee altitude is 17,319 km. The EGA rotates the spacecraft trajectory six degrees into Bennu’s orbit plane, after which the spacecraft will spend another year traveling to the asteroid. An additional small DSM is performed between EGA and asteroid approach. The spacecraft will be between 0.8 and 1.4 AU from the Sun for the majority of the mission (Figure 25).

[Insert Figure 25 here]

The Earth–Moon flyby during the EGA provides an opportunity to calibrate OCAMS, OVIRS, OTES, and TAGCAMS. The EGA will be used to exercise instrument command sequences that will be used at Bennu by collecting images and spectra of deep space and the Earth/Moon system. These data will be used to verify instrument radiometric performance, coalignment between the sensors, and out-of-field response with point source and spatially resolved data. The Earth observations will begin one day after closest approach, when the spacecraft range is 100,000 km with phase angles of 20°–30°. Additional Earth observations will occur three and six days after the EGA. For the MapCam, at a range to target of 100,000 km, the field of view subtends a diameter of around 5000 km, fitting within the Pacific Ocean. For the PolyCam, the field of view subtends 500 km. Lunar observation conditions are not as favorable. The Moon will be observed on EGA + 3 days, when the spacecraft range to target is around 272,000 km. However, low phase angles that are optimum for imaging and spectroscopy do not occur until the range to target approaches 1,000,000 km, a factor of 50–60 greater range than that to the Earth.

### *6.4.3 Approach*

The spacecraft will begin observing Bennu in June 2018, when the asteroid is just bright enough for detection by PolyCam (~15$^{th}$ magnitude). These first science observations mark the beginning of asteroid proximity operations. This Approach Phase is divided into three parts to create a gentle approach to Bennu that affords adequate time to optically acquire the asteroid using the spacecraft's onboard cameras, and also to use the cameras and spectrometers to survey the vicinity of the asteroid for any hazards that may be present, and to characterize the asteroid point-source properties for comparison to ground-based telescopic data. The three parts are delimited by three deterministic Asteroid Approach Maneuvers (AAMs), labeled AAM1, AAM2, and AAM3 (Figure 26).

[Insert Figure 26 here]

**Surveying the Operational Environment**

A primary objective of the Approach Phase is to survey the asteroid operational environment for potential spacecraft hazards. Discovery of either a dusty environment or a natural satellite will trigger a safety assessment and review of the approach strategy. Spitzer telescope observations show no thermal-excess evidence for a dust belt around Bennu (Emery et al. 2014). However, it is possible that dust may be present near the asteroid due to recent activity or at a level below the detection threshold of the Spitzer observations. To ensure spacecraft safety, a dust search will begin when the spacecraft is at a distance of 1 million km. PolyCam will be able to detect dust as an extended source along Bennu's orbital track.

Ground-based characterization of Bennu reveals no evidence for natural satellites. However, it is possible that satellites as large as 20 m are present, based on detection limits from ground-based radar observations. Such objects are of intrinsic scientific interest and also represent a potential spacecraft hazard. The spacecraft will perform a series of natural satellite searches to mitigate this risk. The first search will hunt for >1-m satellites within 35 km of Bennu. The search will be conducted at a spacecraft–Bennu range of ~1000 km. Five search fields will be imaged by MapCam (panchromatic filter) five times with an interval of 45–60 minutes between visits. Searches will be conducted over three consecutive dates. This interval will allow any satellites to move relative to the background star field and enable detection using well-tested moving-object detection techniques. The second search will hunt for >10-cm satellites within 26 km of Bennu. This search will be conducted at a spacecraft–Bennu range of ~200 km. Ten search fields will be imaged by MapCam (panchromatic filter) five times with an interval of 45–60 minutes. Searches will be conducted over three consecutive dates. During each search, one field will be centered on Bennu to measure the position of Bennu and determine its spatial offset with any detected satellite. The solar phase angle relative to the asteroid is <50° during the natural satellite survey to ensure sufficient lighting for visibility.

**Characterizing Bennu Point-Source Properties**

To achieve the objective of providing ground-truth data for telescopic characterization, the team will obtain "disk integrated" photometric and spectral data of Bennu. These data provide an important link between the telescopic data and the spatially resolved data that will be obtained later in the mission. In addition, they will provide important first-look information, constraining the rotation period and photometric properties, identifying important spectral features, and allowing for an initial assessment of the surface's thermal inertia, which is a proxy for the average regolith grain size.

Bennu's rotation rate will first be measured by obtaining a series of asteroid light curves. These data products are photometric measurements taken over two full asteroid rotations (~8.6 hours) along with the parameters that describe the shape and period of the light curve. These light curves will be obtained by MapCam using the b′, v, w, and x spectral filters. Observations will consist of 576 MapCam images taken over the course of 4.5 hours with a cadence of v-b′-v-w-v-x-v and repeat. This results in 360 v images taken once every degree of Bennu rotation, 72 b′, w, and x images taken once every 5 degrees of Bennu rotation. The mission plan includes two dates for Bennu light curve observations.

Bennu's photometric properties will be measured during the Approach Phase by characterizing its point-source phase function. The phase-function data product is a model of the light scattering behavior of the asteroid's average surface as a function of phase angle (sun–asteroid–spacecraft angle). Bennu will be observed in the MapCam panchromatic and b′, v, w, and x filters over the course of the Approach Phase when the signal-to-noise ratio is >5 such that the total range of phase angles sampled spans more than 50° of arc.

Characterization of Bennu's rotationally resolved Vis-NIR spectral characteristics will provide an early assessment of the dominant mineralogy on the surface. The data products that fulfill this requirement are longitudinally resolved lists of minerals and chemicals detected at >5% band depth. The data to be analyzed for each spectrometer will be averaged longitudinally, notionally every 10°, prior to analysis to increase the signal-to-noise ratio. Determination of longitude for the purposes of averaging could be performed in one of at least two ways: (1) using the reconstructed longitude of the boresight for each spot, if the boresight is reasonably near the center of Bennu (which will underfill the spectrometers' fields of view), or (2) by time, where the rotation rate of Bennu is known such that spectra can be averaged over specific intervals equating to the longitude range of interest (e.g., 10° or ~7.15 min). After averaging of spectra within the specified longitude ranges, spectral parameters for mineralogy and chemistry will be calculated for OVIRS I/F spectra and OTES apparent emissivity spectra, and OTES spectra will also be linearly modeled using the mission spectral library.

The OVIRS data from the Approach Phase are also important for constraining global scale factors for the 1064-nm/860-nm color ratios. The safety map uses the reflectance of the asteroid at 1064 nm to ensure that the surface is within the range of reflectance values in which the GN&C LIDAR is designed to operate. Predictions of the variation in intensity are also important to avoid excessively high variations in reflectance that could affect LIDAR ranging performance. Following conversion to units of bidirectional reflectance, the OVIRS data will be used to determine the spectral slope from 860 to 1064 nm using global scaling factors obtained from rotationally resolved Vis-NIR spectral characteristics data. These global scale factors will capture the distribution in 1064-nm/860-nm color ratios across the asteroid as a function of rotational longitude, resulting in minimum, maximum, and nominal scaling factors that will be applied to image mosaics at 860 nm later in the mission.

Additional information about Bennu's mean particle size will be obtained by characterizing the rotationally resolved thermal flux and thermal inertia. The requirement is to measure the thermal flux from Bennu for one full rotation with OTES before it fills the OTES field of view. These observations will provide a direct bridge between higher spatial resolution observations at later mission phases with telescopic observations that have been made of Bennu. Disk-integrated thermal inertia will be derived (as a function of rotational phase) from these observations using the asteroid thermal model. The combination of these full-disk observations of Bennu with space- and ground-based telescopic spectral measurements will provide ground-truth and inform scientific interpretation of telescopic observations of other asteroids.

**Preliminary Shape Model Development**

The asteroid shape model is a mission-critical data product that is required for successful proximity operations. It also provides the framework upon which all other map products are overlain. Images acquired during the final phases of approach will allow the creation of a shape model needed by both FDS and science planners using the technique of stereophotoclinometry (SPC). SPC combines stereo techniques with photoclinometry to derive the slope of an asteroid's surface. The method models surface slopes at each pixel of a given image by initially using stereo data to define a relationship between surface slope and observed albedo. With this relationship in hand, the slopes of a piece of asteroid surface imaged at multiple emission and incidence angles can be obtained via least squares regression that best duplicates the input images. Once the surface slopes are obtained, the height across each map can be determined by integrating over the slopes in a logical manner. These individual topographic maps of the surface can then be collated together to produce a shape model. SPC has been developed and tested over two decades on many different planetary bodies (Gaskell et al. 2008). The NASA Dawn and the ESA Rosetta missions in particular have provided real-time analogs, and many of the procedures the OSIRIS-REx team will use were developed, tested, and verified for that mission. Recent SPC improvements include the use of asteroid limb and terminator data to initially constrain the object's shape. These enhancements were especially useful in determining the shapes of asteroids Lutetia and Steins during the flybys by the Rosetta spacecraft (Sierks et al. 2011; Jorda et al. 2012). The SPC data to be acquired during the Approach Phase will allow the team to get an early start on the development of the 75-cm resolution shape model, the first model delivered from science to FDS. This product will be completed using data obtained in the next phase of the mission, Preliminary Survey.

### 6.4.4 Preliminary Survey

The Approach Phase will be followed by Preliminary Survey, which consists of three hyperbolic trajectories that cross over the North and South poles and the equator at a range of 7 km (Figure 27). The team will obtain data from OLA for the first time, along with additional imagery. These passes will permit radio science measurements to determine Bennu's mass, a prerequisite to planning the maneuvers that place our spacecraft into orbit. The radio science team will process radiometric tracking data, along with LIDAR altimetry and optical-navigation imaging data, in order to determine the mass. The imaging data from the Preliminary Survey passes will provide the final data set to complete the 75-cm resolution global shape model and corresponding rotation state data. Once the shape model is complete, the official asteroid coordinate system will also be developed, which is needed for coregistration of all data products.
[Insert Figure 27 here]

### 6.4.5 Orbit-A

With the asteroid mass, shape, and rotation state constrained, the FDS team will have all the information needed to finalize the design of the orbital insertion maneuver sequence (Figure 28). The spacecraft will be placed into a 1.5-km orbit, beginning the Orbit-A Phase. In this phase the spacecraft enters into a gravitationally bound orbit about the asteroid. The orbital radius will be between 1.0 and 1.5 km, and the orbit is designed such that no maintenance maneuvers will be required for at least 21 days. For stability, relative to solar radiation pressure, these orbits reside in the terminator plane. The nominal orbit radius during Orbit-A is 1.5 km, and the orbit thus has an approximately 50-hour period and a 5.3 cm/s orbital velocity.
[Insert Figure 28 here]
During this phase, the FDS team will transition from star-based to landmark optical navigation. Landmark tracking provides a higher fidelity orbit-determination solution, which is needed to implement the precision maneuvering required for the upcoming mission phases. Once this transition is complete, the routines of orbit determination and maneuver planning that will characterize their activities for nearly a year will be well established. As time and circumstances allow, additional science data collection may take place.

### 6.4.6 Detailed Survey

**The "Baseball Diamond"**
The characterization of Bennu begins in earnest with the Detailed Survey Phase, which requires multiple hyperbolic passes of Bennu to obtain the wide range of viewing angles necessary to characterize the asteroid's global properties. The first series of these passes will target a range of 3.5 km and transit over four sub-spacecraft points near 40° North and South latitude at 10 a.m. and 2 p.m. local time on the surface. Observation will be nadir at 10 a.m. and slightly oblique at 2 p.m., viewing 12 p.m. to ensure good stereo angles for the generation of digital terrain models. The relative locations of these four stations is reminiscent of a "baseball diamond", a term the team has adopted to describe this part of the Detailed Survey phase (Figure 29). Images and OLA spots from each of these points taken through a full 4.3-hour asteroid rotation will provide high resolution, stereo imagery and LIDAR altimetry of the surface.
[Insert Figure 29 here]

One of the key products from this phase is a refinement of the global shape model from 75-cm to 35-cm resolution. These digital terrain maps are generated using SPC processing of OCAMS images combined with range data from OLA. Production of the shape models from these observations produces a rotation state as part of the solution. Images in this phase are of sufficient resolution to produce digital terrain maps of any proposed selected sample sites at 5-cm ground sample distances.

Imaging data will also be used to produce a global image mosaic of Bennu at 21-cm resolution (using a 5 pixel criterion). Additionally, image sequences of the asteroid surface will be released as public engagement products that will provide a sense of spacecraft flight over the asteroid. The global mosaic will be analyzed to generate a thematic map of hazards and ROIs. This product is a key input to the sample-site selection process. To map hazards (objects >21 cm) as well as ROIs (areas of smooth terrain free of hazards), it will be necessary to first generate panchromatic PolyCam mosaics from the global images acquired during this phase. Other inputs to the global mosaic include the PolyCam photometric model, SPICE kernels, and the shape model. The photometric model will initially be based on the Approach Phase data. As more data are acquired later in Detailed Survey, this model will be updated and the global mosaic refined.

Using the global mosaic, Bennu's terrain will be surveyed in bulk and classified as either "safe" or "unsafe," and the results will be visualized in the thematic map of hazards and ROIs. Simultaneously, individual hazards will be mapped, and their location and dimensions will be stored in a hazards/particle database. The database can be queried to provide the precise location and size of all regolith grains that have been counted in a given area, allowing for refinement of the thematic map as the surveying evolves. This information will eventually be thresholded into a binary mask of hazards, which masks out areas of undesirable terrain and serves as an input into the global sampleability map.

**Equatorial Stations**

A second set of 5-km hyperbolic passes will fly through a series of stations targeted to be above the equator, spaced over seven local times of day on the surface (Figure 30). This spacing ensures a number of independent illumination angles and observing geometries to reveal important details about surface properties, including disk-resolved photometric models of Bennu from both OCAMS and OVIRS data. The spacecraft will take four days to travel from south to north for each station. In order to obtain global coverage, the spacecraft boresight will slew from below the south pole to above the north pole and back continuously for 4.5 hours during each equatorial crossing while collecting MapCam, OVIRS, OTES, and OLA data. Because Bennu rotates once every 4.298 hours, these observations will result in global coverage at each of the seven Detailed Survey equatorial stations. Because OVIRS is the science instrument with the narrowest field of view, the slew pattern is designed to ensure complete coverage for this sensor.

[Insert Figure 30 here]

**Global Imaging Products**

During this global characterization phase, PolyCam captures global images for measuring the physical dimensions and orientations of all surface geological features, including craters, boulders, grooves, faults, and regolith. OLA provides an independent means to determine shape and surface texture and provides an absolute range for all other remote-sensing data. Color mapping of the surface of Bennu will be performed using the four wide-band spectral filters on MapCam.

These images will provide spatial context for the high-spectral-resolution data from OVIRS and OTES. These observations will provide spectroscopic and spectrophotometric measurements that will reveal surface reflectance, mineralogical distributions, geology, surface temperatures, and surface thermal characteristics to help complete the global assessment of the asteroid.

The OCAMS data from this phase will be compiled to produce the global MapCam panchromatic photometric model, and four global MapCam color photometric models (one for each color filter). These models are needed to accurately photometrically correct imaging data. For accurate photometric models, it is important to collect observations at a range of incidence angles, emission angles, and phase angles. In particular, surface reflectance is a strong function of the phase angle; hence observations will span the widest phase-angle range possible.

The asteroid surface reflectance will vary with changes in illumination and viewing geometry. The photometric modeling effort will thus include an effort to capture the natural variations in reflectance observed within individual images in the imaging data and in each spot in the spectral data. These variations will be modeled with semiphysical mathematical descriptions of light scattering, to test the models against the data in a least squares sense to determine the best-fit model. Once a photometric model is derived for a particular wavelength, it can be used to correct all the data at that wavelength to a reference viewing and illumination geometry for radiometrically corrected inter-comparisons of surface features, and for mosaicking of surface images. Without a photometric correction, image mosaics can maintain significant brightness differences from image to image, which create an appearance of seams and dominate the visual impact. Spectral maps made without a photometric correction invariably show regions near limbs and terminators to be much redder than they actually are, and spectral changes due to photometric effects will dominate the variations in the spectral data set before they are photometrically corrected. For these reasons, fast photometric modeling and photometric correction are important processes that will be optimized for this mission.

Spectral variations across the surface of Bennu will be visualized by producing color ratio maps. These color ratio maps will be produced from MapCam images in the b′, v, w and x filters. After a photometric correction is performed for images in each color band, ratios in I/F are used to create false-color composite images. The band ratios of interest are b′/v to characterize the ultraviolet slope, the v/x to characterize the visible slope, and a combination of the v, w, and x images to characterize the presence and depth of any 0.7-micron absorption feature, which is diagnostic of hydrated phyllosilicate minerals (Johnson and Fanale 1973; King and Clark 1989; Vilas 1994). Other ratios may be deemed diagnostic and also used. Individual band-ratio images can be combined into RGB composite images and mosaicked into color ratio maps.

Color ratios will provide spectral information at the highest available spatial resolution, and can be diagnostic of properties such as composition (Delamere et al. 2010), particle size distribution (Jaumann et al. 2016), and space weathering (Chapman 1996). One of the main benefits of color ratio maps is their ability to resolve the ambiguity between topographic shading and differences in surface materials. Hence, color data has been a powerful tool (in conjunction with higher spectral resolution information) in geologic interpretations at all spatial scales (Le Corre et al. 2013) and will provide the only spectral information for Bennu at sub-meter scales.

The OCAMS x-filter global mosaic (860-nm wavelength) will be multiplied by the 1064-nm/860-nm global scale factor determined in the Approach Phase to produce the global 1064-nm reflectance map. This map quantifies the asteroid surface reflectance at 1064 nm and zero phase angle, which is needed to predict GN&C LIDAR performance. Since the GN&C LIDAR performs critical ranging measurements during sample acquisition, this data product is a key input to the safety map.

The global imaging and spectral data obtained from the Detailed Survey equatorial stations will be extensively analyzed to study the geology of Bennu. Four distinct geological data products have been identified, though many others will likely be produced, depending on the nature of the asteroid surface. The global crater geology map is a map of the size and physical location of all impact craters on the surface of Bennu with their geological context delineated. The global linear features geology map is a map of geomorphological features on the surface of Bennu that are not impact craters or boulders and are linear in appearance. The global boulder geology map is a map of the integrated geology of boulders located on the surface of Bennu. The boulder geology map presents the locations and sizes of boulders/rocks across Bennu, in conjunction with diagnostic spectral information and relationship to the geopotential of the body. Finally, the global regolith geology map is a map of the global distribution of deposits or patches of loose material (regolith) on the surface of Bennu, along with evidence for mass movement of regolith. Here, regolith will be defined as particles <20 centimeters (or smallest resolvable) in shortest dimension. Supplementary information may include any relevant properties of the regolith (i.e., depth, color, thermal properties, and mineralogy). The boulder map and the regolith map feed into the sampleability and science value maps for site selection. Ultimately, all of these maps, along with any others developed, will be combined into the definitive global geology map of Bennu.

**Spectral Data Products**

Many spectral data products from the Detailed Survey phase can be produced from data acquired at individual stations. The key data products that are derived from single-station Detailed Survey spectral observations are the global mineral and chemical maps. These global maps are a graphical and numerical display of the spatial distribution of minerals and chemicals identified on the surface of Bennu. These values are obtained from linear least squares models of OTES emissivity spectra (phase abundances) and OVIRS reflectance spectra (band strengths), respectively. These data will be binned spatially and mapped to the Bennu shape model to produce mineral and chemical maps that will contribute to assessing science value and long-term science, as well as informing sample-site selection and providing crucial compositional context for the sample.

Spectral data also are analyzed to produce additional higher-level data products. The global space weathering map convolves several possible indicators of surface alteration on Bennu to map the surface regions that exhibit the effects of time-dependent alteration due to exposure to space weathering processes such as micrometeoroid and solar wind bombardment. The global dust cover index map is a measure of the spatial distribution of surfaces on Bennu having a thermal infrared spectral signature indicative of the presence of fine particulates (smaller than ~65 μm). Global maps of this value are derived from binned and averaged values of the dust cover index, which is calculated from OTES emissivity spectra (the average emissivity over a specific spectral range). The heritage for the dust cover index is a product first published by Ruff and Christensen (2002).

Data acquired at multiple dayside stations will be used to produce the Global OVIRS photometric models. The photometric modeling data products describe the scattering behavior of the asteroid surface as a function of viewing geometry, illumination geometry, and wavelength. Photometric models will be derived for each channel (wavelength) of the OVIRS spectral data. These models serve the practical purpose of permitting the OVIRS data acquired at all daytime stations to be photometrically corrected to a common viewing geometry, enabling the spectra to be compared with each other even when illumination or viewing angles differ. This model will facilitate development of many other OVIRS-derived data products.

An important product built from global OVIRS coverage of Bennu that depends on the photometrically corrected spectral data is the global bolometric bond albedo map. The procedure is to use the photometrically corrected OVIRS radiance data and the OVIRS photometric model to compute a normal albedo for each OVIRS spot along with an average phase integral for the entire asteroid. The normal albedo is multiplied by the phase integral to compute the spherical albedo for each spot. The spherical albedo is integrated over wavelength, weighted by the solar-flux model to get the bolometric bond albedo for each of the OVIRS measurements, which are mapped to produce the bolometric bond albedo map.

In the event that plumes are found originating from Bennu (an admittedly low probability event) the team will produce a global dust-and-gas-plume geology map, estimate the plume density distribution in the asteroid environment, and attempt to discern the plume composition. The first product is a map of the geology of plumes and their activity on the surface of Bennu down to 1-m resolution. Plume composition is a list of minerals detected on any spectrally measurable plumes emanating from Bennu. This is a contingency data product, and spectral analysis will be on a best-efforts basis because the team is not developing the software tools and reference spectra that would be required to properly interpret spectra of dispersed particles. In other words, spectra of plumes will be transmission measurements, which are not equivalent to reflectance or emission data of the solid surface for which the team is preparing. This data product would consist of “spot” information about the composition of any spectrally detectable plumes.

**Thermal Data Products**

The thermal state of the asteroid surface is important to characterize for both safety and sampleability assessment. The OTES spot temperature data from the Detailed Survey phase will be used to produce a set of global temperature maps at different times of day. These data will be collected from fixed observing points at different times of day as Bennu rotates beneath the spacecraft. As a result, the variation in local time will be minimized for any particular global map. There will, however, be latitude-dependent effects on temperature due to the variation in solar illumination and the resultant surface heating with latitude. In addition to the temperature maps, several ancillary maps will be produced that give the average local time of the OTES data used in each map element (surface facet), the average emission vector in each map facet, the true orbit anomaly of each facet, and the average wavelength at which the temperature was measured for each map facet.

Bennu is a dark object, with an albedo of ~4.5% (Hergenrother et al. 2013; Emery et al. 2014). At closest approach to the Sun, some regions on its surface may reach 400 K near perihelion. To ensure spacecraft safety the team must measure the temperature distribution across Bennu’s surface upon arrival. Accurate temperature measurements taken over several times of day will provide the team with the information needed to create thermal models that include estimates of thermal inertia. Thermal inertia is determined by observing the response of the surface to changes in energy input—primarily solar insolation. Global thermal-inertia maps will be used to assist in the determination of the physical properties of the surface (including mean effective particle size) and to study anisotropies in thermal emission for the investigation of the Yarkovsky effect. Global thermal-inertia maps will also enable refined predictions of surface temperatures for later mission phases. The results will be reported as a map of single values (with appropriate uncertainty) for each resolution element, which will roughly correspond to the OTES spot size. This data product is the reason that this mission phase is designed to provide observations of most of the asteroid (>80% of the surface) at seven different times of day. Temperatures for each OTES spot measurement will be determined in the emissivity/temperature retrieval step of the OTES data processing. The project-approved thermal model will be used to model the diurnal temperature variation for each surface spot observed. Thermal inertia will be varied in the model to find the best match to the observed diurnal temperature variation for each map facet.

The OTES data will also contribute to producing an updated asteroid thermal model. The requirement is to provide software (i.e., a numerical model) that is capable of ingesting thermal measurements of the asteroid in order to determine the thermal inertia of the surface and, using that thermal inertia, predict the expected amount of Yarkovsky-induced acceleration. This prediction will be compared with measurements of the actual Yarkovsky acceleration measured by the mission team. The model is also used to predict global and site-specific temperatures during sampling.

The wide range of data products developed during this phase of the mission will be integrated and analyzed to produce the integrated global science value map. At the end of detailed survey, the team will have the final information to make an initial selection of up to 12 candidate sample sites. In addition, the work to accomplish the second mission science objective should be complete—mapping the global properties of the asteroid.

### *6.4.7 Orbit-B Phase*

At the end of Detailed Survey, the spacecraft will enter a close orbit around Bennu and commence the Orbit-B Phase with a nominal radius of 1 km (Figure 31). The primary science activities during this phase are the radio science experiment, shape modeling and topographic measurements based on OLA ranging, and the survey of potential sample sites. When site-specific observations are not being made, the spacecraft will be nadir pointed to support OTES spectroscopy and OLA ranging. OLA will map the entire asteroid surface at 7 cm spatially, with vertical precision of ~3 cm.

[Insert Figure 31 here]

**Radio Science Experiment**

The mass of Bennu and its gravity field will be determined with high accuracy in the radio science experiment. During this experiment, the spacecraft's orbit will be monitored with continuous DSN coverage as it naturally evolves over a nine-day period with no thruster firings. The spacecraft will have nadir pointing for the first and last day, and inertial pointing for the intervening 7 days. Radio science uses radiometric tracking data taken through the spacecraft telecommunication system. The solutions for these data products are mainly dependent on the Doppler data, which has noise of 0.1 mm/s for data taken at 60-second intervals at a 1-sigma level. This noise is slightly conservative compared to the requirements: radio science requires that the Doppler accuracy at a 60-second rate shall be 0.0733 mm/s at the 1-sigma level, and the DSN will provide Doppler at 0.05 mm/s 1-sigma over 60-second intervals. The low-gain antenna will have most of the viewing opportunities during the radio science experiment, and is able to obtain these measurements with a half-angle field-of-view around the boresight of 60 degrees (as verified by the spacecraft team). The other important data inputs are the NavCam optical navigation images. The NavCam images are taken once every 10 minutes when Bennu is in view.

The radio science team will use spacecraft-telemetry tracking to determine the mass of Bennu and estimate the gravity field to second degree and order, with limits on the fourth degree and order. Knowing the mass estimate and shape model, the team will compute the bulk density of Bennu. Together, this information constrains the internal structure and possible density inhomogeneity. The gravity-field knowledge provides input to the global surface-gravity-field map, which consists of a table where for each surface facet of the asteroid a value for the local gravity-field vector is defined. In addition, a global surface slope map will be generated, consisting of a value of the slope and its direction for each facet of the asteroid shape model. These maps provide information on regolith mobility and will be used to identify areas of significant regolith pooling as potential sampling sites.

**OLA-Based Shape Modeling**

OLA data will be used to generate a second, independent shape model that is solely based on ranging data. This shape, along with the SPC-derived shape, will allow for intercomparisons that will provide the ability to assess the quality of these datasets and enhance the basis upon which operational decisions are made. OLA will also provide precise measurement of surface tilts. At a global scale, OLA will measure the shape of Bennu to provide insights on the geological origin and evolution of the asteroid. Combined with a carefully undertaken geodesy campaign, OLA-based precision ranges, radio science (two-way tracking) data, and OCAMS images will yield constraints on any global-scale internal heterogeneity of Bennu and hence provide further clues to its origin and subsequent collisional evolution. OLA-derived global asteroid maps of slopes, geopotential altitude (Scheeres et al. 2016), and vertical roughness will provide quantitative insights into how surfaces on Bennu evolved subsequent to the formation of the asteroid (Cheng et al. 2002; Barnouin-Jha et al. 2008). The quality of our products will be sufficient to directly measure flow structure in the regolith if present, including flow fronts, lee-side shadow zones behind blocks, and any evidence for block imbrication (Miyamoto et al. 2007). Establishing any connection between surface lineaments that possess measurable topography, and the evidence that their surface distribution might be related to other geological features (such as craters) will provide additional constraints on the internal nature of Bennu (e.g., Marchi et al. 2015).

The science team is also responsible for generating data products to support the onboard NFT autonomous guidance system. They will produce a catalog of up to 300 NFT features consisting of the following: a position defined in asteroid-center-fixed coordinates; a two-dimensional array of displacement (heights) relative to a reference plane above the asteroid surface to represent the shape; and a two-dimensional array of relative albedo values to capture variations in how light reflects off the asteroid surface. Relative albedo is the average of the relative surface reflectance at the given grid point computed from all images in which the grid point is visible. This requirement ensures the production of a catalog with an adequate number of sufficiently defined features for NFT to perform its functions of navigation state estimates and the time of touch estimate during the final descent to the asteroid surface. Each feature is entered into the NFT catalog only when it has been tested with sufficient fidelity to ensure that the NFT will successfully correlate the feature in flight.

**Site-Specific Surveys**

Analysis of the global maps produced during Detailed Survey and the global topography and gravity-field data collected during the Orbit-B phase will result in identification of up to twelve viable sample sites. In order to down select to the top two sites, all candidate sites will be studied to assess their safety, sampleability, and science value. Site-specific Orbit-B observations will include OCAMS imagery from multiple emission and incidence angles (though at a limited and relatively unfavorable range of phase angles), targeted OLA observations, REXIS X-ray spectroscopy, and OTES thermal emission measurements to confirm predictions for maximum temperatures at the site. The OCAMS images will be combined to produce a local photomosaic and a thematic map of cobbles for each candidate site. The OLA data will be processed to generate site-specific digital terrain maps. By the end of this Orbit-B phase, two of these sites will be prioritized for detailed reconnaissance.

### *6.4.8 Reconnaissance Phase*

The Orbit-B phase will conclude with a down-select to a primary and a secondary sampling site. These two sites will be the target of a series of low-altitude reconnaissance observations (Figure 32). These observations will consist of high-resolution images and LIDAR-ranging measurements taken from 225 m above the surface to resolve objects as small as 2 cm. Spectral observations and context imagery will be obtained during higher reconnaissance passes at 525 m from the surface. Both sites will be fully characterized so that, if a second sampling attempt is required, the team can immediately begin planning to perform the TAG at the secondary site.

[Insert Figure 32 here]

To perform the “recon” flybys, the spacecraft will leave the terminator plane orbit at the equatorial crossing and fly in a prograde sense across the sunlit side of the asteroid at an altitude of 225 m above the surface. The two candidate sampling sites will be observed between 40° and 70° solar phase angles. The spacecraft will perform a maneuver to recapture into the terminator plane orbit 5 hours after it departed. Two weeks of orbital operations will take place before the next 225-m recon flyby, during which maneuvers are designed and orbital phasing maneuvers are performed. For each site, after the first set of two 225-m recon flybys are completed, there will be one 525-m altitude recon flyby to obtain spectroscopic data between 40° and 70° solar phase angles over the expected sampling region. Between sorties, the science data will be processed, analyzed, and interpreted to characterize the primary and backup sample sites.

Reconnaissance data are key to developing the site-specific safety, sampleability, and science value maps at local scales. The cm-scale images will be analyzed to map the variation of particle size frequency distributions (PSFDs). Using site-specific mosaics, regolith grains ≥2 cm will be counted, and the results will be stored in the hazard/particle database. The database can be queried to provide the precise location and size of all regolith grains that have been counted in a given area. This information will be used to derive the PSFD function of any defined area that has been reconnoitered. Spectral data from OVIRS and OTES will be combined to produce site-specific mineral and chemical maps and a site-specific dust-cover-index map. OCAMS (via SPC) and OLA data will be used to produce detailed site-specific tilt and topographic maps. MapCam color images will be mosaicked into site-specific color ratio maps. OTES surface temperature information will be used to produce site-specific temperature maps (at the time of sampling) and site-specific thermal-inertia maps. All of these products are combined using predefined and tested algorithms to generate the final set of site-specific safety, sampleability, and science value maps. At the conclusion of the Reconnaissance phase, the site-selection board will make its recommendations to the Principal Investigator for a final decision on the primary and secondary sample sites. Reconnaissance completes much of the science investigation needed to meet the third mission science requirement—provide high-resolution sample context and document the nature of the regolith and the sample site. It also lays the groundwork for completing the primary mission objective—sample return.

### *6.4.9 Rehearsal*

The mission has adopted a sampling strategy referred to as Touch-And-Go, or TAG. TAG uses the momentum of a slow, descending spacecraft trajectory to maintain contact with the surface for a few seconds, just long enough to obtain a sample, followed by a controlled back-away burn. One of the most obvious benefits of TAG is avoiding the need to develop a system for landing on the surface and collecting a sample. The microgravity environment at Bennu’s surface makes extended contact problematic, requiring a system to secure the spacecraft to the surface to ensure that sampling does not launch the spacecraft away from the surface in an uncontrolled manner.

Because TAG is a critical event, the team has adopted a “go slow” approach that includes at least two rehearsals. A methodical and incremental series of approach events will be executed to safely prepare for the sample collection. The TAG rehearsal maneuvers follow a similar structure to the actual TAG maneuvers, beginning with an orbital phasing maneuver about four days prior to the nominal TAG time. The phasing maneuver leads to the spacecraft being at the proper point in the orbit at the proper time to descend to the desired point on the asteroid’s surface, accounting for the asteroid’s rotational state.

The first rehearsal will exercise the period from departure to a predefined “Checkpoint” before returning to orbit (Figure 33). The orbit departure maneuver places the spacecraft on a prograde trajectory that brings it to a point 125 m above the TAG site. The orbit departure latitude is opposite the TAG site latitude, and approximately four hours of flight time elapse between orbit departure and arrival at Checkpoint. The second rehearsal takes the spacecraft from orbit to a “Matchpoint,” where the spacecraft achieves a hovering state over the desired sampling location before a return to orbit. During each rehearsal, the spacecraft will collect and analyze tracking data, LIDAR ranges (from the GN&C LIDAR system), and OCAMS and TAGCAMS imagery to allow the team to verify the performance of the flight system before proceeding to the next step. Only when the team and NASA are satisfied that the chances of success are acceptable will the team proceed with TAG. The team anticipates the first TAG attempt will take place in July of 2020. The entire event will be imaged by SamCam at a rate of just under one frame per second.

[Insert Figure 33 here]

### 6.4.10 Sample Acquisition

**TAG Implementation**

The mission uses a novel method to collect the sample. The TAGSAM head is a simple annulus with a filter screen on the outside circumference, held at the end of an articulated arm that is extended several meters from the spacecraft. The head is mounted on a wrist assembly with a compliant U-joint that allows it to articulate and make full contact even if the angle of approach is not normal to the surface. During the collection time, which can last for up to 8 seconds for a soft surface, high-purity nitrogen gas is injected into the interior of the annulus where fines and small pebbles up to 2 cm in size are entrained in the gas flow. If the TAGSAM head is firmly seated on the regolith, the only path for the gas and entrained regolith is through the wire-mesh filter, where the material is captured.

Descent to the surface begins with a maneuver from the 1-km orbit (Figure 34). The next maneuver is performed at Checkpoint and places the spacecraft onto a trajectory that intersects the asteroid, with a 16 cm/s approach velocity. Approximately 600 seconds later the spacecraft arrives at the Matchpoint location 55 m above the asteroid's surface, where a final maneuver is performed to adjust the spacecraft velocity such that the vertical velocity at contact will be 10 cm/s and the lateral velocity will be zero (±2 cm/s). Following the last maneuver, the spacecraft approaches the surface along the vector normal to the sampling plane. MapCam and TAGCAMS images are collected to provide additional verification of the trajectory and provide photo documentation of the sample site. OVIRS and OTES continuously acquire data during this phase.

[Insert Figure 34 here]

Because trajectory dispersions following the deorbit maneuver are too great to use a predefined set of Checkpoint and Matchpoint maneuvers, spacecraft state information from either the LIDAR or NFT will be used to update the spacecraft state at Checkpoint. This updated state will then be used to autonomously adjust both maneuvers. The TAG dynamics are driven by the surface topography and response to TAGSAM contact, as well as by the ability of the FDS team and the GN&C subsystem to deliver the spacecraft to the surface with an accurate touchdown velocity and minimum attitude and rate errors. Both LIDAR-guided TAG and NFT are capable of monitoring the descent corridor to ensure the spacecraft is on the proper trajectory for a safe TAG event.

Sample collection occurs when surface contact is sensed. At this point, a pyro valve opens to a bottle of high-purity nitrogen gas. The gas is injected into the regolith on the surface; the regolith is fluidized and directed into the sample collection filter. The surface contact pads also acquire fine-grained material upon touching the asteroid surface. After the spacecraft sampling system has contacted the asteroid's surface and ingested the regolith sample, a 0.7 m/s maneuver is performed to move the spacecraft up and away from the asteroid. The entire TAG event will be documented through imaging with SamCam.

**TAG Reconstruction**

The mass of the collected sample will be determined using a technique based on measurement of a change in the moment of inertia of the spacecraft (May et al. 2014). The TAGSAM arm will be extended orthogonal to the sampling configuration, along the X-axis of the spacecraft. The mass of the sample is determined by comparing the moment of inertia of the spacecraft with the TAGSAM arm extended after the sampling attempt to a similar measurement made before TAG and assessing the change.

After the sample-mass measurement, the articulated arm positions the TAGSAM head for visual documentation by SamCam. Using the diopter lens in its filter wheel, SamCam will refocus to acquire a series of in-focus images of the TAGSAM contact pads. These data will consist of a suite of postsampling images taken of the bottom of the TAGSAM head at a variety of incidence and emission angles. They are analyzed to provide verification of collected surface sample. In addition, it is possible, though not required, that SamCam images will include regolith grains within the bulk collection volume, depending on geometry of the Sun relative to the grains and TAGSAM head at the time of imaging.

A parallel effort to evaluate the success of the sampling event utilizes a dynamical analysis to reconstruct the force profile on the spacecraft resulting from asteroid contact. Asteroid contact forces will be reconstructed using received telemetry from OCAMS, TAGCAMS, and the spacecraft GN&C system. During TAG, the spacecraft reaction wheels and attitude-control thrusters will be enabled to damp out any torques that are imparted on the spacecraft and prevent the spacecraft from contacting the asteroid surface with a component other than the TAGSAM head. Accurate reconstruction of the TAG event must incorporate the nontrivial spacecraft dynamics accurately, and these analyses must also include the sample-head motion and pogo-spring compression in the TAGSAM forearm. Spacecraft telemetry provided from the TAG event is used to reconstruct spacecraft dynamics and solve for asteroid-contact forces. The output is a time history of forces acting on the TAGSAM sample head derived from these inputs. In addition, a video of the sample acquisition event will be produced from a series of SamCam images.

The spacecraft telemetry and images are assessed against the results of discrete-element modeling of TAGSAM interaction with regolith. In these models, TAGSAM-shaped solid objects are simulated penetrating granular material (Figure 35). These simulations include relevant material properties for the granular medium (coefficients of friction, cohesion, and an adjustable size distribution of particles and cohesion), the entire 6-degree of freedom for TAGSAM, and the pogo-spring behavior. The results of these simulations are compiled into an “atlas” of numerical modeling outcomes. The reconstructed spacecraft motion and torque profile during the actual TAG event will be matched to the most similar entry in the force atlas to provide a quick-lookup capability to constrain the regolith properties and response to TAG.

[Insert Figure 35 here]

Validation of the numerical modeling is done by direct comparison with laboratory measurements. In these experiments, the team performed low-velocity impact experiments in 1-*g* and micro-*g* into a wide range of granular materials to calibrate the numerical models. A number of measurements were made on a series of analog materials to serve as calibrations for both numerical-continuum models and the granular-mechanics models. The numerical models were calibrated against these laboratory results and show close matches to the laboratory data.

The TAG reconstruction data product is useful both for critical mission decision making and for long-term scientific study of asteroid regolith properties. First, combination of the TAG force profile along with the sample-mass measurement and image analysis will help determine whether sufficient sample was collected. In the event that insufficient sample was obtained, understanding the details of the TAG will help assess if the problem was due to the properties of the sampling site or a problem with the spacecraft. Second, details of the interaction of the spacecraft and TAGSAM with the surface of Bennu will yield information about the bulk physical properties of the regolith, essentially building the tools to maximize scientific return from the spacecraft–asteroid interaction.

**Sample Stowage**

Confirmation of successful sample acquisition will result in a decision to stow the TAGSAM head in the SRC (Figure 36). Once the Principal Investigator agrees that the baseline amount of sample has been collected, the arm places the head in the SRC for return to Earth. TAGCAMS images are used to verify that TAGSAM head placement is properly aligned within the SRC capture ring. The arm will also be retracted slightly after capture to "tug" on the head and ensure proper seating. Once the team verifies that the TAGSAM head is properly affixed within the SRC, an explosive bolt will fire to sever the connection between the TAGSAM head and arm. After successful stowage, the spacecraft will be put in a slow drift away from Bennu and placed in a quiescent configuration until departure in March 2021 for the Return Cruise phase back to Earth.

[Insert Figure 36 here]

### *6.4.11 Return Cruise*

Nominal departure for Earth will commence with a burn in March 2021 that places the spacecraft on a ballistic trajectory for Earth return in September 2023 (Figure 37). This phase of the mission will be relatively quiescent in terms of spacecraft operations, with no large maneuvers required. However, the science team will benefit from the two-and-a-half-year Return Cruise phase to complete the long-term science objectives of understanding Bennu's geologic and dynamic history and to evaluate the post-encounter knowledge against the parameters in the Design Reference Asteroid document compiled from the pre-encounter astronomical characterization campaign.

[Insert Figure 37 here]

**Geological and Dynamical Evolution**

During the Return Cruise phase the team will compile all remote-sensing data and map products to constrain the nature and history of geologic material on the surface of Bennu and its dynamical history and orbital evolution. The geological analysis includes analyzing the particle sizes, shapes, sorting, and spatial distribution of surface material and inferring the geologic units, surface age, and surface processes that dominate asteroid geology. The team will use this information to reconstruct the geologic history of Bennu and develop geologic hypotheses that are testable with data obtained from analysis of the returned sample.

The hypotheses for the dynamical history of Bennu have been assembled in a series of papers based on the astronomical data set (Campins et al. 2010; Delbo and Michel 2011; Walsh et al. 2013; Bottke et al. 2015; Lauretta et al. 2015). In these studies, the authors developed a hypothetical dynamical timeline of how Bennu formed as a fragment from a large asteroid disruption event and how it traveled from the main asteroid belt to NEA. This timeline will be tested based on: Bennu's crater population; the measured magnitudes of the Yarkovsky and YORP effect on Bennu; and the nature of Bennu's surface properties (e.g., has Bennu's surface been desiccated upon close approach to the Sun, has it been hit by high-velocity impactors). Data products will involve surface images, geologic maps of Bennu's features, and estimates on the strength and magnitude of the Yarkovsky/YORP torques. This study will integrate these data to address the driving questions as to whether or not Bennu is related to any known asteroid family or a lone member of the "background" asteroid population. If a link to a known asteroid family is established, the team will explore the collisional history and dynamical lifetime of the Bennu main-belt family. In addition, they will seek to understand if there are siblings of Bennu in the NEA or main-belt populations that are following a similar dynamical path by comparing Bennu to the broader asteroid population.

Satisfying the fourth mission objective requires a detailed study of the Yarkovsky acceleration and evaluation of future planetary encounters and a reassessment of Bennu's impact hazard. To achieve this objective, the team will measure the thermal-recoil acceleration on the asteroid based on tracking of the asteroid in its orbit and the observed deviation from a gravity-only trajectory. Given a joint global solution of the asteroid gravity and ephemeris, and the spacecraft ephemeris from radio science, the team will derive a pseudo-range measurement from the Earth to the asteroid center of mass. During encounter, precision tracking of the spacecraft, in combination with modeling of the spacecraft motion relative to Bennu, will provide the most accurate determination of the Yarkovsky effect ever accomplished, achieving a signal-to-noise ratio >400. This investigation will strongly improve upon the existing measurements derived by Chesley et al. (2014).

The rotation state and possibly the shape of Bennu are the result of YORP radiative torques. The spinning-top shape of Bennu is shared by many objects in the near-Earth population. Although these shapes can be the product of other processes (e.g., Statler 2015), many believe these are the result of a rubble-pile asteroid response to being spun up to fast rotation speeds via YORP torques. The YORP effect is a "windmill"-like phenomenon related to radiation pressure (from incident, reflected, and reemitted photons) acting on the asymmetrical shape of an asteroid (Figure 38). The more propeller-like shape that an asteroid has, the faster it spins up or down. YORP also creates a torque that can modify the spin vectors of small bodies like Bennu. These torques frequently drive asteroid obliquities to common end states near 0° or 180° (e.g., Bottke et al. 2006; Vokrouhlický et al. 2015). Modification of rubble-pile asteroids occur when rotational angular momentum is added to or subtracted from the body, causing blocks and particles to move in response to the resulting centrifugal forces (e.g., Walsh and Richardson 2006). Thus, YORP may add enough angular momentum to produce downslope movement, mass shedding, and shape changes (e.g., Scheeres et al. 2016). In addition to establishing the spinning-top shape, the resulting wide-scale resurfacing may have brought fresh, unweathered material to Bennu's surface.

[Insert Figure 38 here]

The remote-sensing data will be combined with ground-based and Hubble Space Telescope observations of Bennu's rotation state to measure the YORP effect. Using the pre-encounter astronomical observations, the team has determined both the sidereal rotation period and orientation of rotation poles for Bennu. Inclusion of the rotation state characterization from the asteroid encounter may reveal secular effects due to the YORP torques over the twenty-year arc of rotation-state information. Detailed measurement of the rotation state during the encounter, compared to the period determined from ground-based light curve measurements, will allow us to detect the YORP effect to within $10^{-3}$ °/day/year. If this information is combined with previous light curve observations, the precision of the detection limit may improve by a factor of ~5.

The science team will also develop a comprehensive thermophysical model of the asteroid using data obtained during the asteroid encounter. Comparison of the Yarkovsky and YORP effects predicted from this first-principles approach to the direct measurement of the resulting asteroid acceleration and change in rotation state will test our understanding of these phenomena and lead to a substantial improvement in our knowledge of the fundamental parameters that give rise to these effects.
The team will also focus on the short-term dynamical evolution of Bennu. They will constrain the timeline for Bennu–Earth close approaches in the recent past and future and address whether any of these are likely to have resurfaced Bennu. The team will also address the impact hazard and define measurement requirements to refine the probability of impact. During encounter, precision tracking of the spacecraft, in combination with modeling of the spacecraft motion relative to Bennu, will provide the most accurate determination of the Yarkovsky effect. This increase in position knowledge leads to better understanding of the possible impact threat from Bennu specifically and the physics of the Yarkovsky effect in general.

**Ground-Truth for Asteroid Astronomy**

The science team is responsible for comparing the properties of asteroid Bennu determined from ground- and space-based telescopes with data obtained by the spacecraft instruments during the asteroid encounter. The measurements made at the asteroid during encounter will allow the team to critically test the pre-encounter understanding of Bennu, built from astronomical observations. A key data product resulting from the asteroid encounter is the Design Reference Asteroid scorecard. This document will track how well (or poorly) our pre-encounter understanding of Bennu matched reality. In cases where our telescopic observations provided accurate information, we will be able to continue to confidently apply these techniques to other asteroids. However, the most interesting results will be obtained in areas where the ground-based observations proved inaccurate. In these instances, we will be able to use the additional encounter observations to thoroughly review and refine our techniques. The resulting knowledge will improve our ability to characterize small bodies throughout the Solar System and complete the fifth objective of OSIRIS-REx: improve asteroid astronomy by providing ground-truth data for Bennu observations.

## 6.5 Data Archiving

The OSIRIS-REx team is dedicated to long-term preservation of the mission science data in a form that will be useful to future researchers. The mission PDS archive is organized into bundles for each detector (OCAMS, OTES, OVIRS, OLA, REXIS, TAGCAMS), DSN data, SPICE, higher-order data products bundled by scientific discipline (altimetry, astrometry and photometry, image processing, radio science, spectral processing, regolith development, and thermal analysis), and two special products bundles, a sample-site selection bundle that contains the data products used to make the site selection and a remote-observations bundle that contains data and pointers to Bennu observational data collected by team members from ground- or space-based telescopes. Each bundle contains data collections grouped by data processing level and by time interval. Each PDS bundle also contains a document collection, a context collection, and a schema collection to provide the appropriate ancillary information to properly interpret and use the data. The OSIRIS-REx archive is designed to and implemented in the PDS4 standard.

## 6.6 Sample Return

On September 24, 2023, the SRC will land at the UTTR west of Salt Lake City, UT. Four hours before entry, the SRC will be released from the spacecraft bus, and a divert maneuver will be executed to place the spacecraft into a heliocentric orbit. The SRC will enter Earth's atmosphere at more than 12 km/s. The majority of SRC deceleration occurs via drag against the SRC heatshield; the remaining deceleration comes from a pair of parachutes, first a drogue and then a main parachute. The SRC will soft land at UTTR.

After recovery of the SRC at the UTTR, it will be partially disassembled under controlled conditions in a temporary clean room set up at the UTTR to remove the sample canister and the enclosed TAGSAM head. SRC de-integration will be done by Lockheed personnel assisted by NASA Johnson Space Center (JSC) curation and Science Team members. The sample canister, SRC heatshield, SRC backshell, and environmental samples and materials from the recovery site will be packaged in separate containers.

After recovery teams ensure that the SRC is safe to handle, it will be airlifted to a custom-built curation facility. Stardust-heritage procedures will be followed to transport the SRC to JSC, where the samples will be removed and delivered to the OSIRIS-REx curation facility. After Earth return, samples will be made available to the worldwide scientific community, who will perform precise analyses in terrestrial laboratories to understand the nature of Bennu's regolith and further constrain its geologic and dynamic history.

## 6.7 Curation and Sample Distribution

After receipt at JSC, the samples will be characterized by the preliminary-examination team, consisting of science team members and JSC curation staff. The science team will be allocated no more than 25% of the sample, with the rest archived for long-term study or transferred to our international partners. The spacecraft carries four types of material that will be curated and analyzed by the mission science team: bulk Bennu samples from the TAGSAM head; surface samples from the surface contact pads; witness plates for contamination monitoring purposes; and spacecraft hardware, which may also have adhering asteroid dust. Two cleanroom laboratories at JSC are dedicated to support these activities: the new ISO-5 sample laboratory to be built in 2019, and the existing ISO-7 Microparticle Impact Collection (MIC, previously known as the Space Exposed Hardware Lab).

The OSIRIS-REx preliminary-examination team (PET) will complete a sample catalog of the returned sample, which will be placed online within six months of Earth return. Representative samples will also be sent to the Canadian Space Agency in exchange for their contribution of the LIDAR, and to the Japanese Aerospace and Exploration Agency as part of the cooperative agreement with NASA for collaboration between the OSIRIS-REx and Hayabusa2 teams. As with most of NASA's sample collections, a portion of representative sample will also be stored at a NASA remote storage facility at White Sands Test Facility, NM (Allen et al. 2011). Based on this information, investigators from around the world may apply for material and witnesses using an established astromaterials loan request process.

At the completion of mission activities in 2025, primary responsibility for samples will be transferred to the JSC curation staff for long-term curation. Bennu regolith samples will be available for the worldwide sample analysis community to study for decades to come. OSIRIS-REx asteroid samples will be allocated to researchers as individual particles, groups of particles, and sliced or subdivided particles, as dictated by research requirements. Witness plates and portions of spacecraft hardware will be available and allocated for analysis in response to external requests. The Curation and Analysis Planning Team for Extraterrestrial Materials will provide critical oversight for all curation activities.

## 6.8 SRC Analysis

There are a number of important reasons to carry out analytical studies of the SRC components. These studies will be multifold and will address issues of scientific and curatorial importance. The materials to be analyzed include: (1) any asteroid regolith found outside the sample canister; (2) the heatshield and backshell of the SRC; (3) thermal sensors; (4) the sample canister air filter; (5) the sample canister; (6) witness plates; and (7) UTTR air, soil, water, and other environmental samples collected from the landing site. The primary science interest with the interior of the heatshield and backshell is to use interior surfaces to assist with contamination control and assessment. The SRC and sample-canister thermal indicators are in place to provide verification that the interior of the SRC remained <75°C since sample stowage. Analysis of the contents of the air filter will provide critical information about the nature of any volatile materials present, including materials derived from the asteroid sample, atmospheric contaminants acquired during reentry, and spacecraft outgassing products. The canister interior can serve as additional surface area where sample outgassing products or spacecraft contamination could have adsorbed. The UTTR environmental samples provide knowledge of any adhering soils, water, or other materials that may have made it into the SRC through the backshell vents.

## 6.9 Sample Analysis

The OSIRIS-REx science team continues to develop the sample analysis plan (SAP), which will serve as the governing document for the analysis of the returned samples. The primary purpose of the SAP will be to ensure that the sample analyses performed by the preliminary-examination team will effectively and efficiently achieve the science requirements relevant to sample analysis. The SAP will provide guidelines for determining which portions of the returned samples, contact pads, and witness plates will be allocated to the PET and which will be archived for later study. The SAP will outline the types and amounts of samples required for each analysis, outline optimal analytical sequences for coordinated analyses, and identify the necessary sample preparation and sample mounting procedures. The final version of the SAP will not be delivered until 2021, after the spacecraft has departed Bennu and the mass and nature of the returned sample is better constrained.

The SAP focuses on addressing hypotheses related to the origin and geological and dynamical history of Bennu (Figure 39). A set of 70 preliminary hypotheses has already been drafted by the science team that is based upon our best understanding of Bennu and its most likely counterparts among meteorites and cosmic dust samples. Some of these hypotheses will undoubtedly be revised based on the detailed geological studies of Bennu and high-resolution imaging of the sampling site. These hypotheses consider the full historical arc of asteroid Bennu, beginning with stellar nucleosynthesis and dust formation in ancient stars, through chemical processes in the Galaxy and nascent solar nebula, planetary formation, and recent geological history.

[Insert Figure 39 here]

The major thematic areas of the SAP are: the Pre-Solar Epoch, the Protoplanetary Disk Epoch, the Geological Activity Epoch, the Regolith Evolution Epoch, the Dynamical History Epoch, and ending with the OSIRIS-REx Epoch. Sample analyses related to the Pre-Solar Epoch will quantify and characterize materials that predate the formation of the Solar System, such as circumstellar dust grains and interstellar organic materials. Hypotheses related to the Protoplanetary Disk Epoch include determining where and how mineral and organic components within Bennu (and its parent body) may have formed in the hot inner nebula and outer Solar System. The Geological Activity Epoch considers ancient planetary processes within Bennu (and its parent body), such as the timing and mode of aqueous and thermal alteration and whether telltale clues in the samples tell us where Bennu was housed in the parent body prior to its disruption. The Regolith Evolution Epoch will address how Bennu's surface materials formed after Bennu was liberated from its parent body and were subsequently altered by exposure to space. The Dynamical History Epoch will address how Bennu came to be in its current orbital configuration from processes such as planetary close encounters, YORP, and Yarkovsky effects. Finally, the OSIRIS-REx Epoch will address the interaction of the spacecraft with the asteroid.

The many and varied hypotheses relating to Bennu's origin and history will be addressed through a series of detailed and extremely sensitive chemical, mineralogical, spectral, and isotopic studies in terrestrial laboratories. Microscale and petrological studies will reveal whether Bennu resembles meteorites or cosmic dust in our collections and will constrain histories of geological processes. Precise isotopic measurements will constrain the age of Bennu regolith, while high spatial resolution ($<1$ μm) isotopic measurements will be used to identify presolar dust grains. Atomic-scale crystallographic studies by transmission electron microscopy will be used to identify signatures of shock from impacts and cosmic ray radiation damage. Studies of the nature and origin of organic phases will include molecule-specific isotopic measurements, characterization of amino acids and other molecules of astrobiological importance, and even measurements at micrometer scales. Finally, the thermal conductivity and heat capacity of the returned samples will be directly measured in the laboratory. These fundamental physical parameters, combined with the state of the regolith on the asteroid surface, drive the thermal inertia and the resulting strength of the YORP and Yarkovsky effects. Thus, the OSIRIS-REx mission will benefit future studies of near-Earth objects as well as main-belt asteroids in many different ways.

# 7 Conclusion

The OSIRIS-REx mission will return samples that present critical links among often disparate studies of asteroids, meteorites, and Solar System formation. The returned samples will furnish important insight into the formation and dynamical evolution of the Solar System and the links between a known asteroid body and meteoritic samples. The measured properties of asteroid regolith will provide important clues on the energy balance of the surface, and the influence this has on the orbital and rotational evolution of Bennu. OSIRIS-REx, together with other small-body missions and telescopic observations, will contribute fundamental insights into asteroids and their past and present roles in planetary systems. Ongoing sample analysis by generations of scientists using cutting-edge tools and methods will guarantee an enduring scientific treasure that only sample return can provide.

# Acknowledgments

The launch of OSIRIS-REx was the culmination of over a decade of hard work by thousands of people across the globe. This event marked the beginning of one of the greatest scientific expeditions of all time—sample return from asteroid Bennu. We thank the multitude of team members and their families for the dedication and support of this mission. This material is based upon work supported by NASA under Contracts NNM10AA11C, NNG12FD66C, and NNG13FC02C issued through the New Frontiers Program. Copy editing and indexing provided by Mamassian Editorial Services.

## FIGURE CAPTIONS

**Fig.1** The OSIRIS-REx spacecraft launched aboard a ULA Atlas V 411 rocket on September 8, 2016 from Cape Canaveral Air Force Station, Florida (United Launch Alliance)

**Fig.2** Orbit diagram of the OSIRIS-REx spacecraft from launch to asteroid arrival, as of the beginning of Approach Phase on August 13, 2018. Included are the spacecraft's positions during Deep Space Maneuver 1 (DSM-1), the Earth Trojan Asteroid Search, the Earth Gravity Assist, Deep Space Maneuver 2 (DSM-2) and the November 2018 rendezvous with Bennu

**Fig.3** OSIRIS-REx mission operations timeline from spacecraft launch in 2016 through the end of sample analysis in 2025

**Fig.4** The Yarkovsky effect causes the orbit of a small body to change over time. As the heat absorbed from the sun during the day is emitted from the night side of the asteroid, a small but continuous thrust is created which pushes the asteroid out of its normal orbit

**Fig.5** Selection diagram for the OSIRIS-REx mission's target asteroid. From the more than 500 thousand asteroids known in 2010, mission scientists down-selected for asteroid accessibility, size, and composition to finally choose the asteroid Bennu

**Fig.6** Radar-derived shape models of Bennu depicting the asteroid from the north, south, east and west hemispheres (Nolan et al. 2013)

**Fig.7** Diagrams demonstrating Bennu's orbit and position as of September 21, 2018. The top diagram depicts Bennu's Earth-crossing orbit and the bottom diagram depicts the asteroid's approximately 6-degree inclination from the ecliptic plane

**Fig.8** The OSIRIS-REx spacecraft installed on a rotation fixture in a NASA Kennedy Space Center cleanroom in preparation for launch (NASA)

**Fig.9** CAD drawing of the OSIRIS-REx spacecraft depicting its principal components

**Fig.10** Progression of the OSIRIS-REx spacecraft's assembly at Lockheed Martin Space Systems in Littleton, Colo. (a) Final machining of the Aft Deck on the 5-axis router. (b) SARA panel post machining spool and insert bonding. (c) The first flight hardware built in manufacturing, the composite sandwich structure of the core cylinder after cure, ready for machining. (d) Mission Principal Investigator Dr. Dante Lauretta and Lockheed Martin ATLO Floor Manager Ben Bryan inspect the core cylinder before assembly. (e) Structural assembly of the aft deck and alignment with the core cylinder. Alignments and assembly work are done on a flat granite table. (f) Composite structure assembly post-alignment verification, ready for static testing. (g) Lifting the assembly to the rotation fixture for initial

testing. (h) First operations in ATLO, the harness routing and installation. (i) Final SARA/TAGSAM installation. (j) Final HGA installation and alignment. (k) Lifting spacecraft into the Lockheed Martin Thermal Vacuum Chamber for Environmental testing. (l) Lifting spacecraft into the shipping container in preparation for transport to NASA Kennedy Space Center in Cape Canaveral, Fla. (m) Completed spacecraft installed on a rotation fixture at Lockheed Martin Space Systems. The balloon in the background was used to simulate weightlessness during testing of the TAGSAM arm. (n) Completed spacecraft prior to fairing encapsulation in the NASA KSC Payload Hazardous Servicing Facility. (Lockheed Martin)

**Fig.11** Labelled view of the OSIRIS-REx instrument deck. The spacecraft, mounted onto a rotation fixture in a NASA Kennedy Space Center cleanroom, has been rotated 90 degrees to place the instrument deck on its side (NASA)

**Fig.12** The Touch-and-Go Sample Acquisition Mechanism (TAGSAM) was created by Lockheed Martin to gather a sample of regolith from Bennu. (a) Top down view of the TAGSAM head (Lockheed Martin). (b) Ben Bryan checks the extended TAGSAM arm during testing in a Lockheed Martin cleanroom. The TAGSAM head is covered to protect it from contamination (Lockheed Martin). (c) Cross-section of TAGSAM operation. Nitrogen gas is forced down from the arm, through the sides of the head to escape from the bottom of the head. The gas fluidizes the regolith on the asteroid's surface and forces it upwards, where it is then captured and collected in the TAGSAM head.

**Fig.13** The Sample Return Capsule is the only part of the OSIRIS-REx spacecraft that will return to Earth at the end of the mission. (a) A technician inspects the open capsule during testing. (b) The closed SRC sits in a Lockheed Martin cleanroom (Lockheed Martin)

**Fig.14** The three cameras of the OSIRIS-REx Camera Suite (OCAMS) were provided by the University of Arizona. MapCam (left), PolyCam (center), and SamCam (right) are responsible for most of the visible light images taken by the spacecraft (University of Arizona)

**Fig.15** The Canadian Space Agency contributed the OSIRIS-REx Laser Altimeter (OLA), a scanning LIDAR (Light Detection and Ranging). OLA will emit laser pulses toward the surface of Bennu and measure the timing of the return pulses to compute the distance between the spacecraft and the surface of Bennu. OLA will use these data to provide high-resolution topographical information about Bennu during the mission (NASA)

**Fig.16** The OSIRIS-REx Visible and Infrared Spectrometer (OVIRS) will measure visible and near infrared light from Bennu, which can be used to identify water and organic materials on the asteroid. OVIRS was provided by NASA Goddard Space Flight Center (NASA)

**Fig.17** The OSIRIS-REx Thermal Emissions Spectrometer, provided by Arizona State University, determines mineral and temperature information by collecting infrared (from 5 to 50 microns) spectral data from Bennu (University of Arizona)

**Fig.18** The Regolith X-ray Imaging Spectrometer (REXIS), a student-built instrument, will determine which elements are present on the surface of Bennu (MIT)

**Fig.19** The TAGCAMS flight unit built by Malin Space Systems. The two Navcams are at the left and center and the Stowcam is on the right. The digital video recorder is behind the center camera head, containing two separate recorder boards. The pocketknife is for scale (Malin Space Systems)

**Fig.20** The OSIRIS-REx mission's fifteen Level-1 science requirements

**Fig.21** Diagram outlining the OSIRIS-REx ground system, its products and its dependencies

**Fig.22** Site selection process to choose the sampling location on the asteroid Bennu

**Fig.23** Timeline of science products to be produced during each phase of the mission from Launch in 2016 though Sample Return in 2023

**Fig.24** United Launch Alliance provided the OSIRIS-REx launch vehicle, an Atlas V 411 rocket. (a) A crane lifts the Atlas V rocket's first stage booster, placing it into position at Space Launch Complex 41 at Cape Canaveral Air Force Station (NASA). (b) Team members guide the solid rocket motor into position for attachment to the Atlas V rocket's first stage booster at Space Launch Complex 41 (NASA). (c) Team members prepare to attach the Centaur upper stage to the Atlas V rocket's first stage booster (NASA). (d) The OSIRIS-REx spacecraft, encapsulated in its fairing, is lifted to the top of the Atlas V rocket assembly (NASA). (e) The OSIRIS-REx mission's Atlas V rocket is ready for launch on Launch Complex 41 (United Launch Alliance)

**Fig.25** Chart depicting the distance of the OSIRIS-REx spacecraft from the Earth (blue) and Sun (orange) throughout the mission

**Fig.26** The OSIRIS-REx spacecraft's trajectory during Asteroid Approach Maneuvers (AAMs) 1-3, which will occur over 28 days during the mission's Approach Phase

**Fig.27** The OSIRIS-REx spacecraft's Preliminary Survey orbit path. The spacecraft will complete five legs to allow for flybys of Bennu's North Pole, Equator and South Pole, using trajectory changes at the Maneuver 1 Preliminary Survey (M1P) though Maneuver 5 Preliminary Survey (M5P) points

**Fig.28** The OSIRIS-REx spacecraft's Orbit A insertion and orbit trajectories. The spacecraft will be put into an orbit around Bennu with a 1.5 km radius on the terminator plane

**Fig.29** Trajectory diagram for the "Baseball Diamond" maneuver, which begins the mission's Detailed Survey Phase. The spacecraft will progress through a series of hyperbolic passes around Bennu, with trajectory changes at the Maneuver 1 Detailed Survey (M1D) through Maneuver 5 Detailed Survey (M5D) points

**Fig.30** Diagram of the spacecraft's Equatorial Stations hyperbolic passes, which will provide the spacecraft's instruments global coverage of the surface of Bennu

**Fig.31** The OSIRIS-REx spacecraft's Orbit B insertion and orbit trajectories. The spacecraft moves through four trajectory changes at the Maneuver 0 Orbit B (M0B) through Maneuver 3 Orbit B (M3B) points

**Fig.32** Diagram of the spacecraft's Reconnaissance Phase trajectories that will provide greater detail on the selected primary and secondary sampling sites

**Fig.33** Diagram of the spacecraft's Checkpoint Rehearsal and Matchpoint Rehearsal trajectories, which will be executed around the selected sample site on the asteroid Bennu

**Fig.34** Diagram of the spacecraft's Sample Acquisition trajectory, during which the spacecraft's TAGSAM head will make contact with the asteroid Bennu to collect a sample of its regolith

**Fig.35** A model TAGSAM (visualized with some slight transparency) penetrates a bed of ~150,000 particles (seen in cross section) with an initial downward vertical speed of 10 cm/s in the microgravity environment of Bennu (g = 20 μm/s2). The cylindrical container has a height and radius of 60 cm (Ballouz 2016 Ph.D. Thesis)

**Fig.36** The spacecraft's TAGSAM head is placed into the Sample Return Capsule during a stowage verification test executed at Lockheed Martin's cleanroom (Lockheed Martin)

**Fig.37** Orbit diagram of the OSIRIS-REx spacecraft's Return Cruise, as of the mission's Sample Return on Sept. 24, 2023. Included are the spacecraft's positions during Baseline Departure from Bennu (Mar. 3, 2021) and Sample Return

**Fig.38** The YORP effect changes the rotation period of a small body according to sunlight's interaction with the body's shape. Sunlight that is absorbed by and emitted from an asymmetrical asteroid will create more torque on the body than sunlight does on a symmetrical asteroid

**Fig.39** Hypotheses developed and considered by the OSIRIS-REx Science Team regarding the journey of the asteroid Bennu from its origin, through its formation and geological activity, to the visit of OSIRIS-REx spacecraft

Figure 1

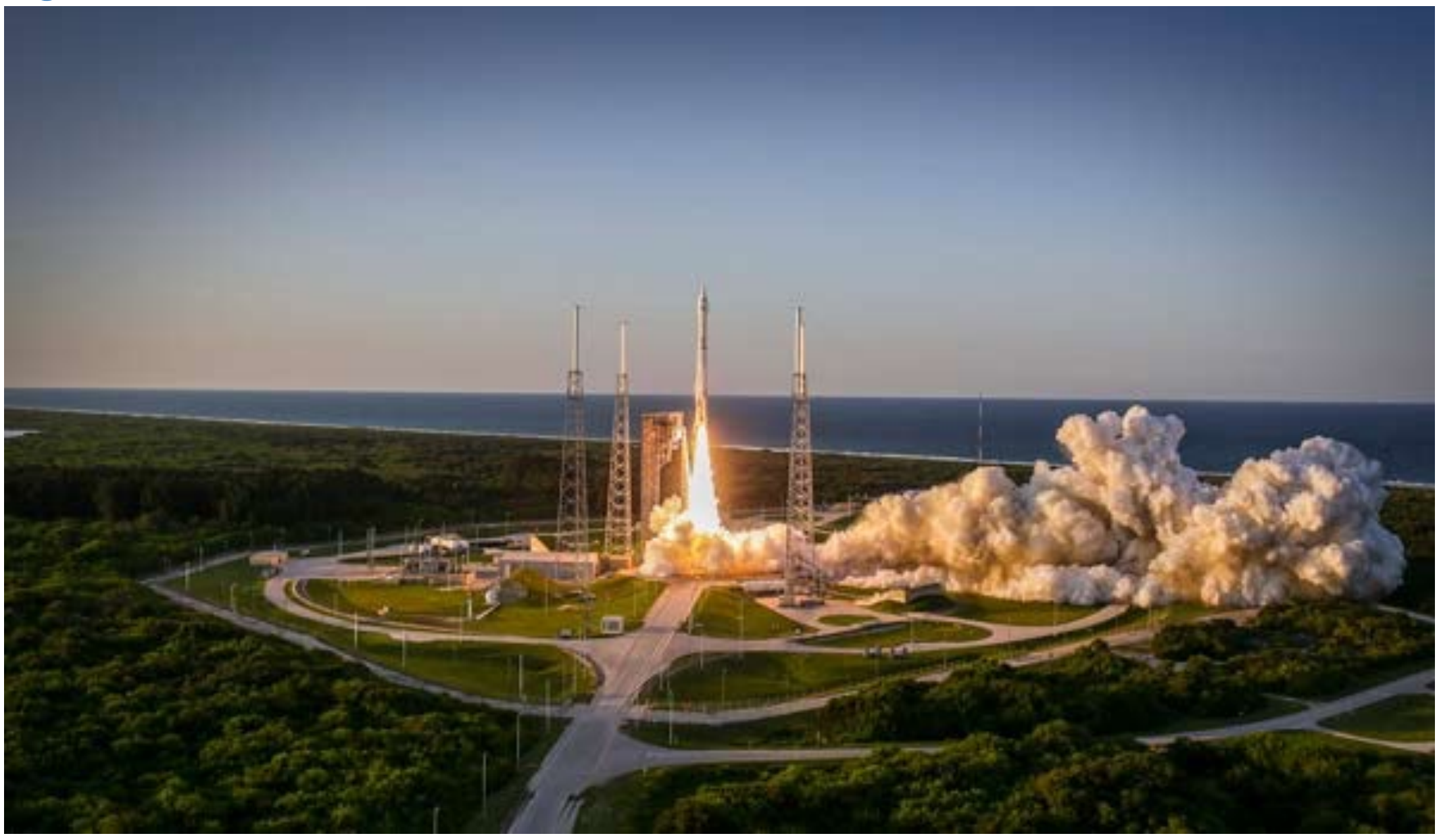

Figure 2

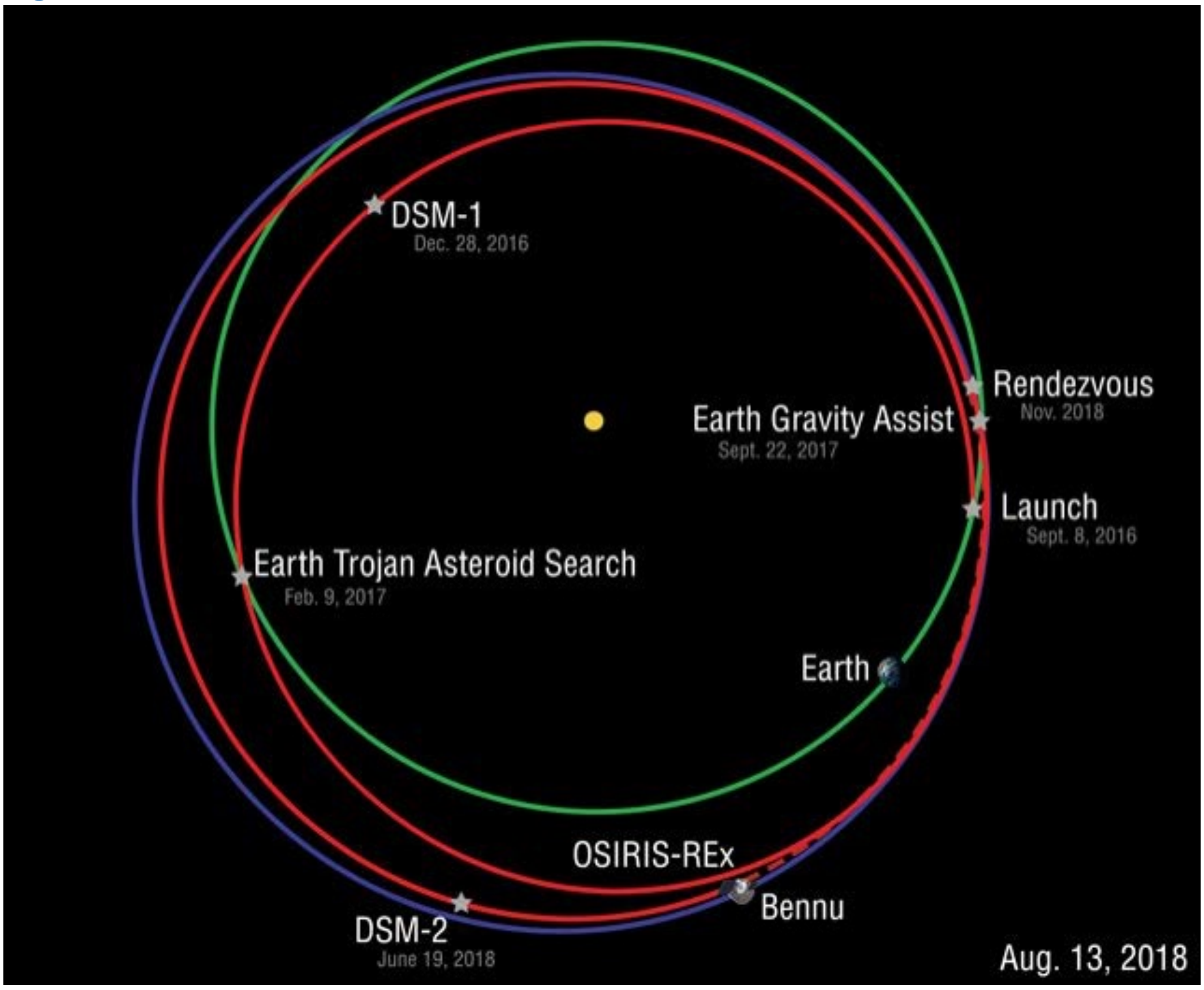

Figure 3

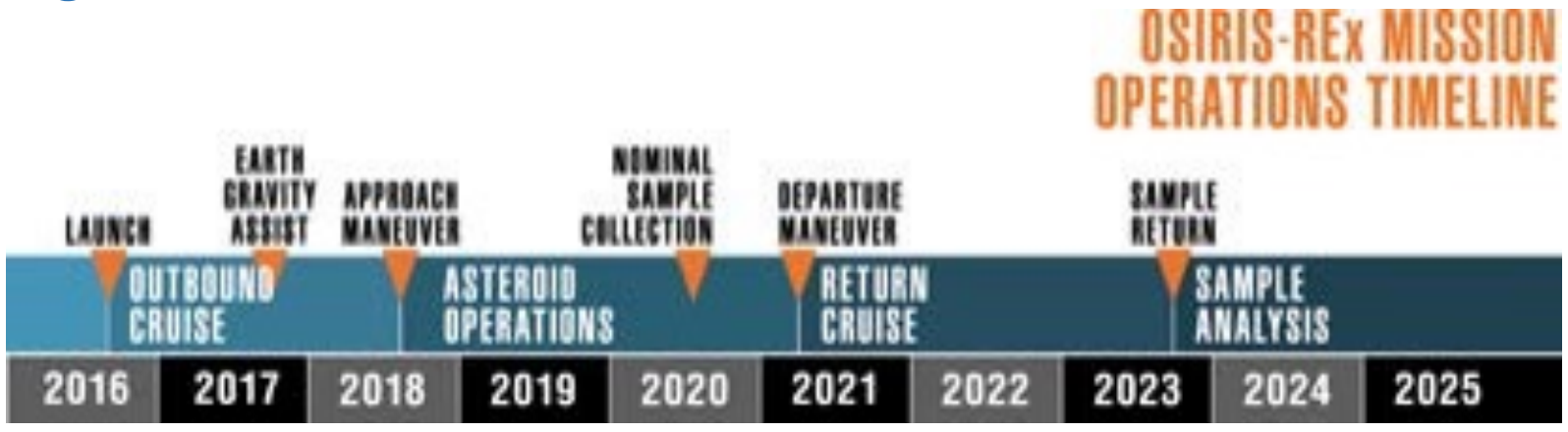

Figure 4

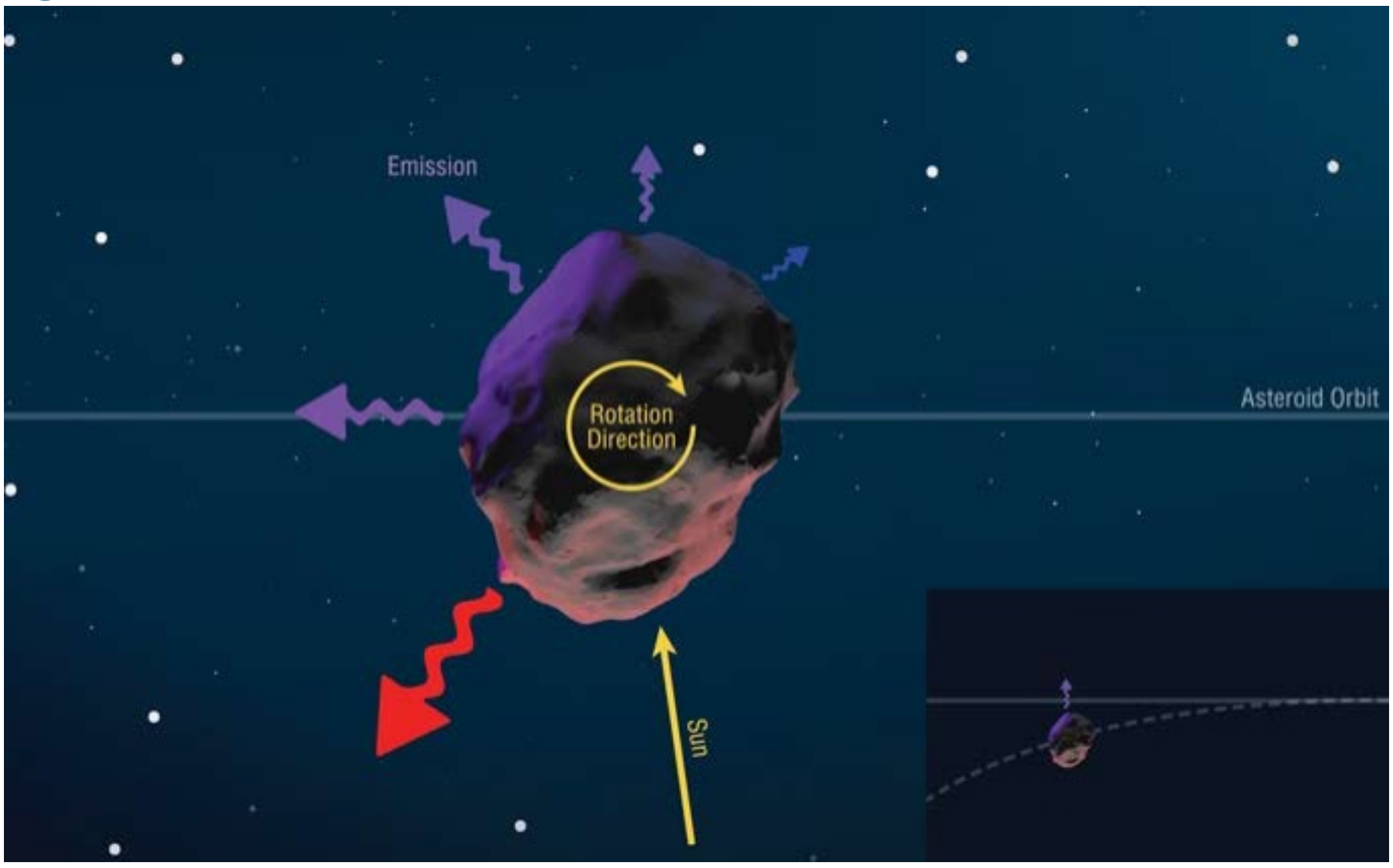

Figure 5

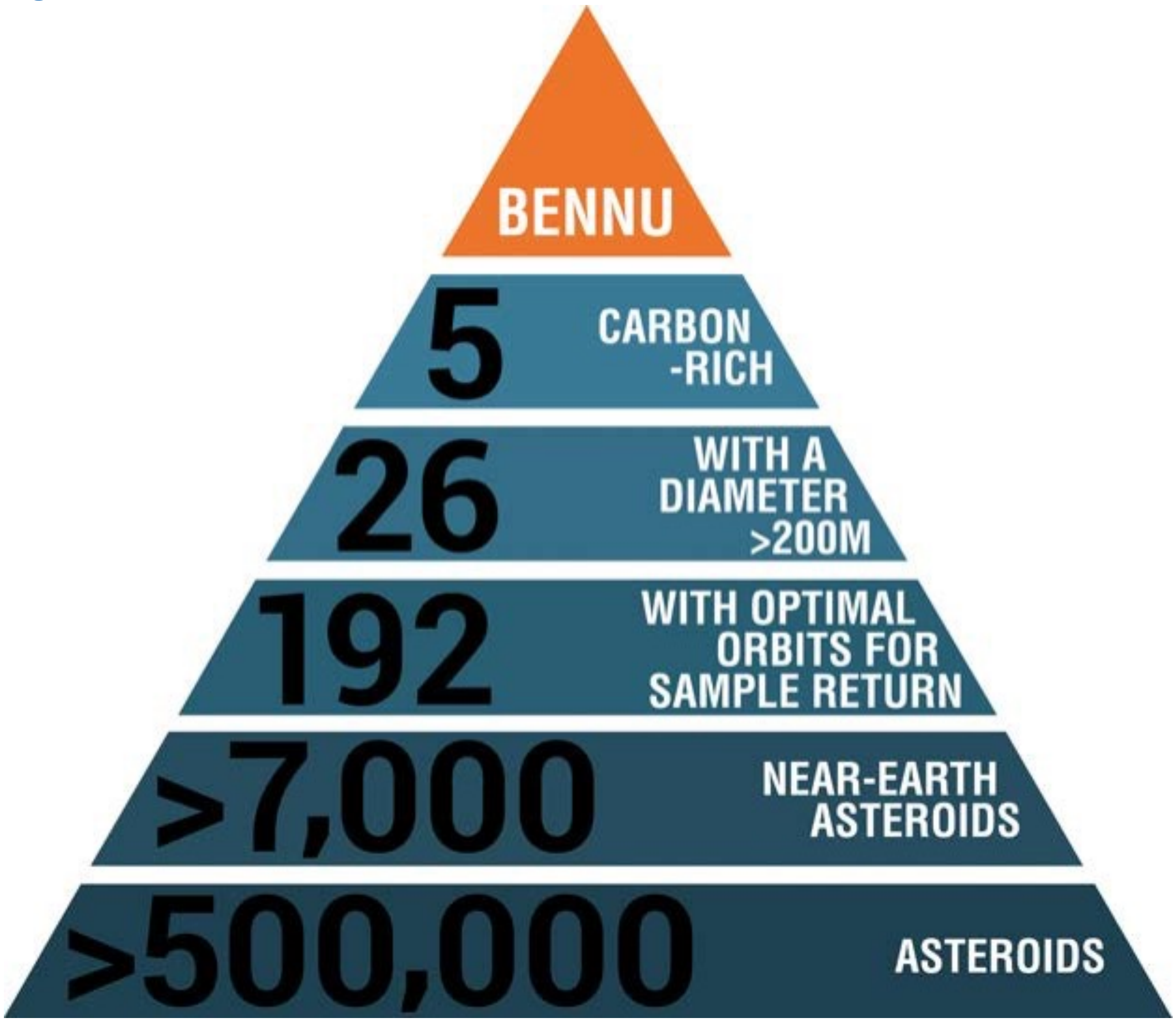

Figure 6

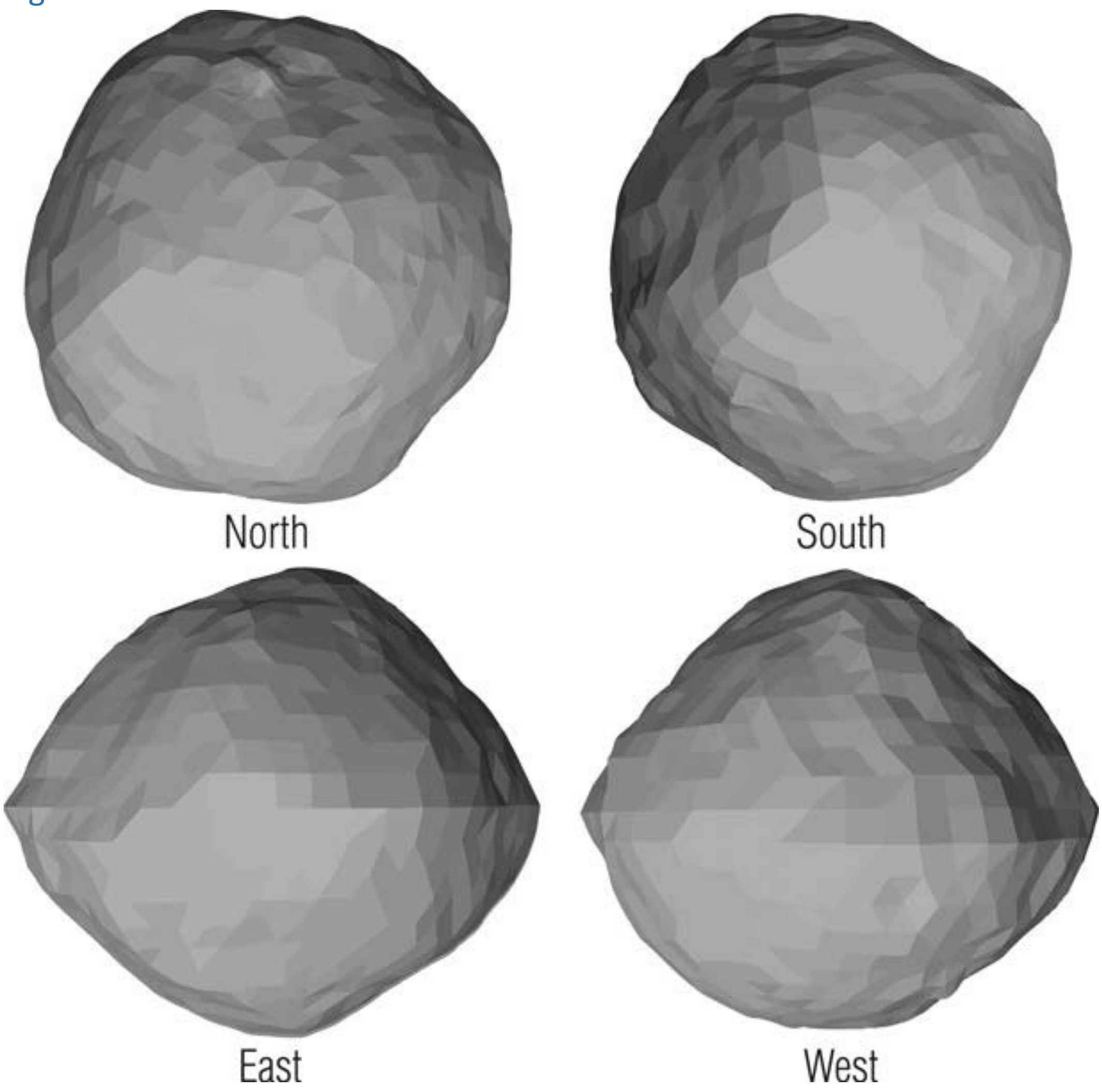

Figure 7

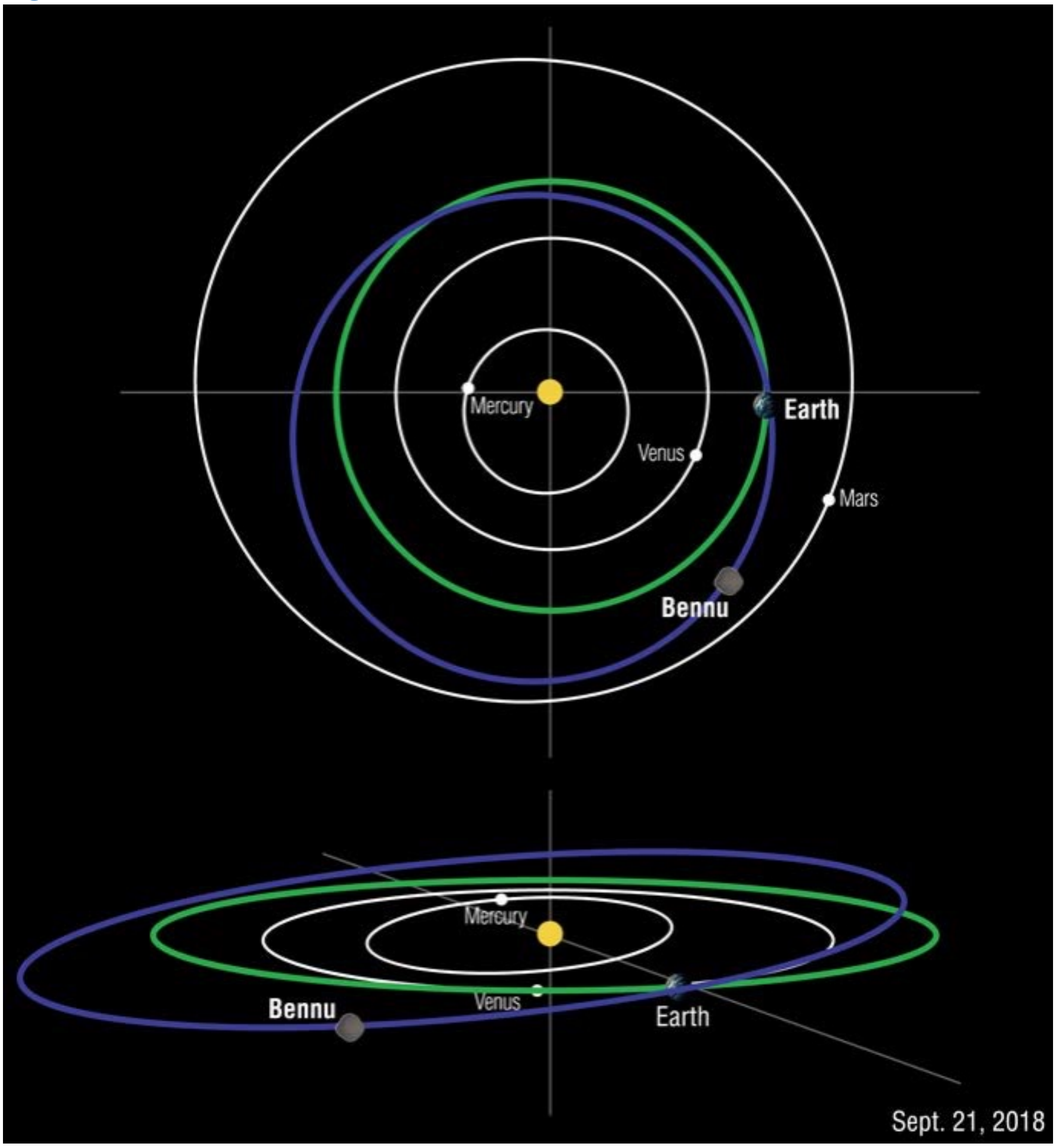

Figure 8

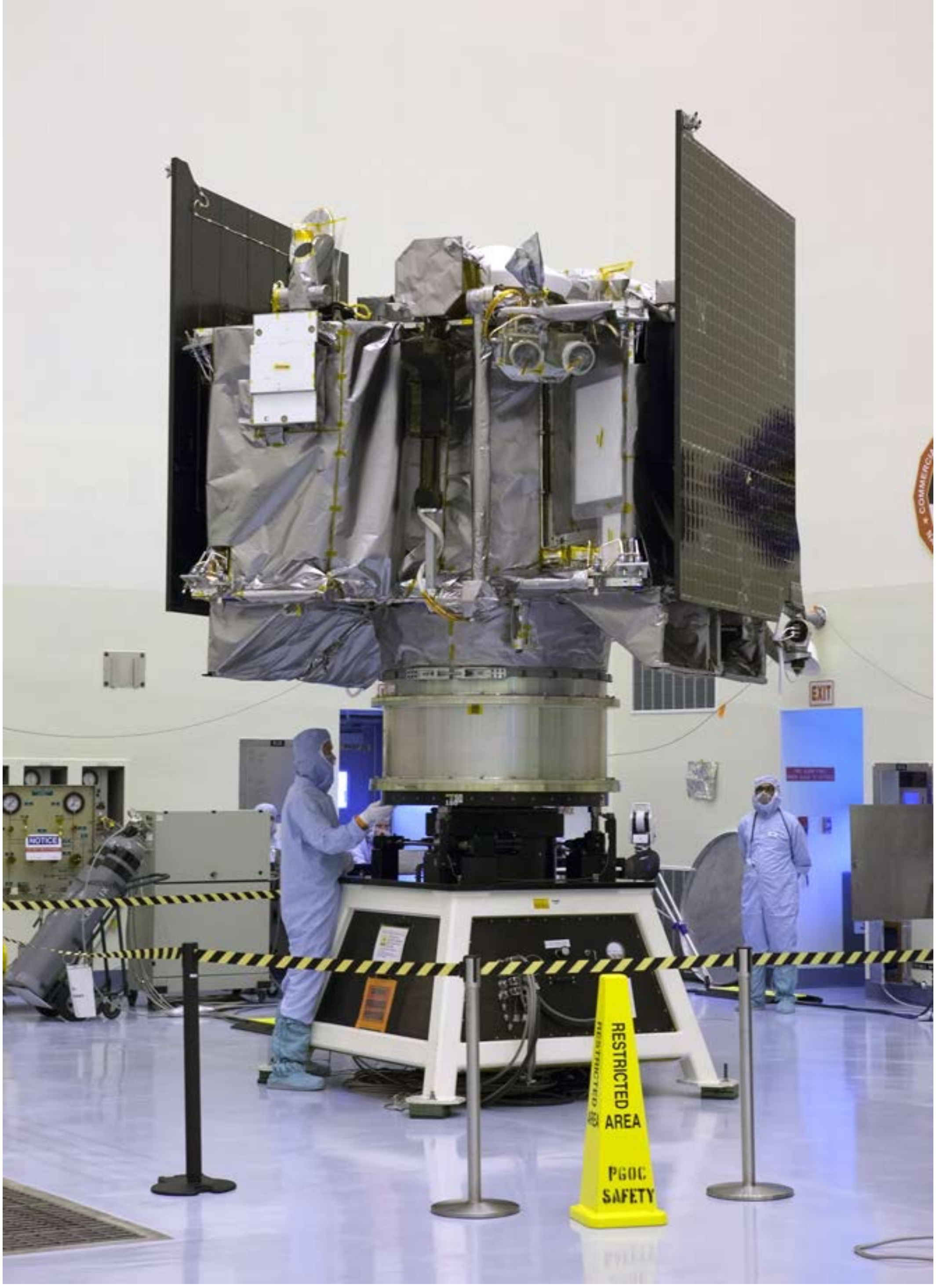

## Figure 9

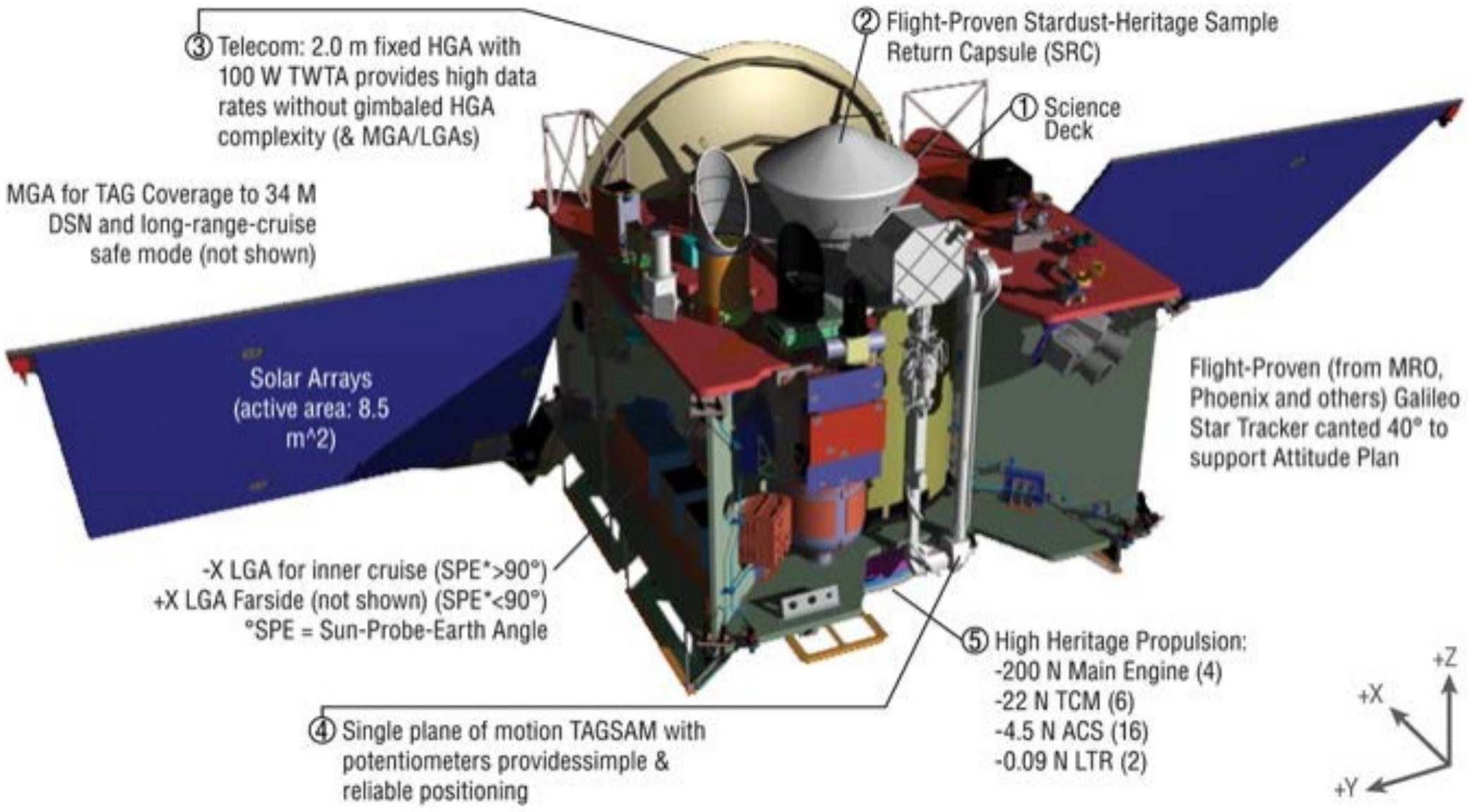
③ Telecom: 2.0 m fixed HGA with 100 W TWTA provides high data rates without gimbaled HGA complexity (& MGA/LGAs)
② Flight-Proven Stardust-Heritage Sample Return Capsule (SRC)
① Science Deck
MGA for TAG Coverage to 34 M DSN and long-range-cruise safe mode (not shown)
Solar Arrays (active area: 8.5 m^2)
Flight-Proven (from MRO, Phoenix and others) Galileo Star Tracker canted 40° to support Attitude Plan
-X LGA for inner cruise (SPE*>90°)
+X LGA Farside (not shown) (SPE*<90°)
*SPE = Sun-Probe-Earth Angle
⑤ High Heritage Propulsion:
-200 N Main Engine (4)
-22 N TCM (6)
-4.5 N ACS (16)
-0.09 N LTR (2)
④ Single plane of motion TAGSAM with potentiometers providessimple & reliable positioning
+Z
+X
+Y

Figure 10

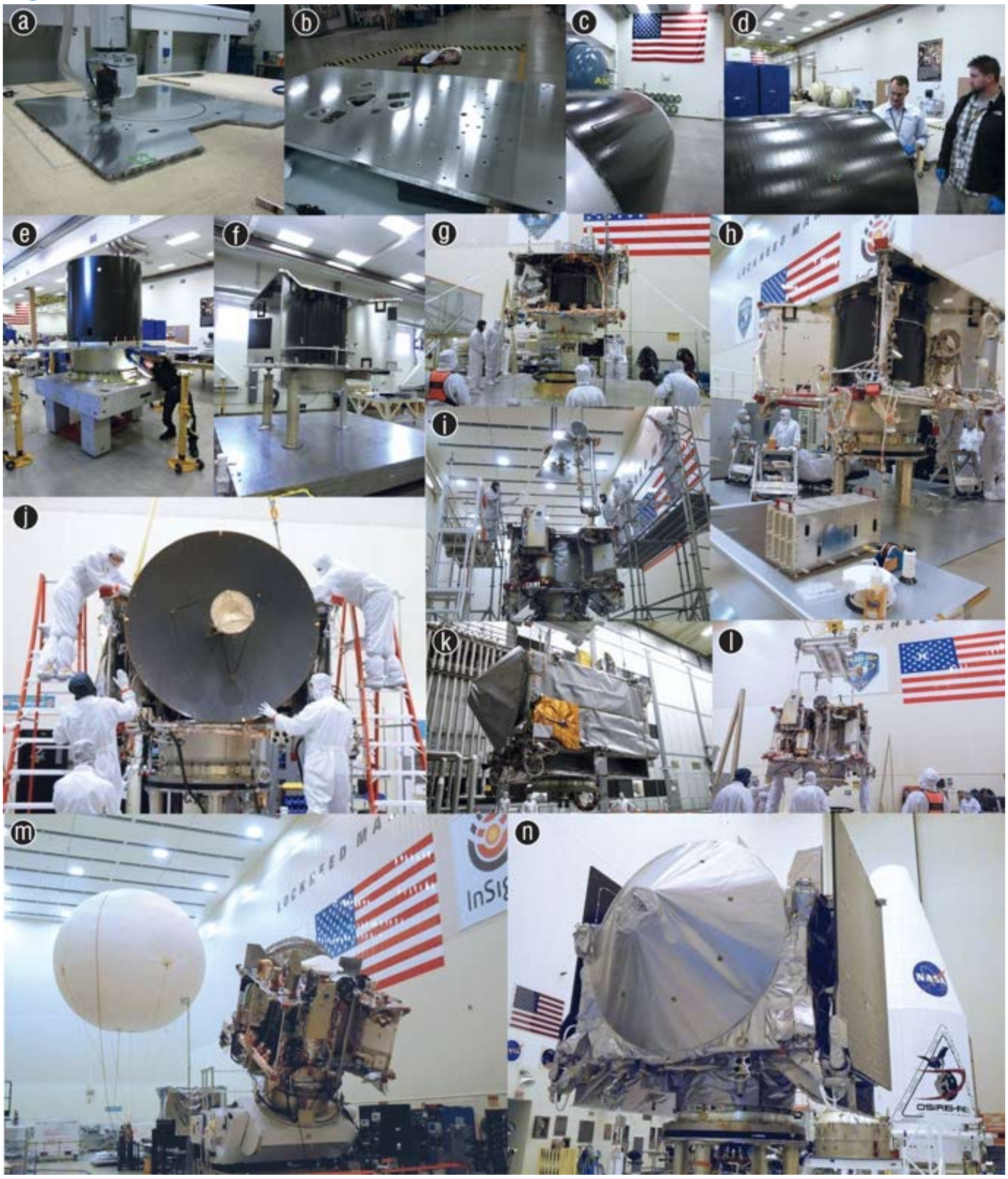

Figure 11

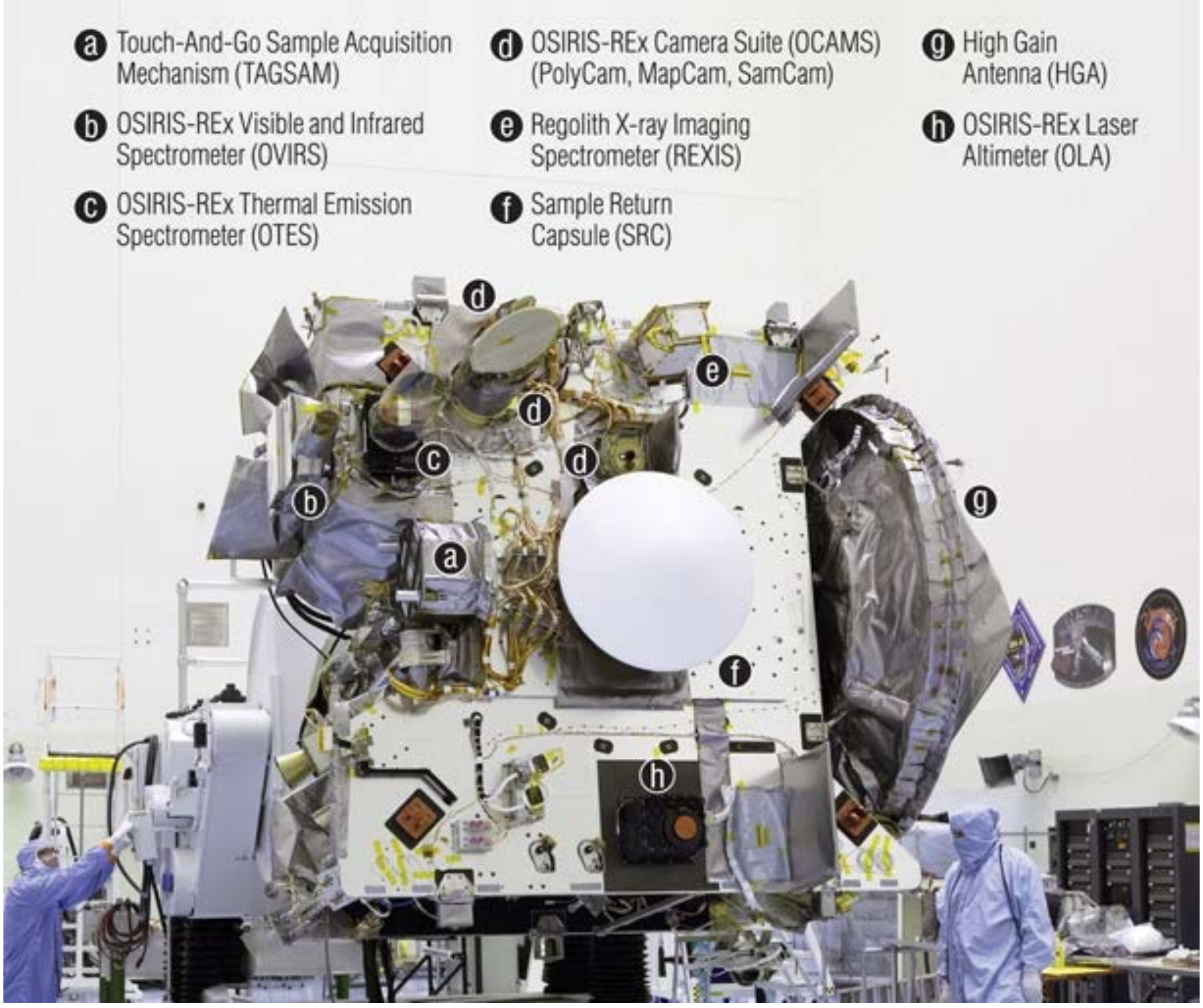

a Touch-And-Go Sample Acquisition Mechanism (TAGSAM)
b OSIRIS-REx Visible and Infrared Spectrometer (OVIRS)
c OSIRIS-REx Thermal Emission Spectrometer (OTES)
d OSIRIS-REx Camera Suite (OCAMS) (PolyCam, MapCam, SamCam)
e Regolith X-ray Imaging Spectrometer (REXIS)
f Sample Return Capsule (SRC)
g High Gain Antenna (HGA)
h OSIRIS-REx Laser Altimeter (OLA)

Figure 12

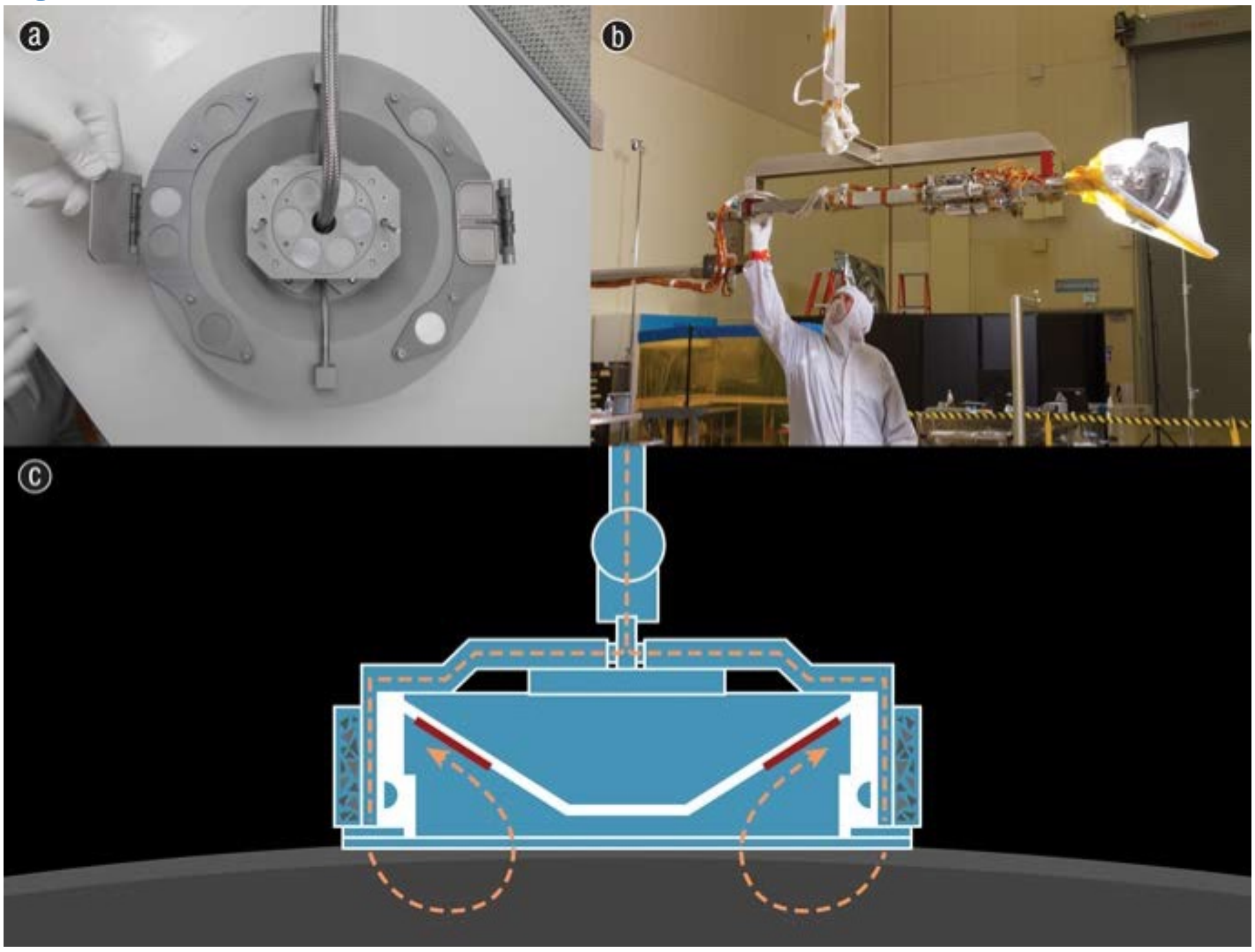

Figure 13

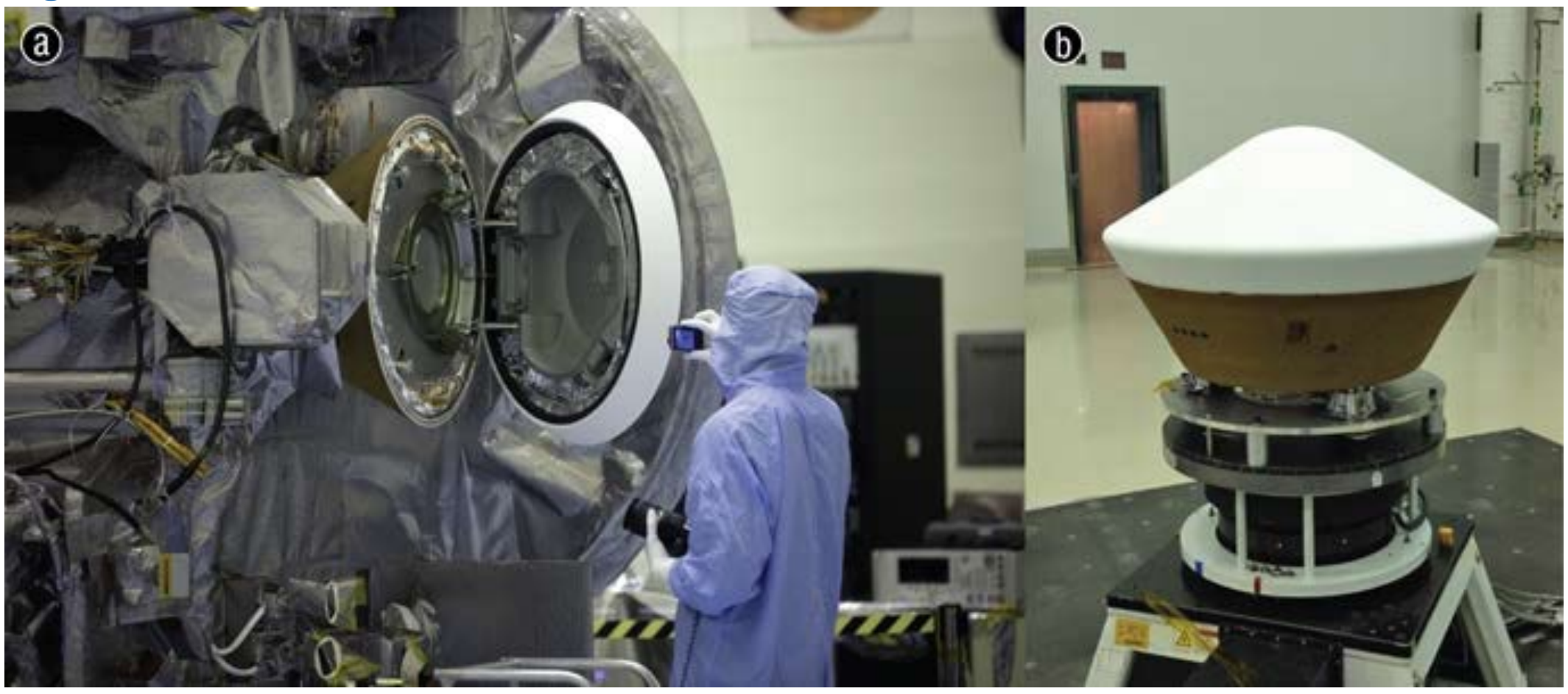

Figure 14

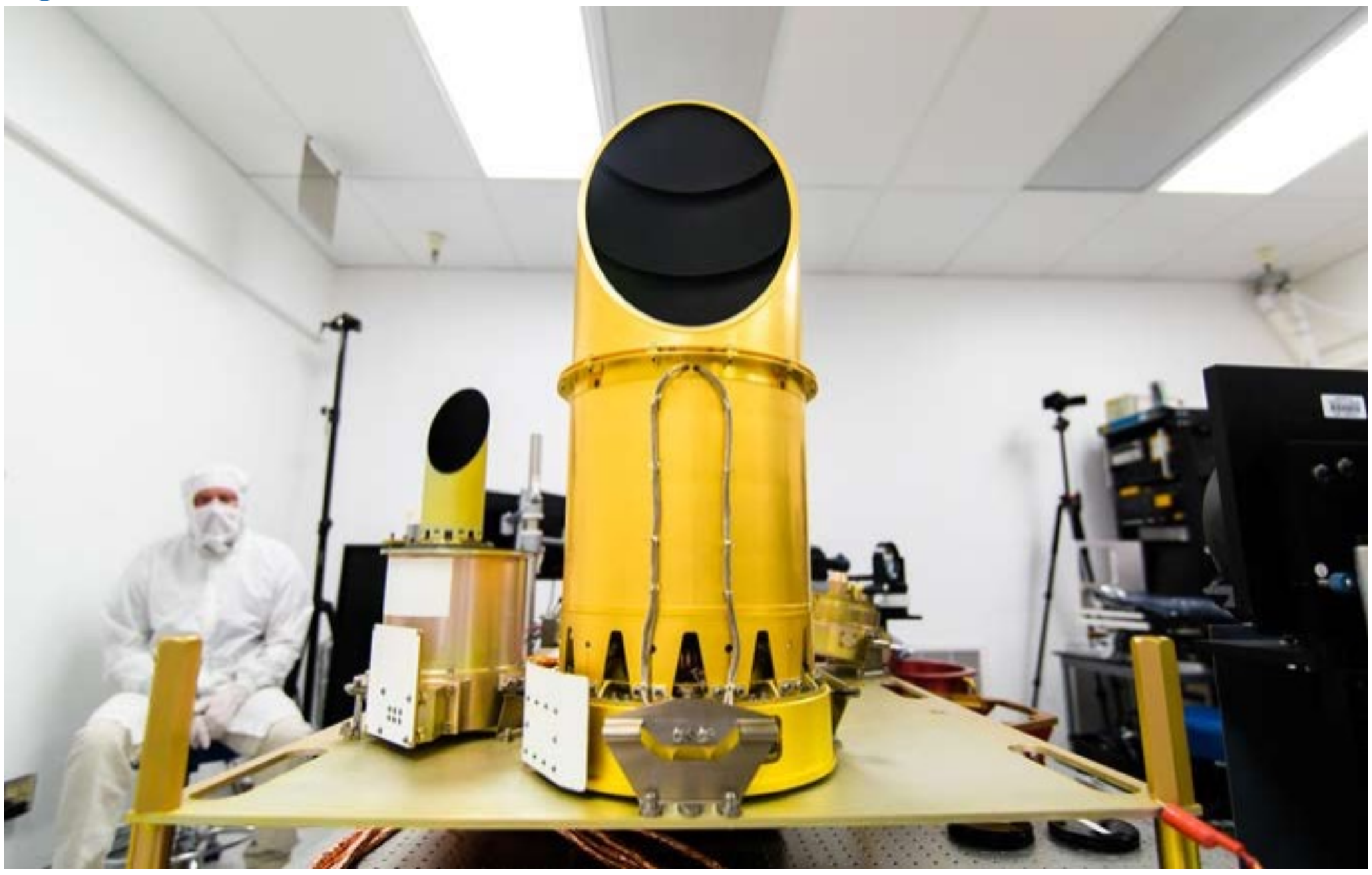

Figure 15

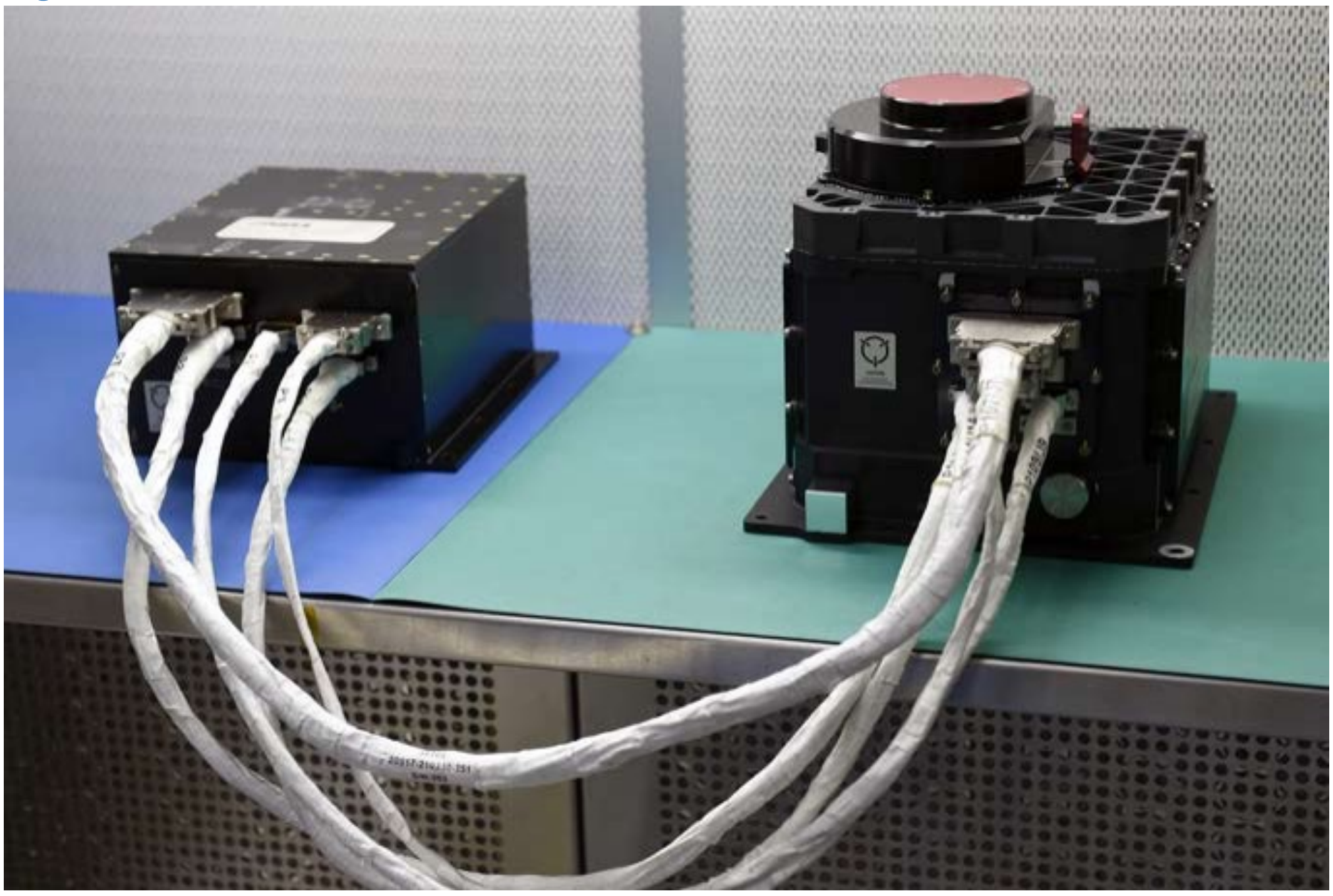

Figure 16

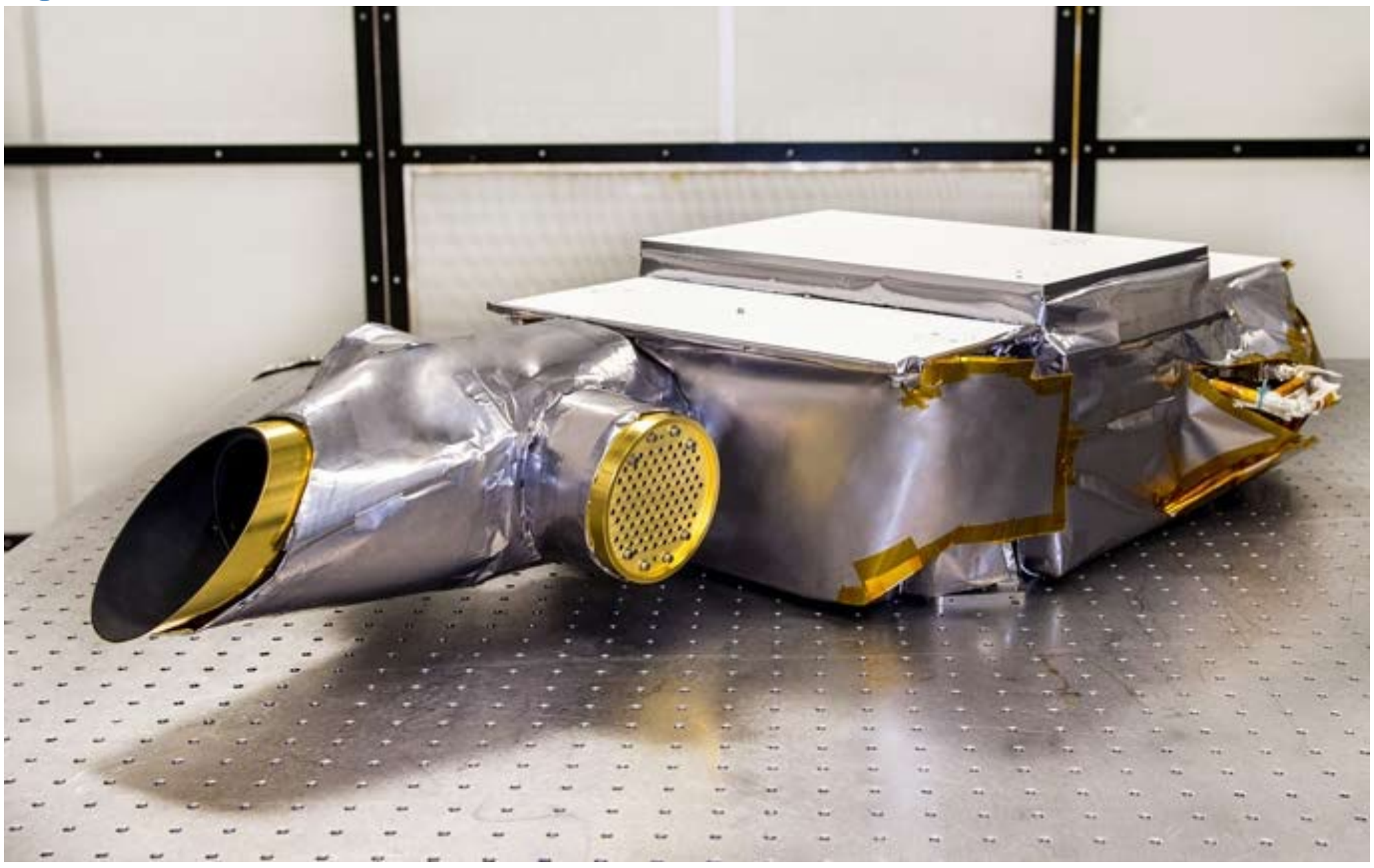

Figure 17

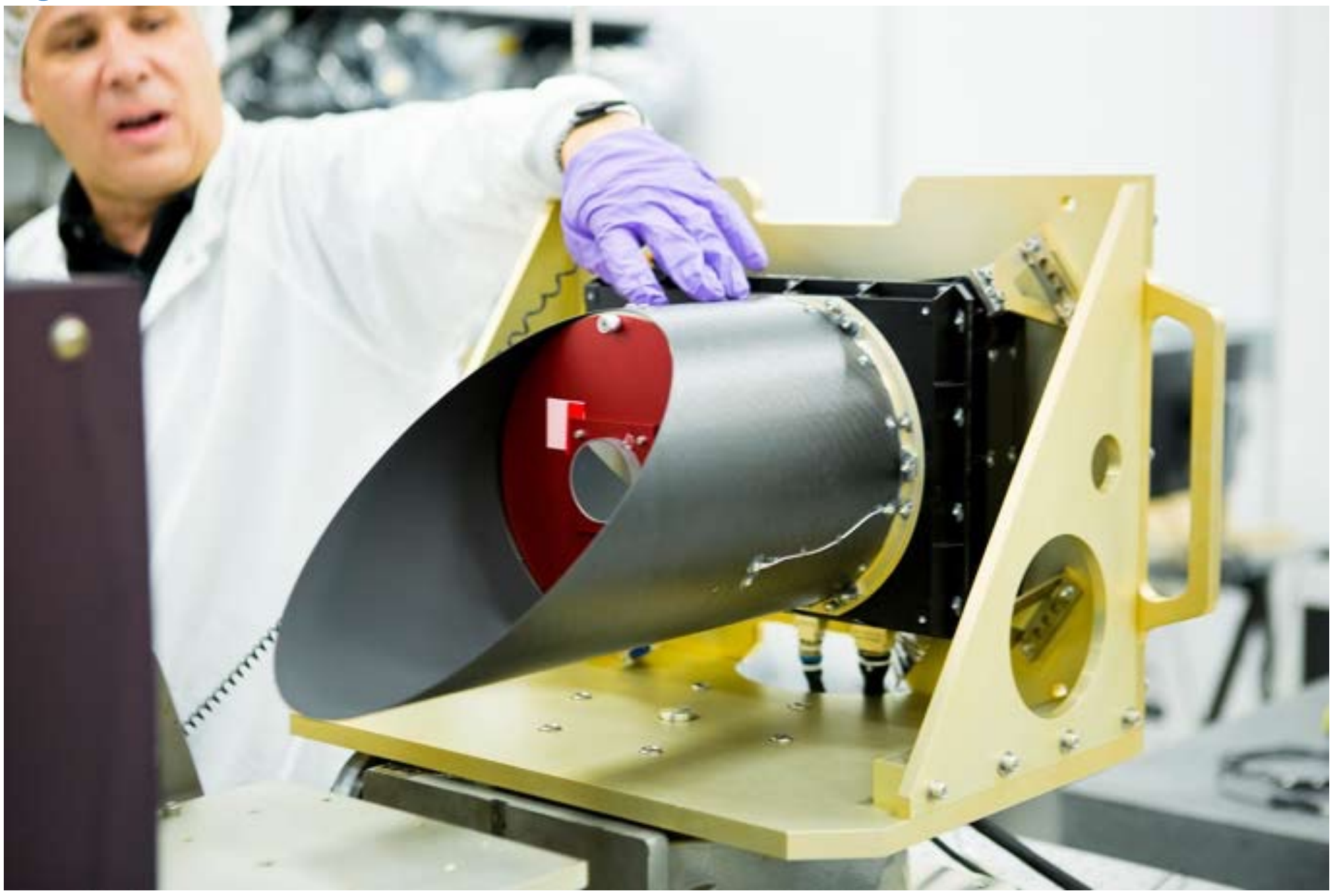

Figure 18

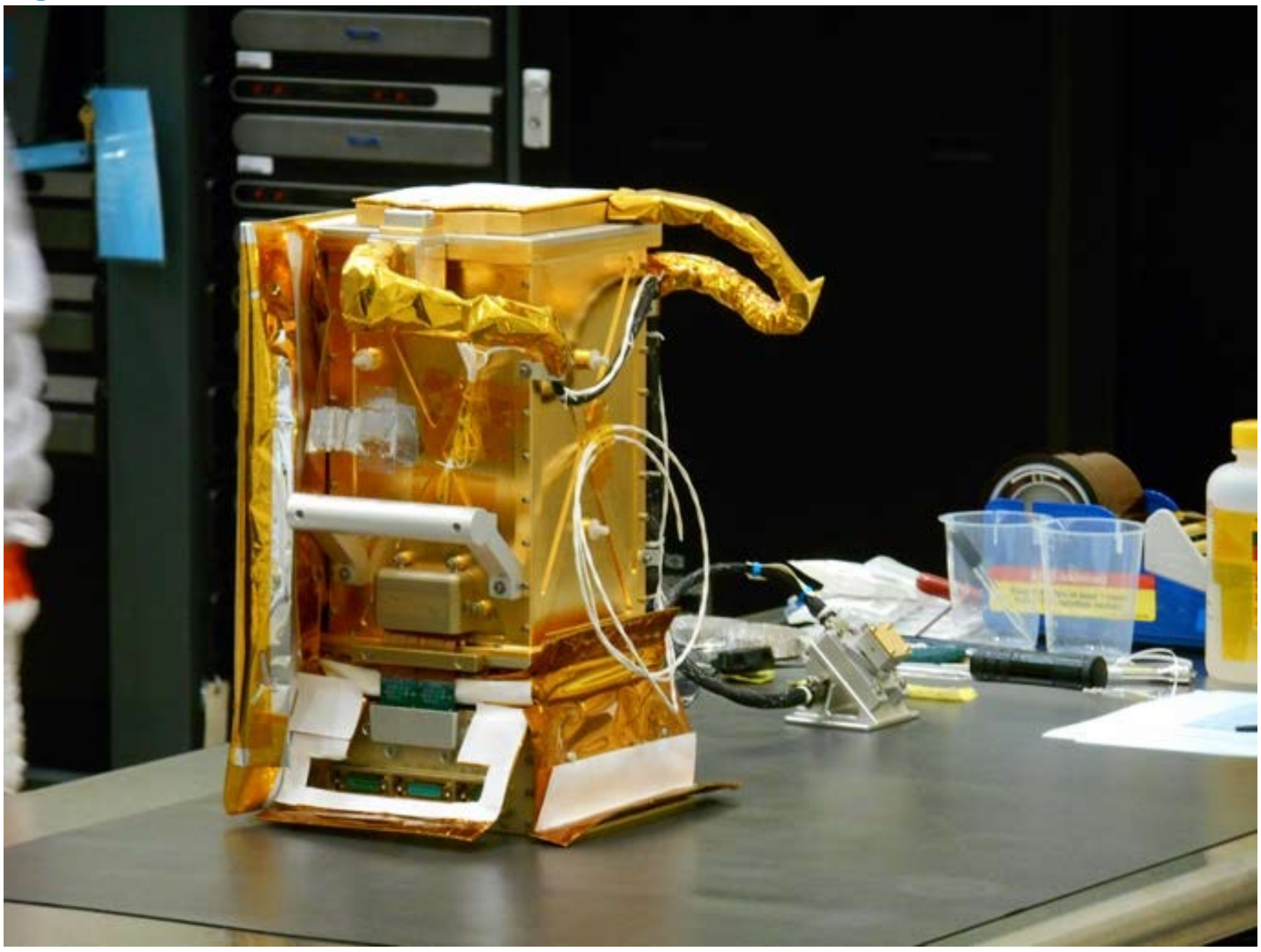

Figure 19

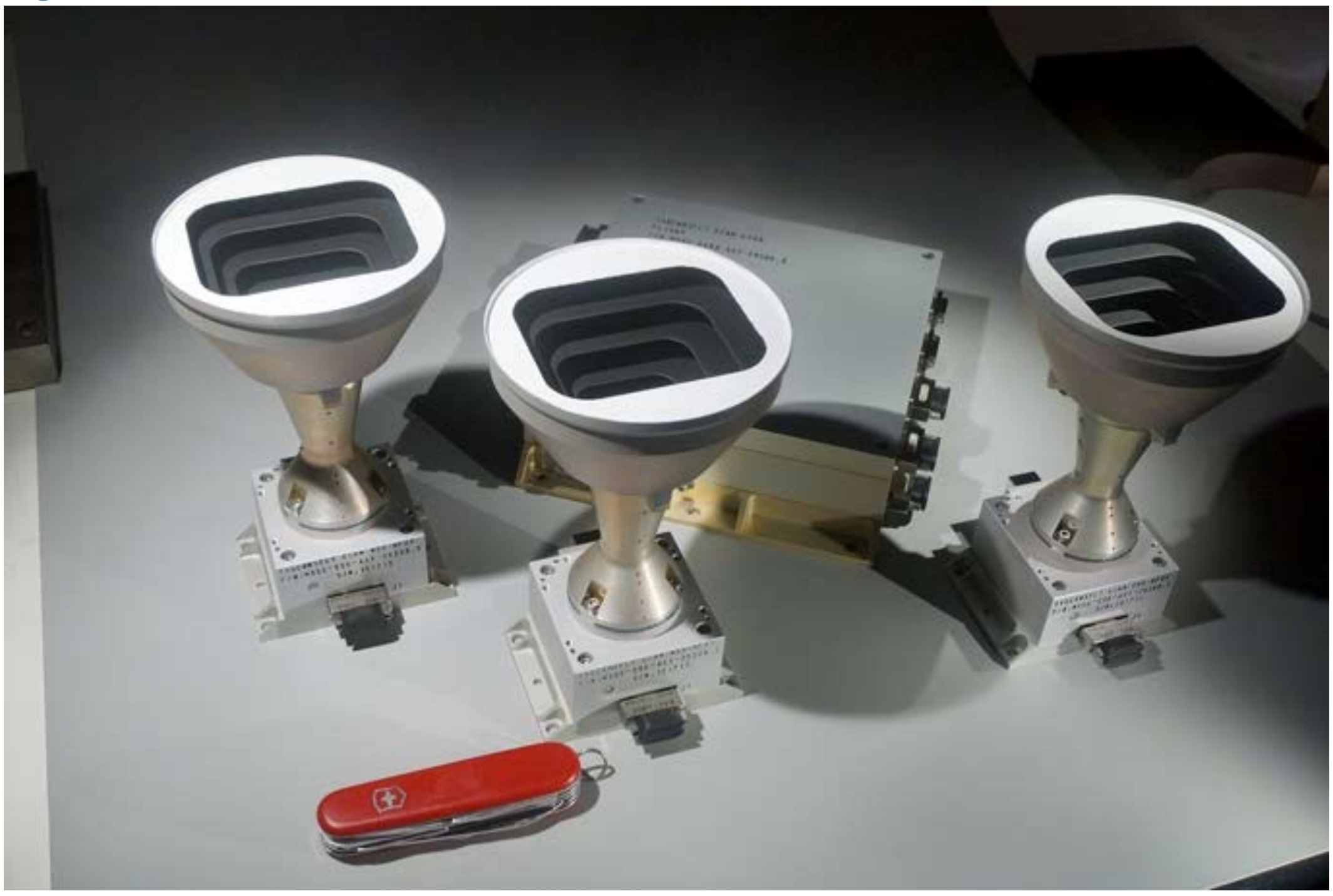

Figure 20

| | Level-1 Requirement | TAGSAM | PolyCam | MapCam | SamCam | Radio Sci. | OLA | OVIRS | OTES | REXIS |
|---|---|---|---|---|---|---|---|---|---|---|
| 4.1.1.1 | Return ≥60 g of bulk sample | ● | | | | | | | | |
| 4.1.1.2 | Document sample contamination | ● | | | | | | | | |
| 4.1.1.3 | Contact ≥26 cm$^2$ of surface material | ● | | | | | | | | |
| 4.1.1.4 | Document the sampling site for: | | | | | | | | | |
| | Texture to sub-cm | | ● | ○ | ○ | | ○ | | | |
| | Morphology | | ● | ● | ● | | ● | | | |
| | Geochemistry | | | | | | | ● | ● | ○ |
| | Spectral Properties | | | ○ | | | | ● | ● | |
| 4.1.1.5 | Produce a sample catalog and analyze samples | Lab requirement | | | | | | | | |
| 4.1.1.6 | Produce a shape model with 1-m resolution. | | ● | ● | ○ | | ● | | | |
| 4.1.1.7 | Determine global geophysical parameters | | | | | | | | | |
| | Slopes | | ◒ | ◒ | ○ | ◒ | ◒ | | | |
| | Surface accelerations | | ◒ | ◒ | ○ | ◒ | ◒ | | | |
| | Geopotential | | ◒ | ◒ | ○ | ◒ | ◒ | | | |
| 4.1.1.8 | Determine density to 1%, gravity to 4$^{th}$ order. | | ◒ | ◒ | ○ | ◒ | ◒ | | | |
| 4.1.1.9 | Surface geology to 1 m spatial resolution | | ● | ● | ○ | | ● | | | |
| 4.1.1.10 | Map minerals & organics to 50 m spatial res. | | | ○ | | | | ● | ● | ◒ |
| 4.1.1.11 | Search for & characterize volatile outgassing | | ◒ | ◒ | | | | ◒ | ◒ | |
| 4.1.1.12 | Search for & characterize satellites | | ◒ | ◒ | | | | ◒ | ◒ | |
| 4.1.1.13 | Search for & characterize space weathering | | | ○ | | | | ● | ○ | |
| 4.1.1.14 | Constrain contributions and measure the Yarkovsky effect | | | | | | | | | |
| | Surface albedo at 0.4 - 2 μm to 5% | | ○ | ○ | ○ | | | ● | | |
| | Thermal emission at 5 - 25 μm to 3% | | | | | | | ○ | ● | |
| | Spin rate to within 10 s | | ● | ● | ○ | ◒ | ● | | | |
| | Obliquity within 1° | | ● | ● | ○ | ◒ | ● | | | |
| | Measure Yarkovsky accel with SNR >400 | | ◒ | ◒ | | ◒ | ◒ | | | |
| 4.1.1.15 | Measure point-source properties | | ◒ | ● | | | | ◒ | ◒ | |

LEGEND

● Capable of fulfilling requirement by itself
◒ Contributes to fulfilling requirement
○ Provides degraded backup capability

Sample Site
Shape
Geophysical Properties
Density and Gravity
Surface Characteristics
Operational Environment
Yarkovsky Effects
Point-Source Properties

Figure 21

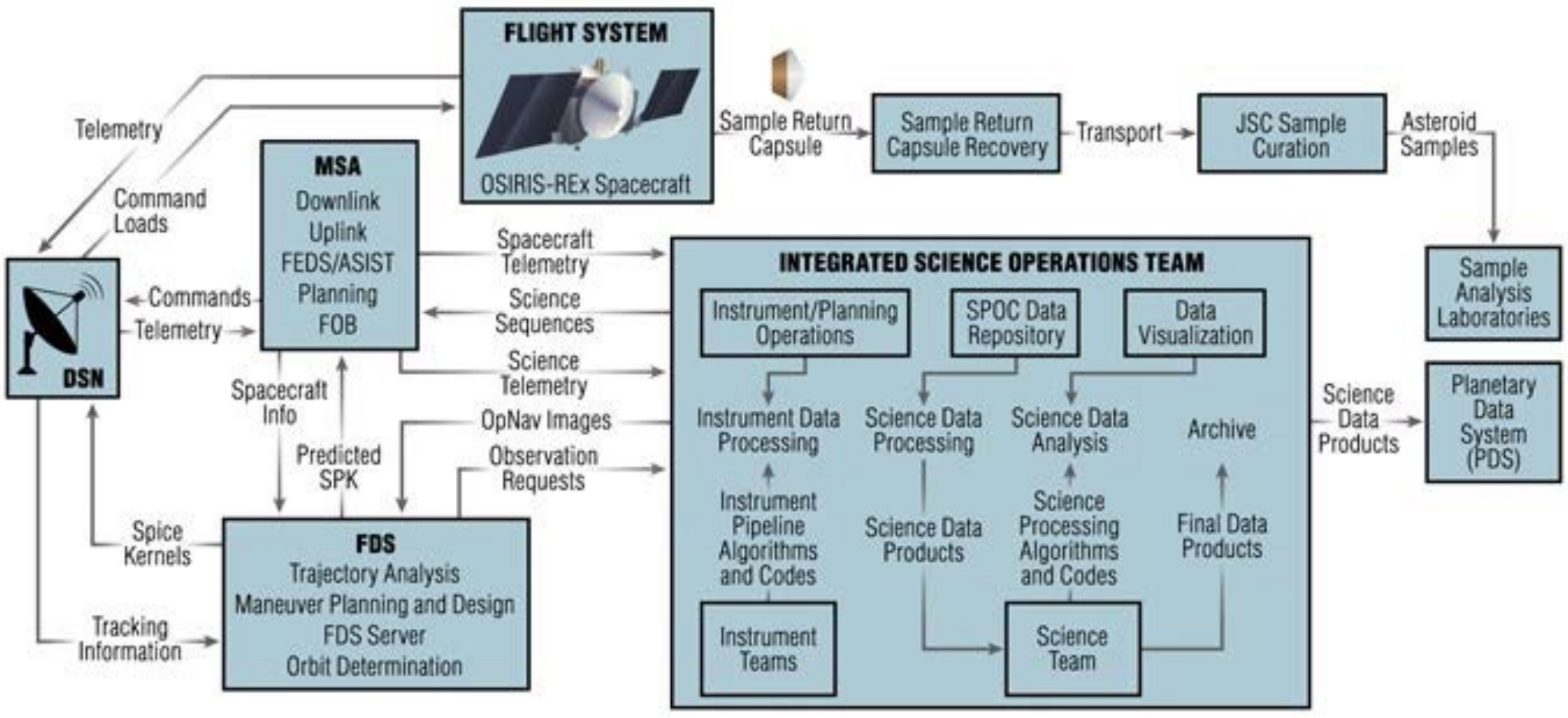

Figure 22

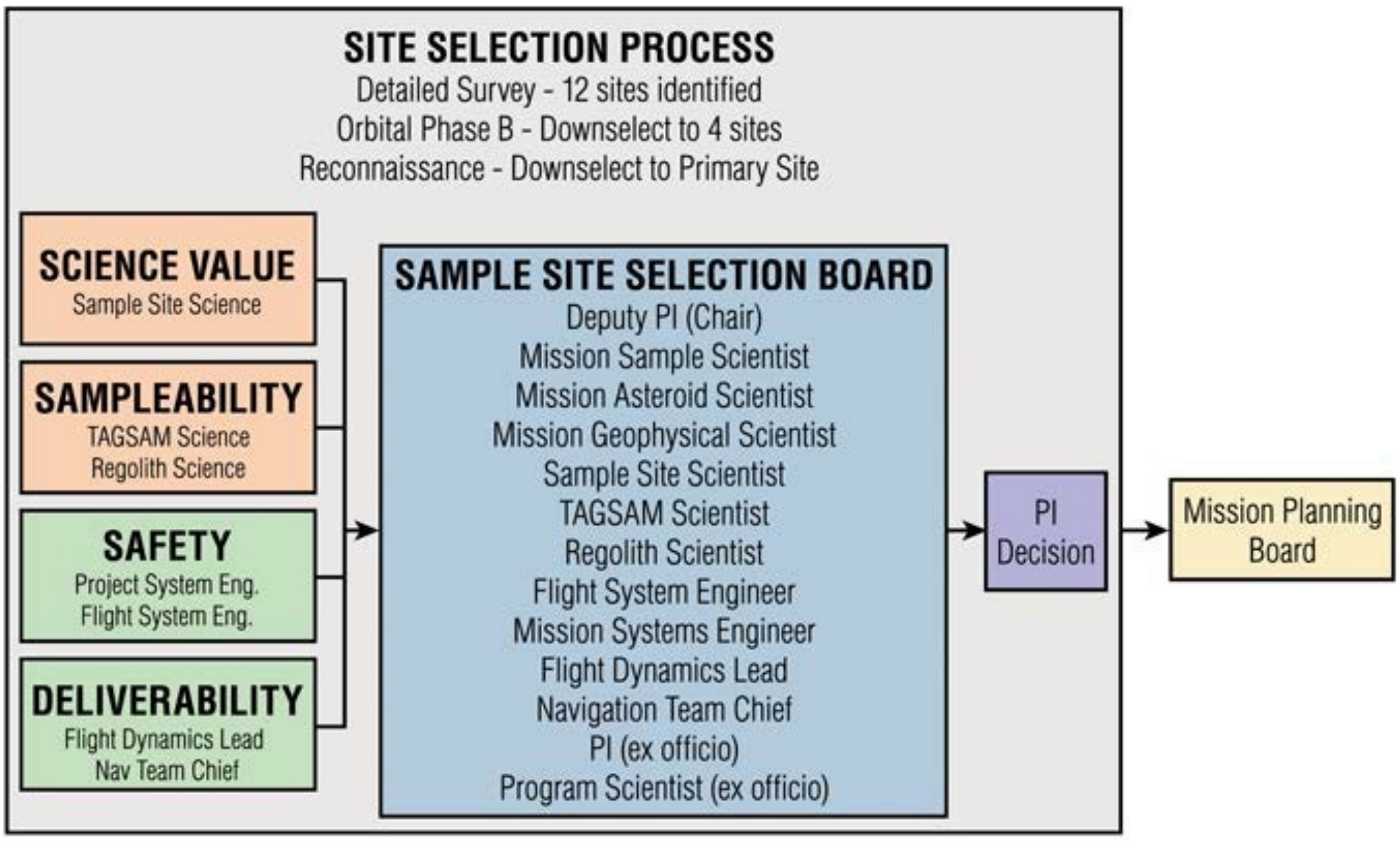
SITE SELECTION PROCESS
Detailed Survey - 12 sites identified
Orbital Phase B - Downselect to 4 sites
Reconnaissance - Downselect to Primary Site
SCIENCE VALUE
Sample Site Science
SAMPLEABILITY
TAGSAM Science
Regolith Science
SAFETY
Project System Eng.
Flight System Eng.
DELIVERABILITY
Flight Dynamics Lead
Nav Team Chief
SAMPLE SITE SELECTION BOARD
Deputy PI (Chair)
Mission Sample Scientist
Mission Asteroid Scientist
Mission Geophysical Scientist
Sample Site Scientist
TAGSAM Scientist
Regolith Scientist
Flight System Engineer
Mission Systems Engineer
Flight Dynamics Lead
Navigation Team Chief
PI (ex officio)
Program Scientist (ex officio)
PI
Decision
Mission Planning
Board

Figure 23

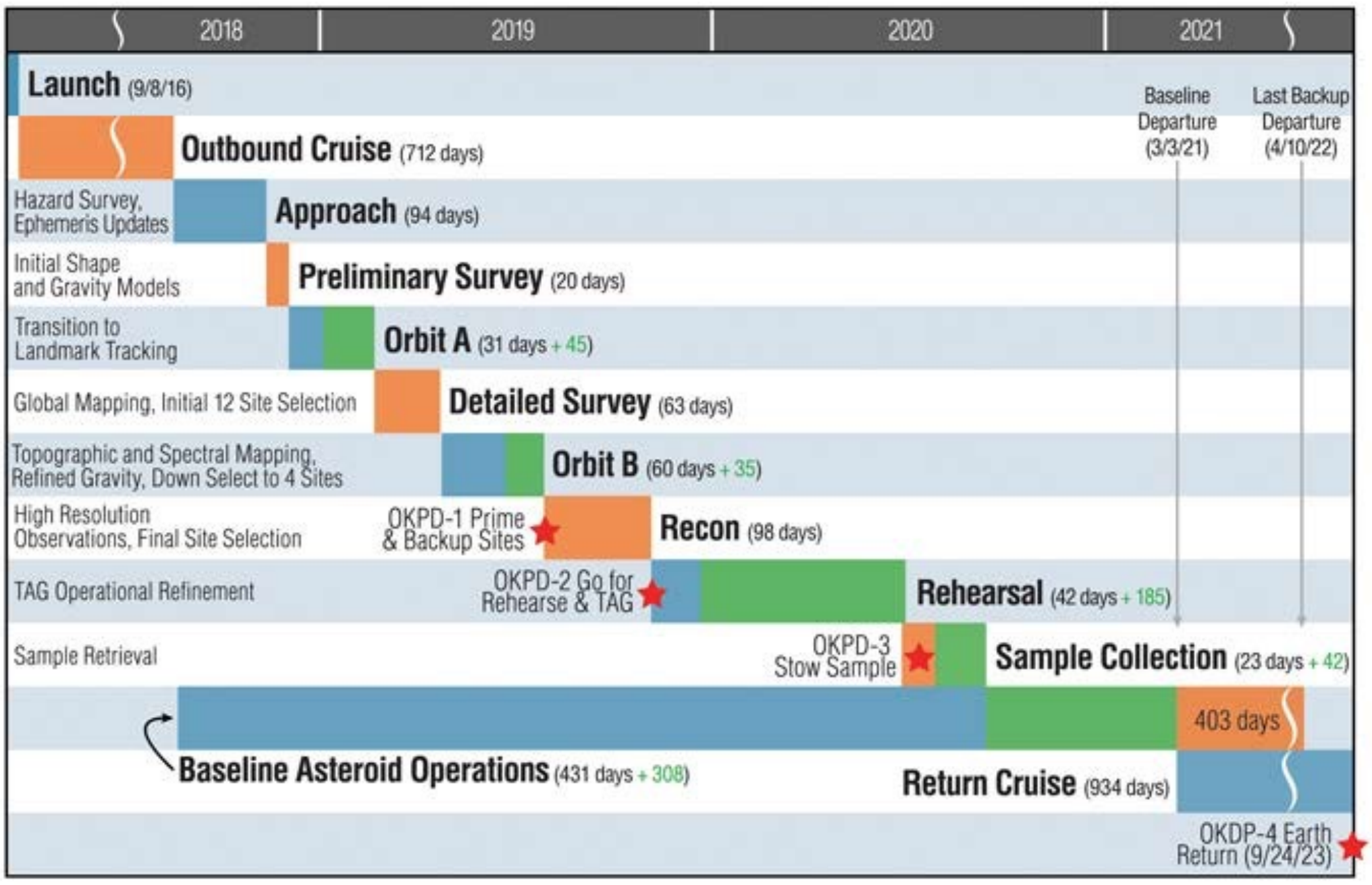

2018
2019
2020
2021
Launch (9/8/16)
Baseline Departure (3/3/21)
Last Backup Departure (4/10/22)
Outbound Cruise (712 days)
Hazard Survey, Ephemeris Updates
Approach (94 days)
Initial Shape and Gravity Models
Preliminary Survey (20 days)
Transition to Landmark Tracking
Orbit A (31 days + 45)
Global Mapping, Initial 12 Site Selection
Detailed Survey (63 days)
Topographic and Spectral Mapping, Refined Gravity, Down Select to 4 Sites
Orbit B (60 days + 35)
High Resolution Observations, Final Site Selection
OKPD-1 Prime & Backup Sites
Recon (98 days)
TAG Operational Refinement
OKPD-2 Go for Rehearse & TAG
Rehearsal (42 days + 185)
Sample Retrieval
OKPD-3 Stow Sample
Sample Collection (23 days + 42)
403 days
Baseline Asteroid Operations (431 days + 308)
Return Cruise (934 days)
OKDP-4 Earth Return (9/24/23)

Figure 24

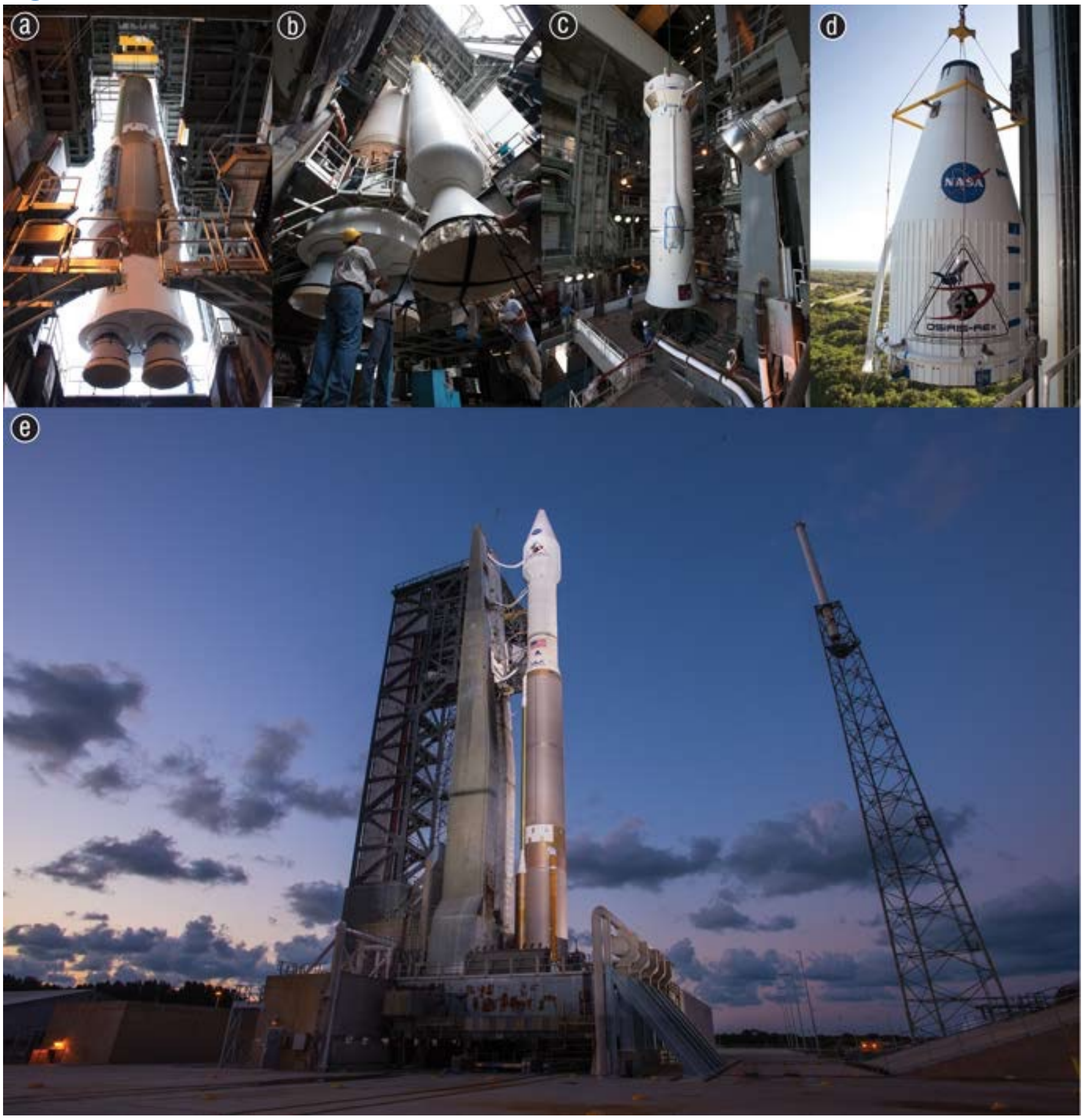

Figure 25

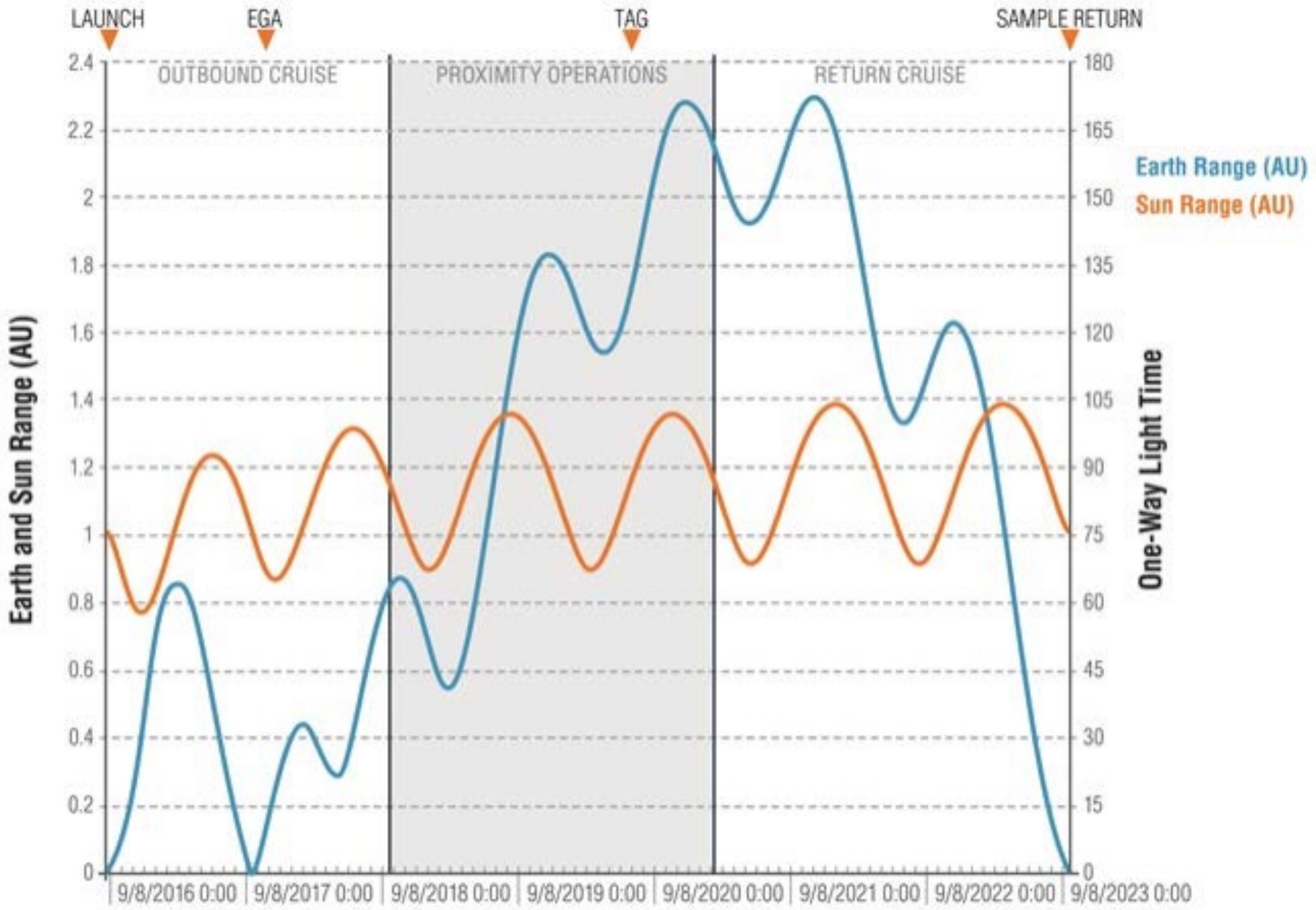

Figure 26

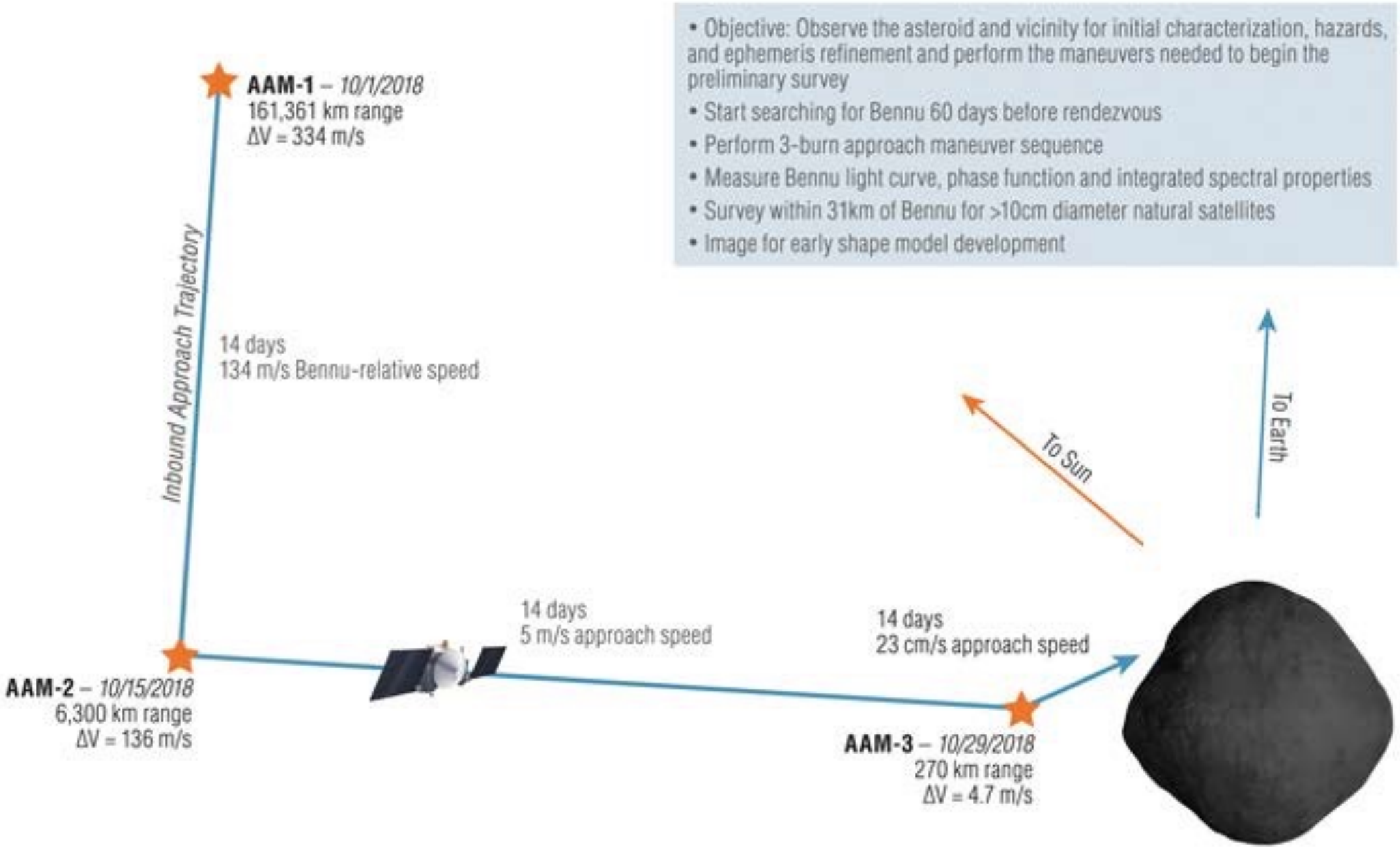

• Objective: Observe the asteroid and vicinity for initial characterization, hazards, and ephemeris refinement and perform the maneuvers needed to begin the preliminary survey
• Start searching for Bennu 60 days before rendezvous
• Perform 3-burn approach maneuver sequence
• Measure Bennu light curve, phase function and integrated spectral properties
• Survey within 31km of Bennu for >10cm diameter natural satellites
• Image for early shape model development
AAM-1 – 10/1/2018
161,361 km range
ΔV = 334 m/s
Inbound Approach Trajectory
14 days
134 m/s Bennu-relative speed
To Sun
To Earth
AAM-2 – 10/15/2018
6,300 km range
ΔV = 136 m/s
14 days
5 m/s approach speed
14 days
23 cm/s approach speed
AAM-3 – 10/29/2018
270 km range
ΔV = 4.7 m/s

Figure 27

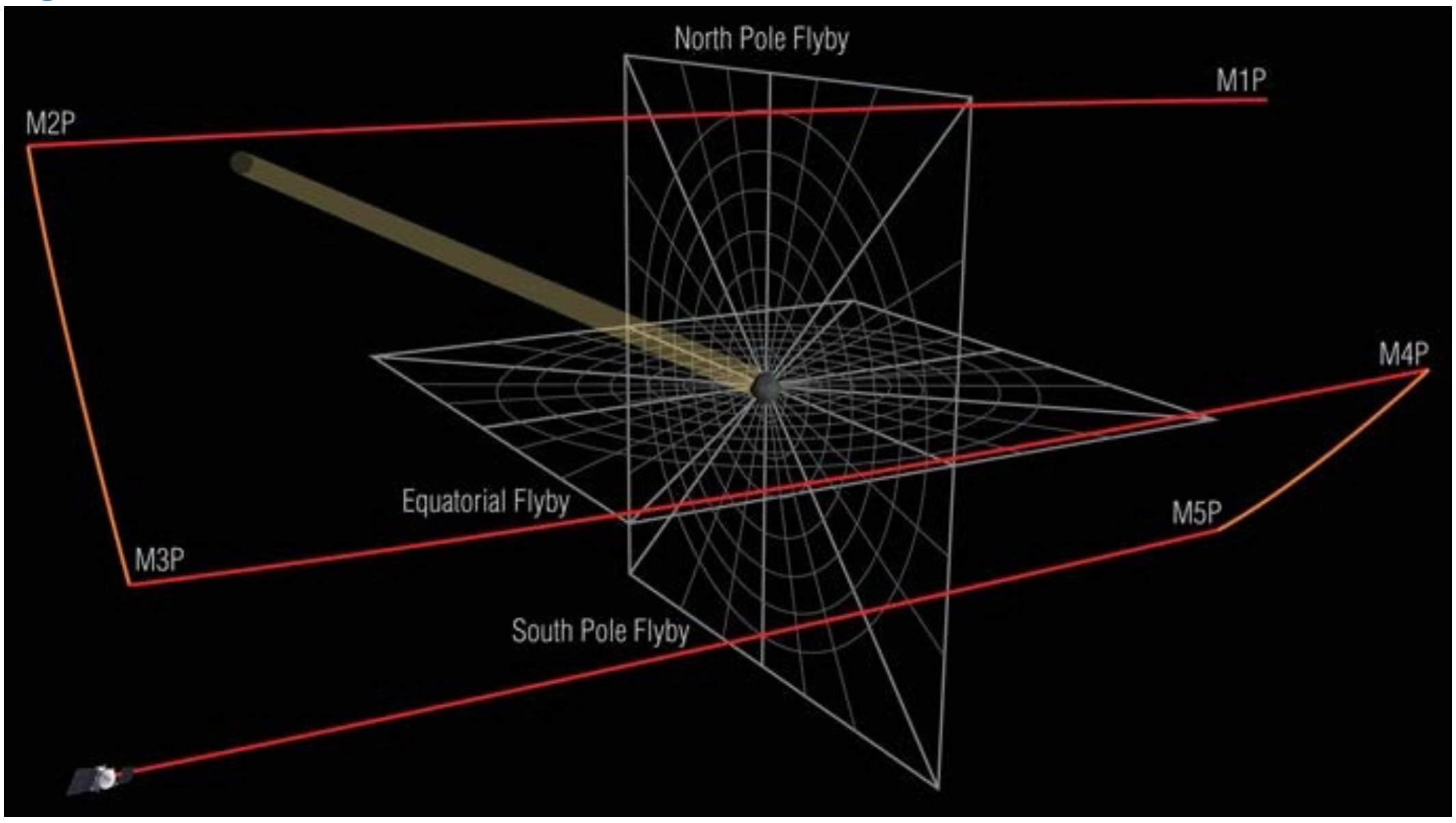

Figure 28

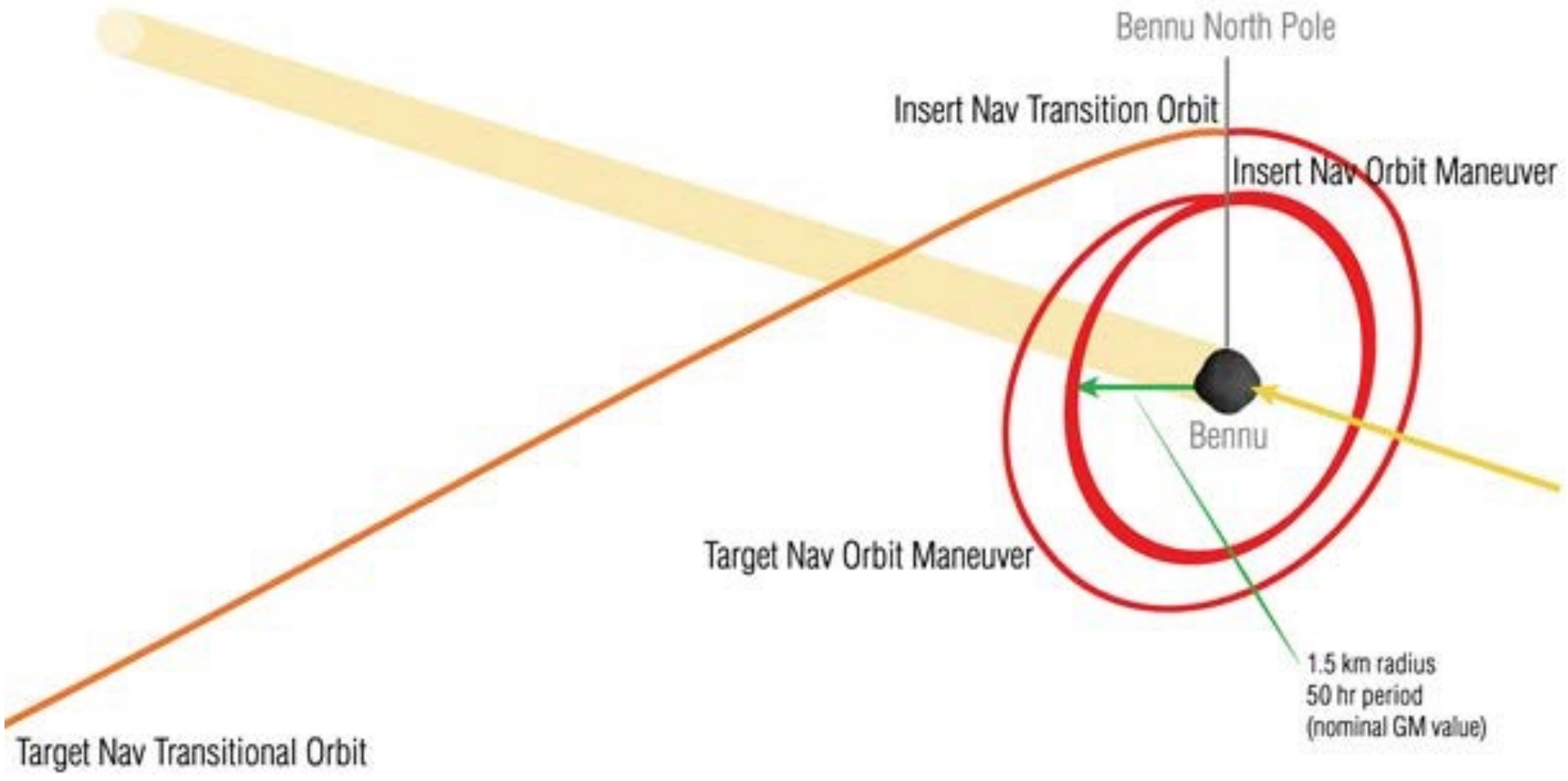
Bennu North Pole
Insert Nav Transition Orbit
Insert Nav Orbit Maneuver
Bennu
Target Nav Orbit Maneuver
1.5 km radius
50 hr period
(nominal GM value)
Target Nav Transitional Orbit

Figure 29

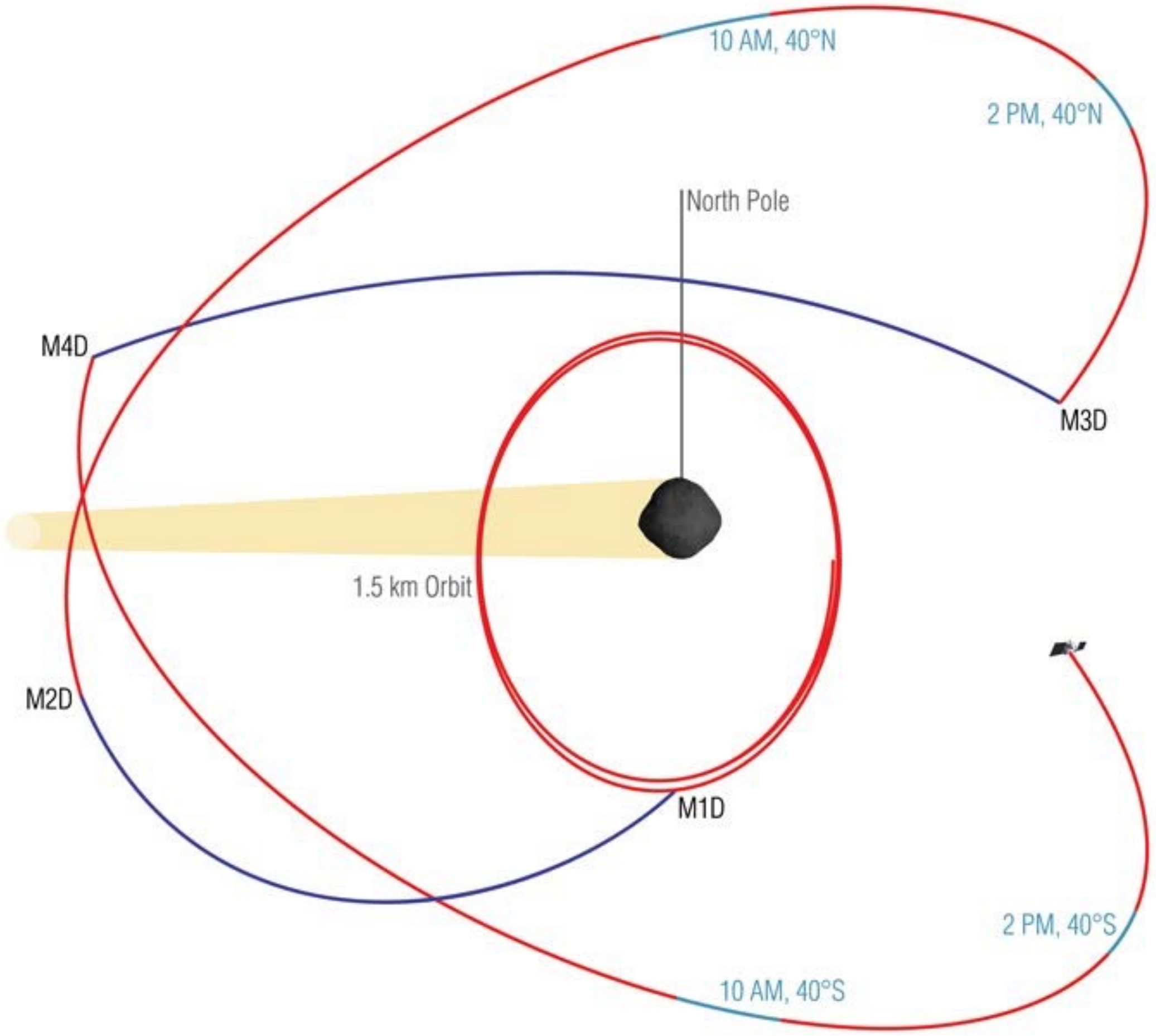

Figure 30

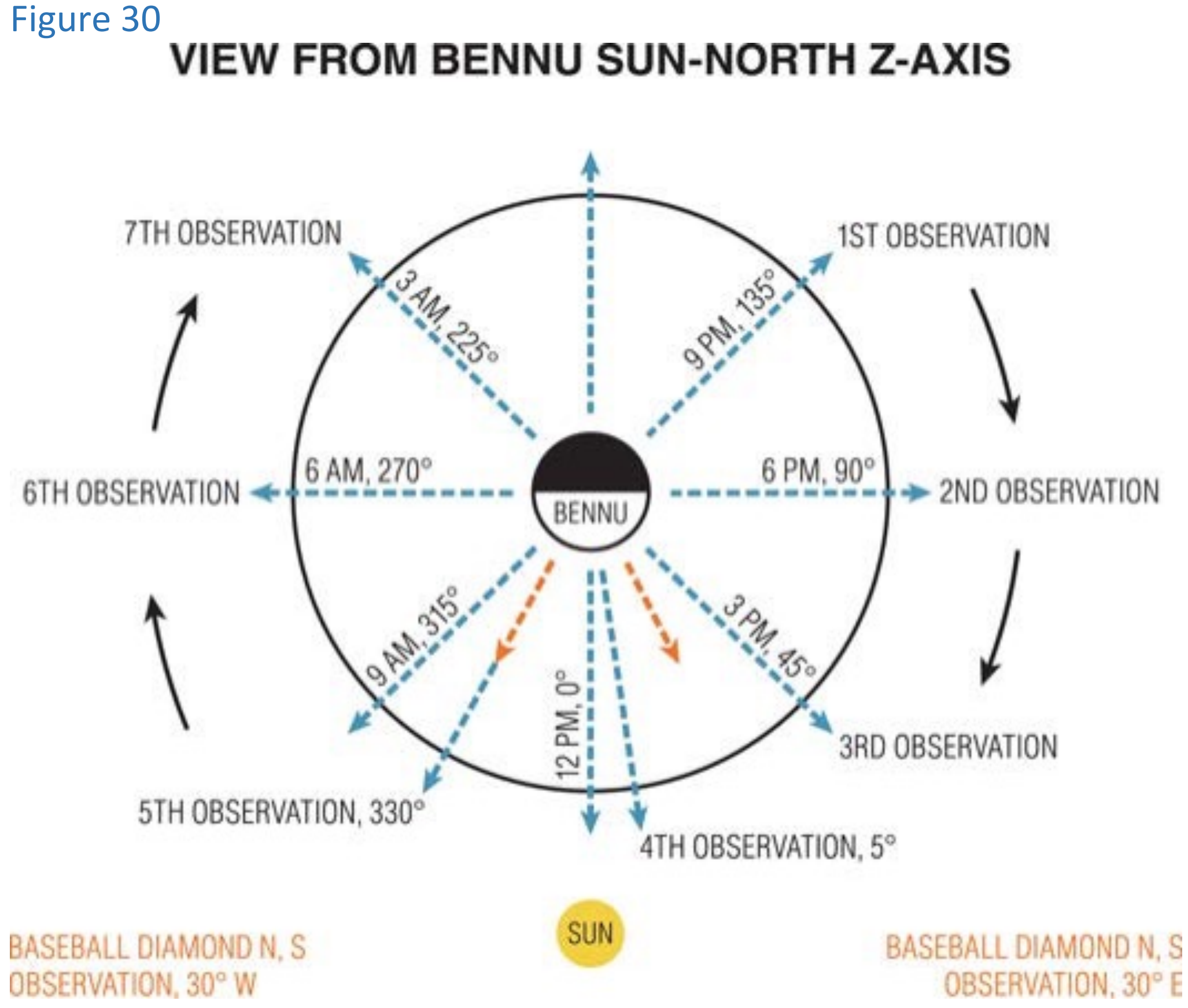

Figure 31

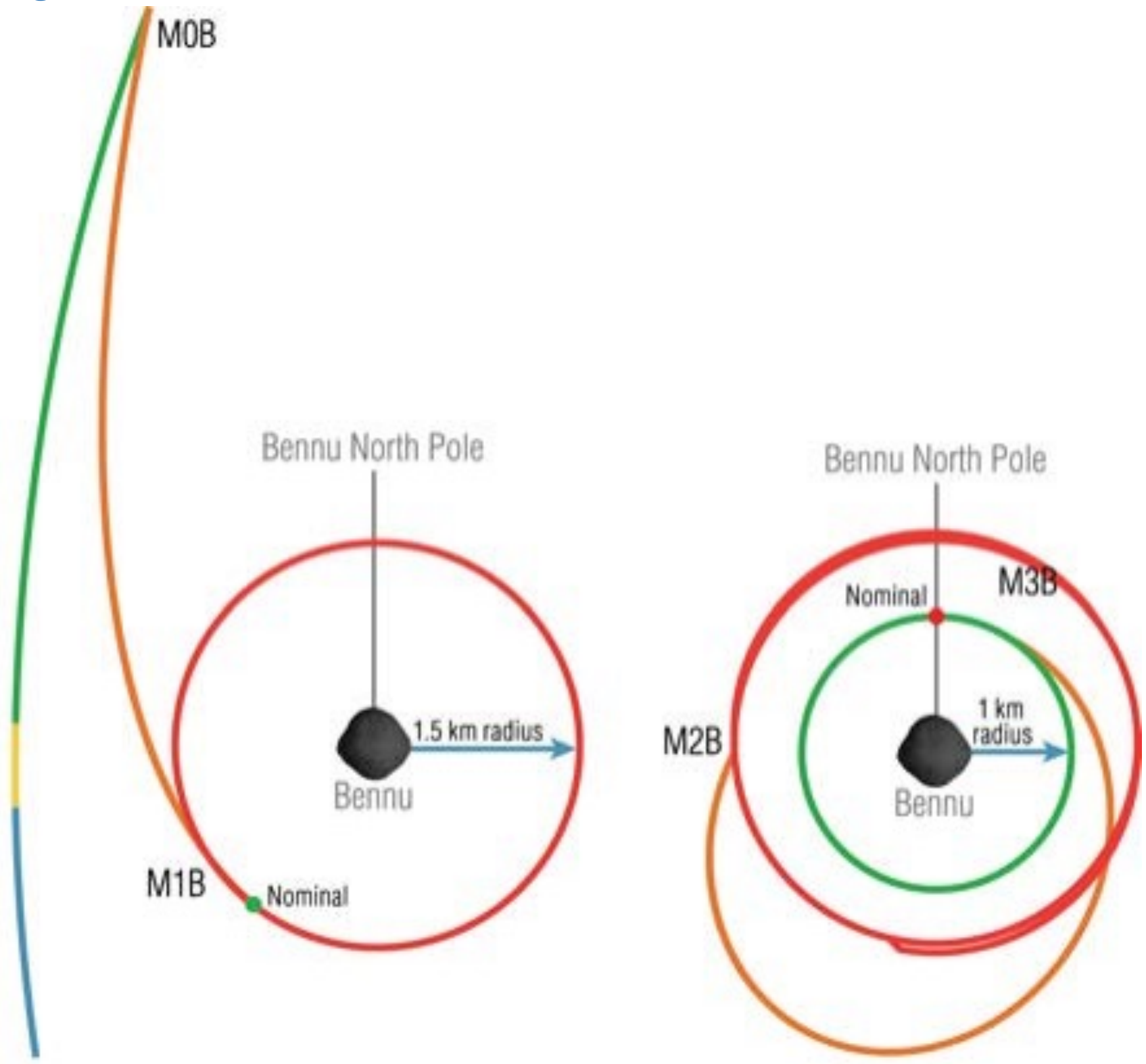

Figure 32

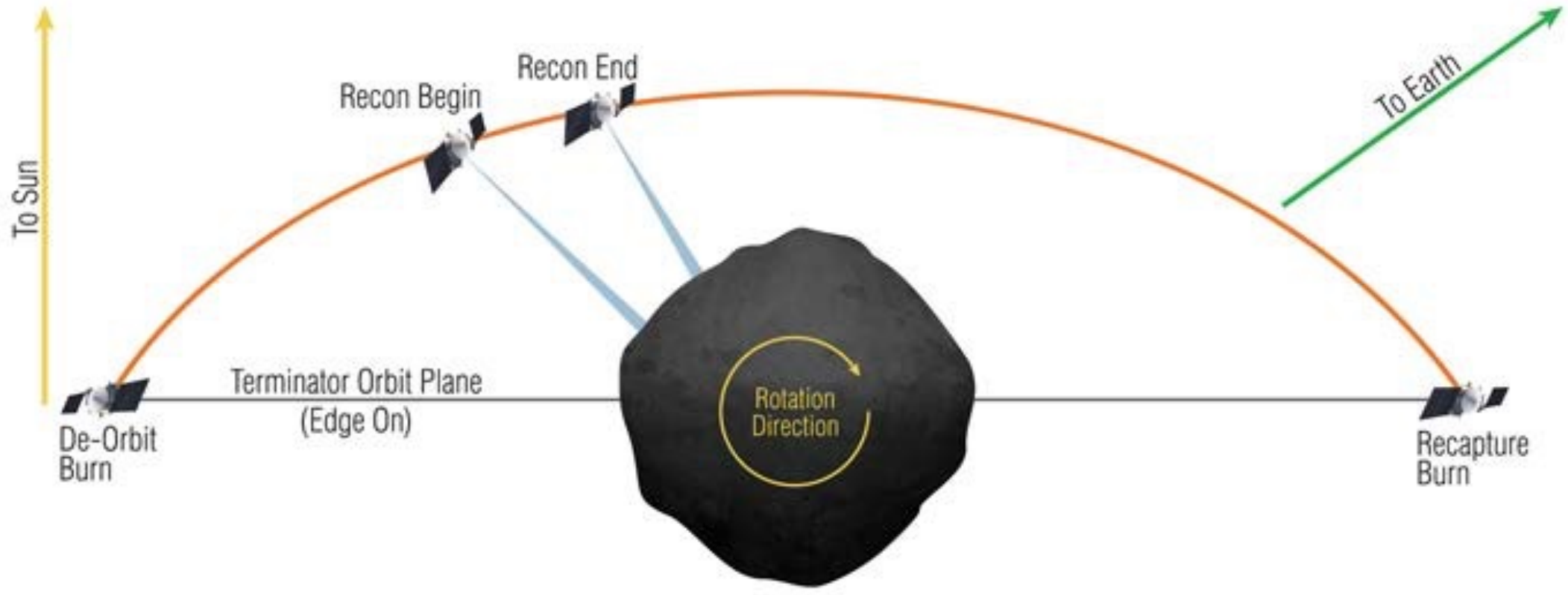

Figure 33

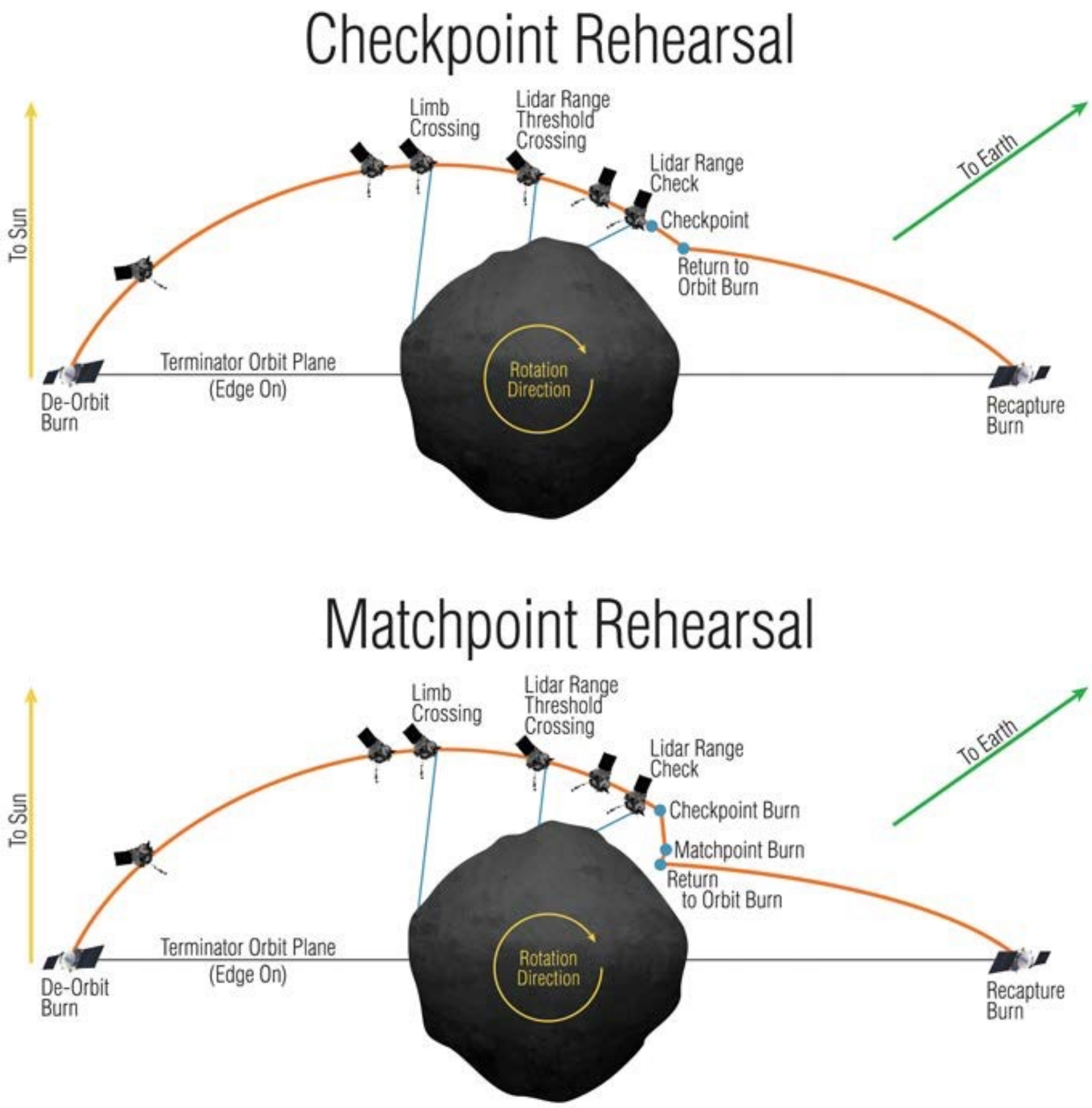
Checkpoint Rehearsal
Limb Crossing
Lidar Range Threshold Crossing
Lidar Range Check
Checkpoint
Return to Orbit Burn
To Earth
To Sun
Terminator Orbit Plane (Edge On)
Rotation Direction
De-Orbit Burn
Recapture Burn
Matchpoint Rehearsal
Limb Crossing
Lidar Range Threshold Crossing
Lidar Range Check
Checkpoint Burn
Matchpoint Burn
Return to Orbit Burn
To Earth
To Sun
Terminator Orbit Plane (Edge On)
Rotation Direction
De-Orbit Burn
Recapture Burn

Figure 34

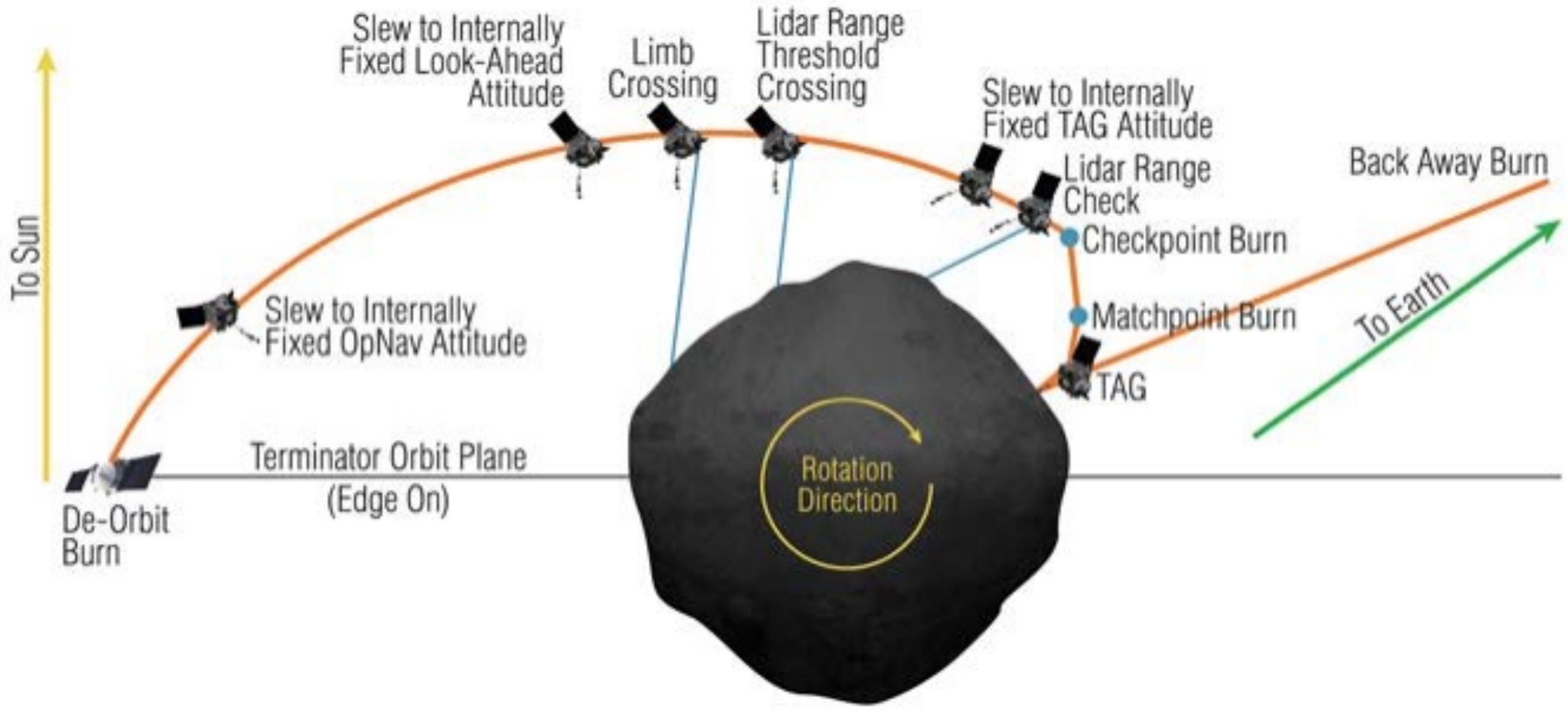

To Sun
De-Orbit Burn
Slew to Internally Fixed OpNav Attitude
Slew to Internally Fixed Look-Ahead Attitude
Limb Crossing
Lidar Range Threshold Crossing
Slew to Internally Fixed TAG Attitude
Lidar Range Check
Checkpoint Burn
Matchpoint Burn
TAG
Back Away Burn
To Earth
Terminator Orbit Plane (Edge On)
Rotation Direction

Figure 35

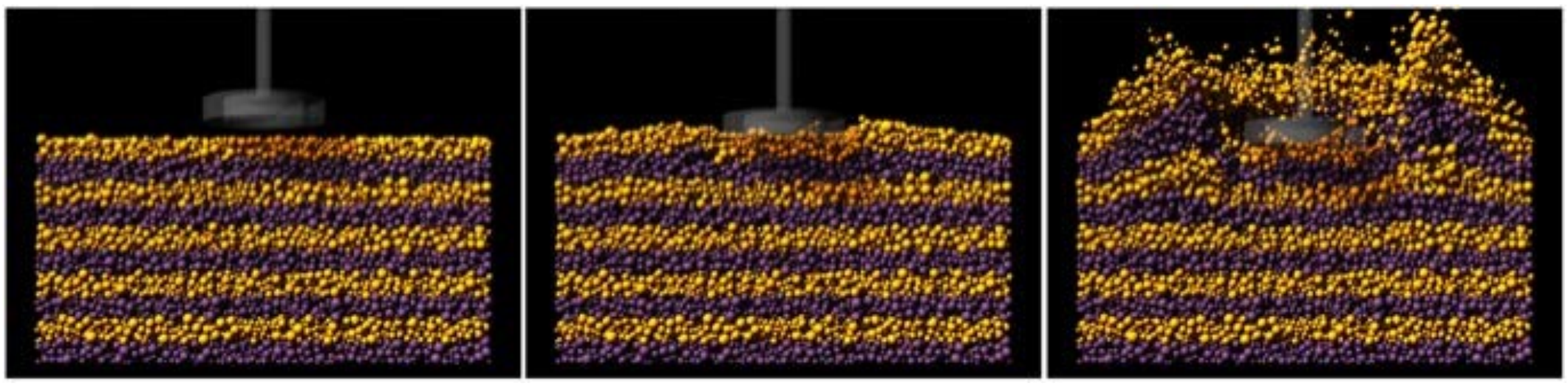

Figure 36
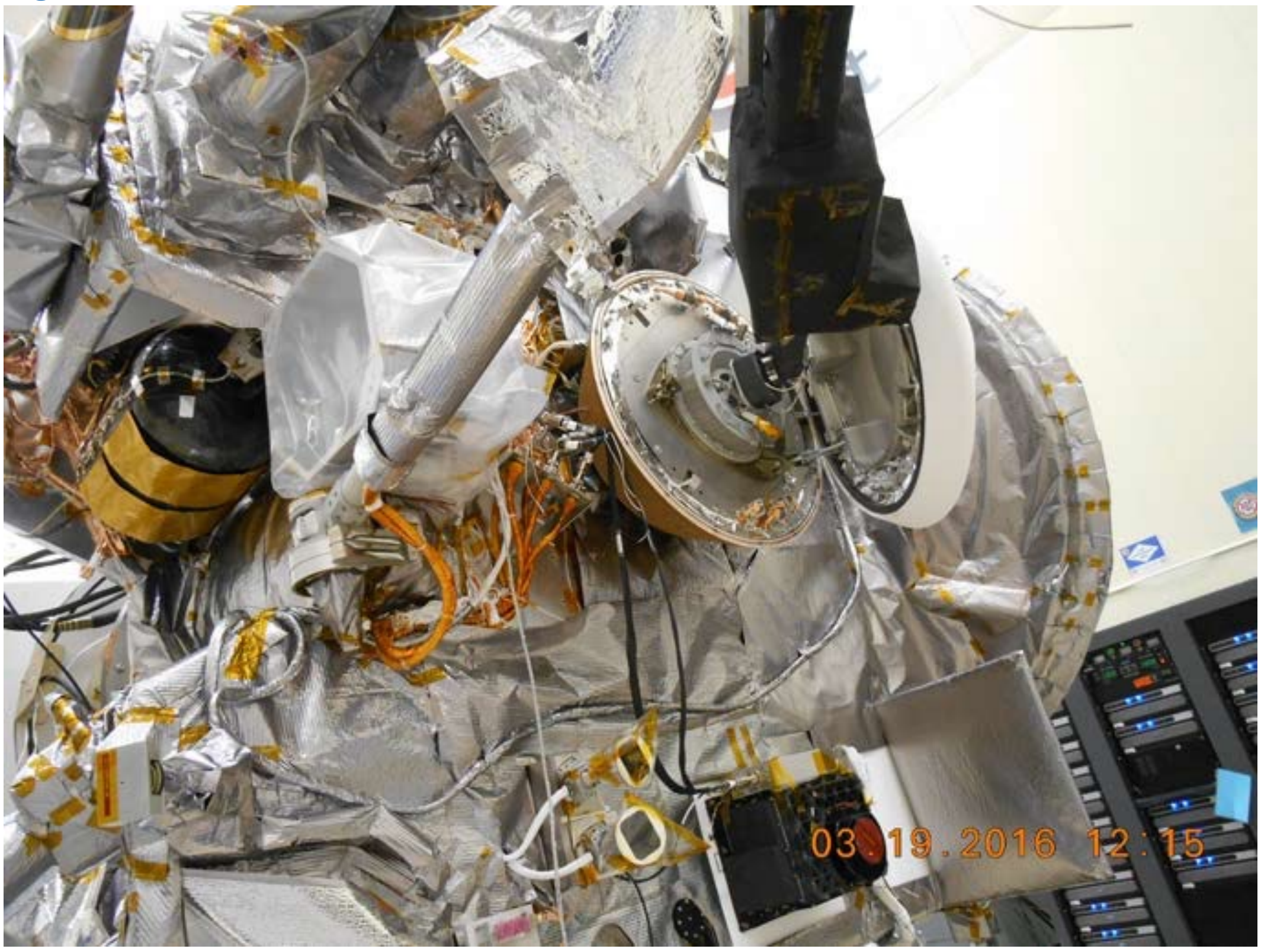

Figure 38

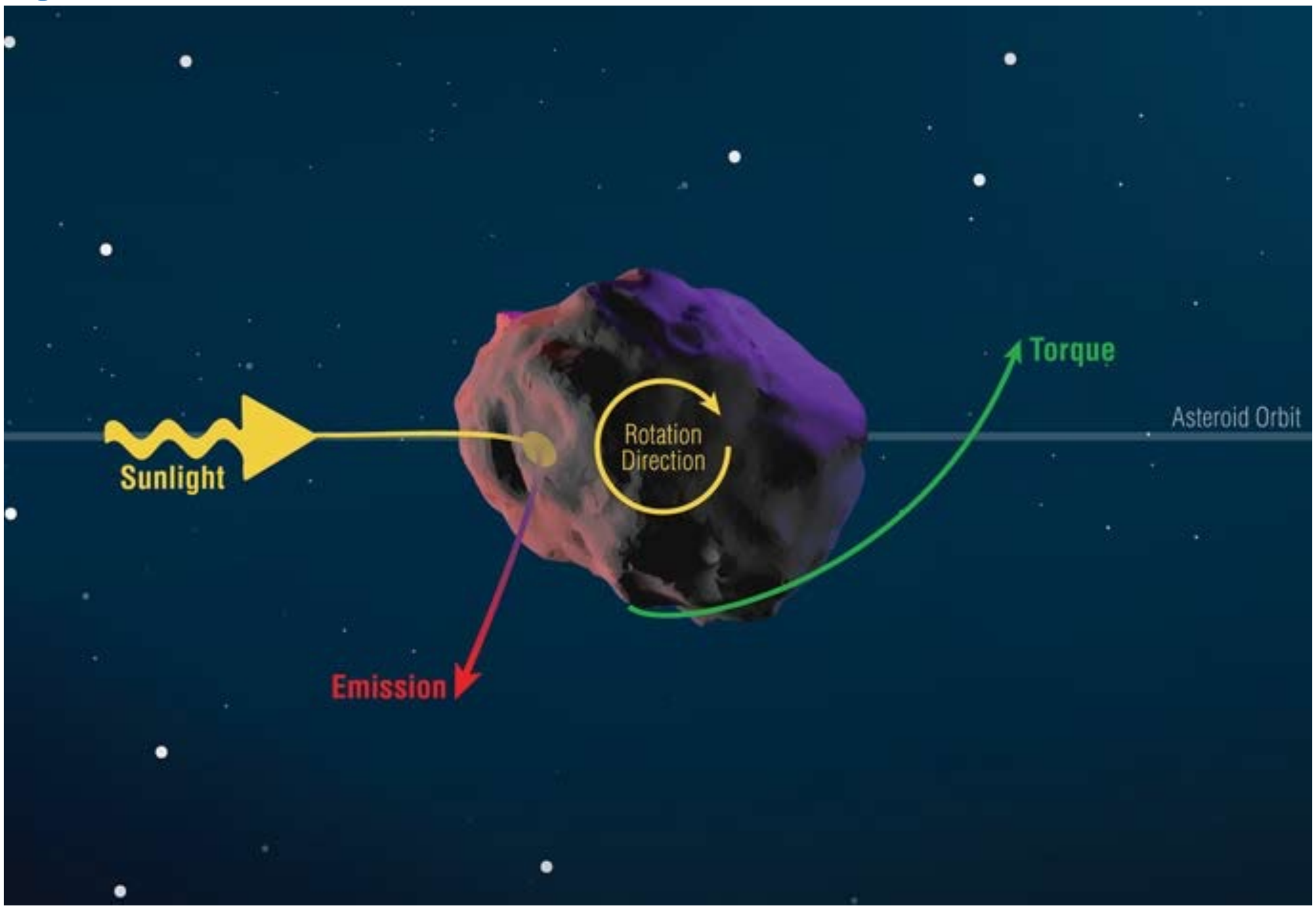
Torque
Asteroid Orbit
Rotation Direction
Sunlight
Emission

Figure 39

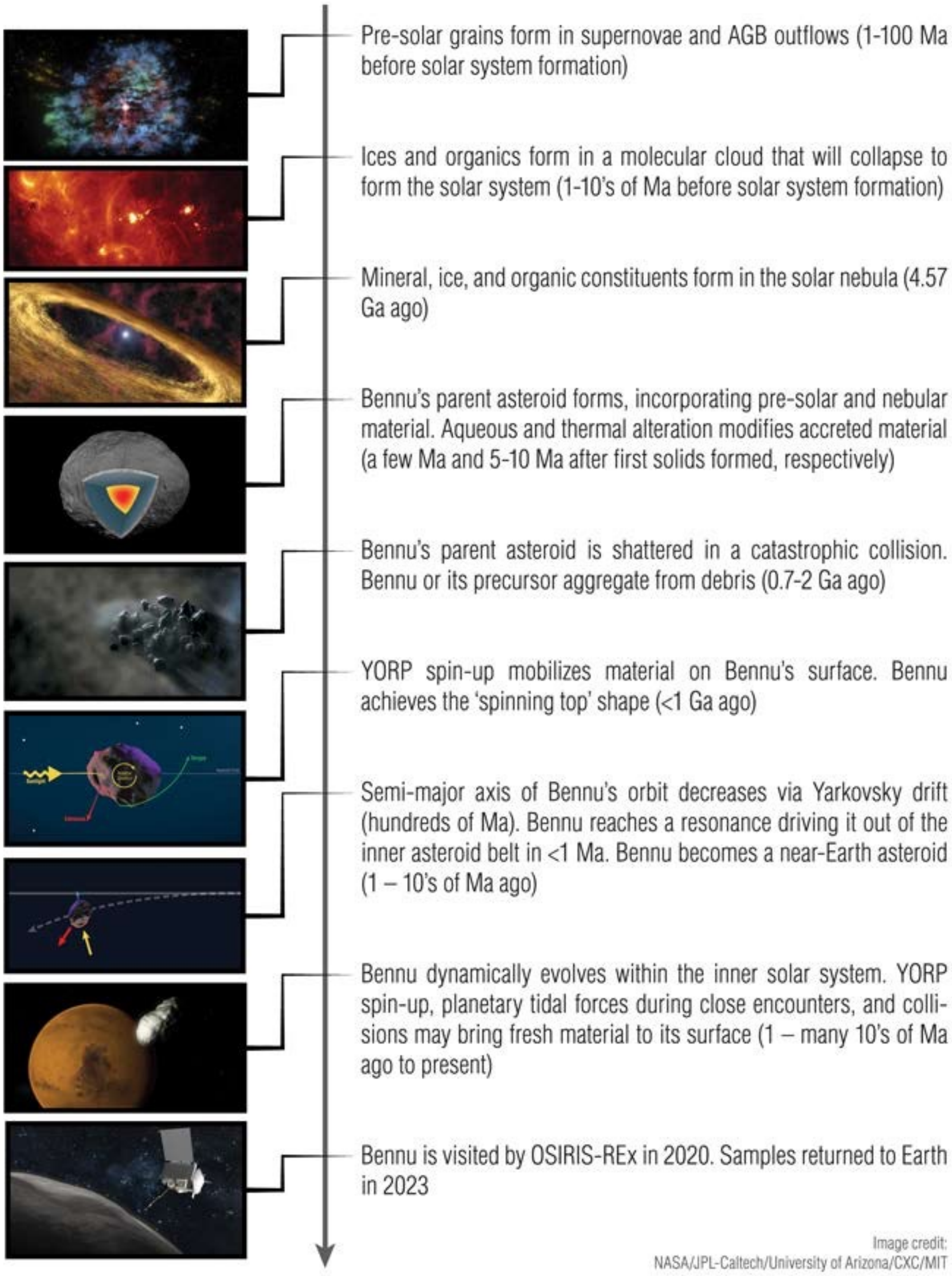
Pre-solar grains form in supernovae and AGB outflows (1-100 Ma before solar system formation)
Ices and organics form in a molecular cloud that will collapse to form the solar system (1-10's of Ma before solar system formation)
Mineral, ice, and organic constituents form in the solar nebula (4.57 Ga ago)
Bennu's parent asteroid forms, incorporating pre-solar and nebular material. Aqueous and thermal alteration modifies accreted material (a few Ma and 5-10 Ma after first solids formed, respectively)
Bennu's parent asteroid is shattered in a catastrophic collision. Bennu or its precursor aggregate from debris (0.7-2 Ga ago)
YORP spin-up mobilizes material on Bennu's surface. Bennu achieves the 'spinning top' shape (<1 Ga ago)
Semi-major axis of Bennu's orbit decreases via Yarkovsky drift (hundreds of Ma). Bennu reaches a resonance driving it out of the inner asteroid belt in <1 Ma. Bennu becomes a near-Earth asteroid (1 – 10's of Ma ago)
Bennu dynamically evolves within the inner solar system. YORP spin-up, planetary tidal forces during close encounters, and collisions may bring fresh material to its surface (1 – many 10's of Ma ago to present)
Bennu is visited by OSIRIS-REx in 2020. Samples returned to Earth in 2023
Image credit:
NASA/JPL-Caltech/University of Arizona/CXC/MIT